\documentclass[aps,pra,twocolumn,nofootinbib,floatfix,longbibliography,superscriptaddress]{revtex4-2}

\usepackage{amsmath,amssymb,amsthm,bbm,bm,color,dsfont,float,graphicx,hyperref,makecell,mathrsfs,mathtools,nicefrac,pgfplots,physics,tikz,times,txfonts}
\usepackage{comment}
\usepackage[normalem]{ulem}  

\definecolor{equationcolor}{RGB}{222,94,100}
\definecolor{alecolor}{RGB}{238,33,80}

\hypersetup{urlcolor=equationcolor, colorlinks=true,citecolor=equationcolor,linkcolor=equationcolor} 
\pgfplotsset{compat=1.18} 

\DeclareFontFamily{U}{mathb}{\hyphenchar\font45}
\DeclareFontShape{U}{mathb}{m}{n}{
	<-6> mathb5 <6-7> mathb6 <7-8> mathb7
	<8-9> mathb8 <9-10> mathb9
	<10-12> mathb10 <12-> mathb12
}{}
\DeclareSymbolFont{mathb}{U}{mathb}{m}{n}
\DeclareMathSymbol{\ggcurly}{\mathrel}{mathb}{"CF}

\makeatletter
\def\blfootnote{\gdef\@thefnmark{}\@footnotetext}
\makeatother

\theoremstyle{plain}

\def\>{\rangle}
\def\<{\langle}

\def\cblue{\color{blue}}
\def\cred{\color{red}}
\def\blk{\color{black}}

\newlength\myindent
\begin{document}
%% ==== Affiliations =====  
\newcommand{\iftia}{Institute of Theoretical Physics and Astrophysics, Faculty of Mathematics,
Physics and Informatics, University of Gda\'nsk, 80-308 Gda\'nsk, Poland}
\newcommand{\lanl}{Theoretical Division (T-4), Los Alamos National Laboratory, Los Alamos, New Mexico 87545, USA.} 
\newcommand{\ictqt}{International Centre for Theory of Quantum Technologies, University of Gda\'nsk,
Jana Bażyńskiego 1A, 80-309 Gda\'nsk, Poland}

\author{Marcin {\L}obejko}
\email{marcin.lobejko@ug.edu.pl; \textcolor{black}{M.\L. and T.B. contributed equally.}}
\affiliation{\iftia}
\author{Tanmoy Biswas}
\email{tanmoy.biswas23@lanl.gov}
\affiliation{\lanl}
\author{Micha{\l} Horodecki}
\email{michal.horodecki@ug.edu.pl}
\affiliation{\ictqt}
\title{
%Quantum advantage in work extraction due to steerability of quantum correlations
%\\
Equivalence of Discrete and Continuous Otto-Like Engines assisted by Catalysts: \\Mapping Catalytic  Advantages from the Discrete to the Continuous Framework
%\\or\\
%\cred The equivalence between discrete and continuous \\  Otto-like heat engines assisted by the catalyst \blk
%\\
%Quantum Thermodynamic Advantage from Steerable Quantum Correlations
}

\date{\today}
\begin{abstract}
The catalytic extension of a discrete two-stroke engine employs a cyclic auxiliary system—the catalyst—that remains decoupled from the baths and performs no work, yet enhances power and efficiency beyond the corresponding non-catalytic counterpart. Theoretical models of discrete engines are relatively easy to analyze but remain challenging for experimental implementation due to the required control over individual strokes. In contrast, externally driven engines that are simultaneously coupled to both heat baths — the so-called continuous engines — are more experimentally feasible. Here, we establish an equivalence between discrete and continuous machines, both with and without a catalyst, by mapping the discrete unitary processes and thermalization steps onto an interaction Hamiltonian and a Markovian model of dissipation. As a result, by replacing probability flows with probability currents, we construct an analogous continuous machine corresponding to previously demonstrated catalytic schemes that generalize Otto engines. We illustrate this mapping for the simplest catalytic extension of the Otto engine, demonstrating catalytic enhancement in the continuous regime.
\end{abstract}

\maketitle
\begin{figure*}[htbp]
    \centering
    \includegraphics[width=14cm]{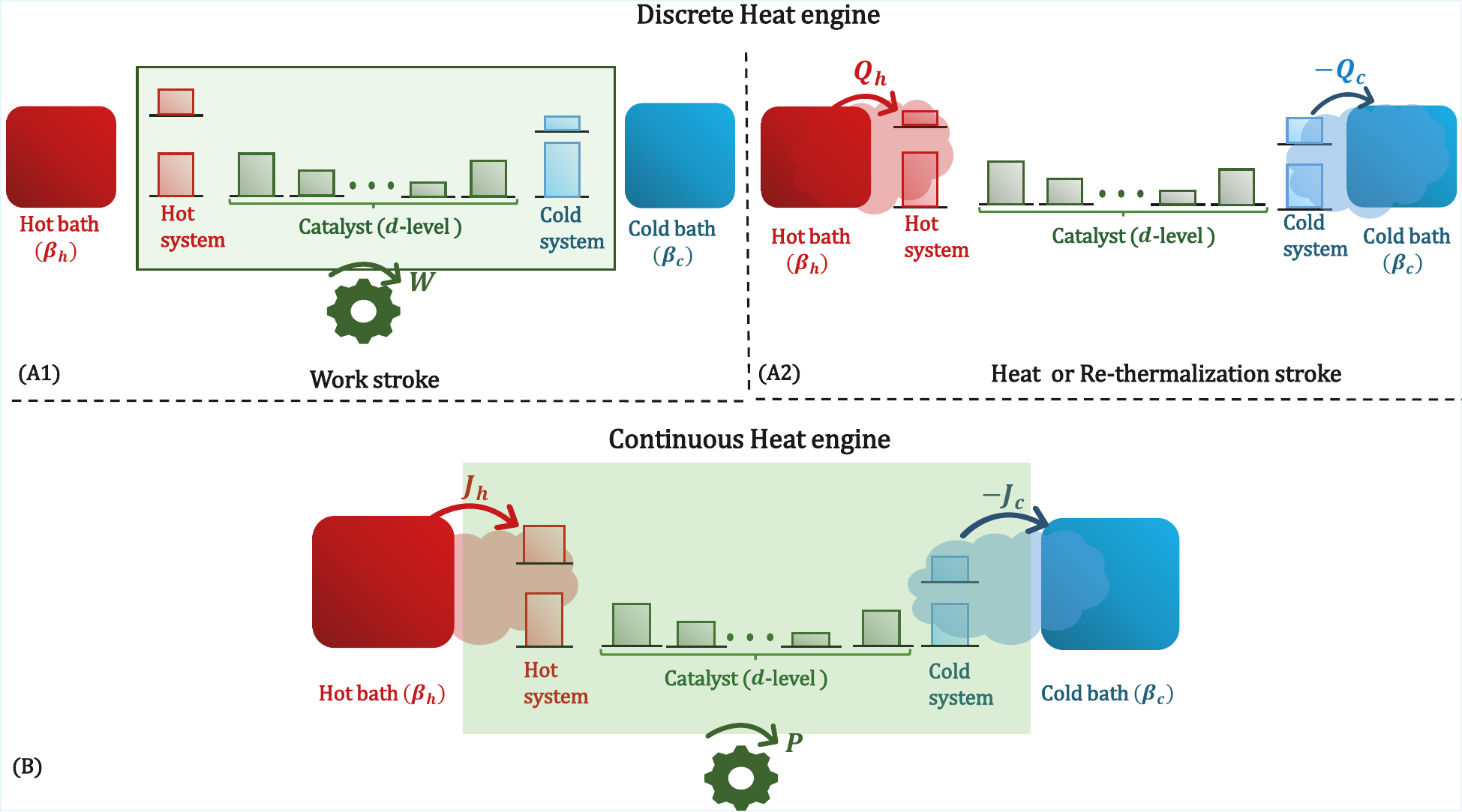}
    \caption{\label{Fig:Discrete_vs_continuous}
    The upper panel illustrates a discrete Otto-like heat engine with a qudit catalyst. The left subpanel (A1) shows the work stroke, where the engine extracts work, while the right subpanel (A2) depicts the heat stroke, during which the qubits rethermalize to their initial temperatures, completing the cycle. The catalyst is chosen so that its marginal state remains unchanged after the work stroke, and correlations with the system are erased during the heat stroke, ensuring the engine is re-initialized. The lower panel (B) shows a continuous Otto-like engine assisted by a qudit catalyst. Unlike the discrete engine, it operates without distinct strokes, remaining continuously coupled to the driving field for work extraction and to the heat baths for thermalization. Dissipative interactions drive the engine toward a stationary state, resulting steady heat currents \(J_h\), \(J_c\), and power \(P\).
  }
\end{figure*}

%\section{Introduction}
\textit{Introduction: }
A microscopic Otto engine can be conceptualized as a machine operating at two distinct frequencies, say $\omega_h$ and $\omega_c$, such that the work extraction process is separated from the thermalization of the energy levels at the corresponding temperatures ($\beta_h$ and $\beta_c$) \cite{Scovil1959,Geusic1967,Alicki_1979,KosloffRezek,Zhang_meystre_PRL,Zhang_PRA,XiaoOtto,QuanPRE,Quan2_PRE,deffner_entropy,Kosloff_entropy,QuanPRE,Quan2_PRE,GevaKosloff1992,Feldman_Kosloff,Kosloff_entropy,Feldman_Kosloff2003}. In other words, the Otto engine consists of two types of strokes: the work stroke and the heat stroke. In this scenario, it can be shown that the optimal performance of the engine is given by the Otto efficiency \cite{CampisiPRB,Campisi4,BiswasPRE}: 
\begin{equation}\label{Defn_Otto_efficiency} 
\eta_{\rm Otto} = 1 - \frac{\omega_c}{\omega_h}. \end{equation} 

Recent works \cite{BiswasPRL,BiswasPRE,fu2025extendinglimitedperformancequantum,Henao2021catalytic,HenaoUzdin} have extended the simplest Otto setup by introducing an ancillary system that acts as a \textit{catalyst}, a well-established concept in quantum information and quantum thermodynamics \cite{brandao2015second,Wilming_catalsysis,Shiraishi_sagawa_PRL,CDatta_Review,Bartosik_review}. In the context of thermal machines, a catalyst must satisfy two key conditions: (i) it does not introduce any new energy scale into the machine and can, in principle, have a degenerate spectrum; (ii) it is restored to its original state after the completion of the thermodynamic cycle without thermal intervention, i.e., it remains thermally uncoupled.

Within this generalized\textit{ catalytic Otto-like} framework, it is possible to design protocols that achieve any efficiency value between the Otto efficiency $\eta_{\text{Otto}}$ and the Carnot efficiency $\eta_{\text{Carnot}} = 1 - \beta_h/ \beta_c$, which constitutes a \emph{catalytic advantage} in efficiency \cite{BiswasPRL,BiswasPRE}. In particular, with the aid of a catalyst, we can design a protocol whose efficiency represents a generalization of the Otto formula, given by  
\begin{equation}
    \eta = 1 + \frac{\sum_i \Delta \varepsilon_i^c}{\sum_i \Delta \varepsilon_i^h}.
\end{equation}
As for the standard Otto scenario, $\eta$ depends solely on the ratio of the energy differences $\Delta \varepsilon_i^h$ and $\Delta \varepsilon_i^c$ associated with degrees of freedom coupled to the two distinct heat baths. If only a single energy difference is associated with each heat bath, the expression for $\eta$ naturally reduces to the Otto efficiency $\eta_{\text{Otto}}$. %When the efficiency $\eta$ of the 

Previous studies on the Otto-like engines, including catalytic advantages, were derived using the simplest theoretical model: an engine operating in discrete strokes that alternate between work extraction and thermalization \cite{Kosloff1984,KosloffLevy2014,MitchisonContemp,two_stroke_Allahverdyan,HenaoUzdin,KosloffRezek,AntonioGuzmán_2024}. While this discrete framework is highly convenient for analytical treatment \cite{SilvaPRE1,SilvaPRL,SilvaPRE3,Lobejko2020thermodynamicsof,Biswas2022extractionof,BiswasDatta}, it is not necessarily the most natural choice for experimental implementation: discrete engines require \emph{external control} to coordinate the sequence of strokes, which can be technically demanding \cite{DannKosloff_2023,PopescuSmallestHeatEngine2010,Popescu_PRL}. In contrast, a more experimentally realistic setup involves engines that are continuously coupled to both heat baths (for thermalization) and a driving field (for work extraction), commonly referred to as \emph{continuous} engines (see, e.g., \cite{Alicki_1979,Klimkovsky2013,KosloffLevy2014,cangemni_Levy_engines,Popescu_PRL,PopescuSmallestHeatEngine2010,PopescuSmallestHeatEngineprinciple2010,SilvaPRL,SilvaPRE1}). Since these engines do not involve discrete strokes, they can operate without external control to manage the sequence of operations, making them more practical and easier to implement in a laboratory.

This raises an important question: To what extent do the results obtained for discrete models of Otto-like engines remain valid for continuous engines that are more realistic? In particular, can such models accurately predict the same efficiency, and how do the two approaches compare in terms of power/work output? 
Addressing this question naturally leads to an important follow-up: to what extent do the catalytic advantages observed in the discrete Otto-like scenario translate to the continuous case? 

These questions point to a broader issue in the field. Much of the theoretical work has concentrated on recovering the dynamics of a continuous engine as the limiting case of a discrete engine, achieved by taking the cycle duration to be infinitesimally small. Most of these studies have focused on Otto engines (without a catalyst) \cite{UzdinLevyKosloff2015,Dechiara_2018,Molitor_Landi,Melo_Landi,DannKosloff_2023,Campbell_2021,Strasberg,Bettman,Piccione2021,Saha_2024,Cusumano_2024,CICCARELLO20221}. This approach inherently relies on discrete engines where the work extracted per cycle vanishes in the limit of infinitesimally small cycle time . Although theoretically motivated, such discrete engines again have limited practical relevance because the work extracted per cycle vanishes in the limit of infinitesimal cycle duration.

In this Letter, rather than taking the continuous limit of a discrete engine, we introduce a continuous analog by implementing weak coupling to external baths along with an appropriate interaction Hamiltonian. This allows us to establish a direct correspondence between key quantities in the discrete and continuous frameworks, where the mapping is simply realized by replacing changes in probability flow with the dynamical probability current. Within this unified picture, we analyze efficiency and power in both paradigms. Notably, for catalytic Otto-like engines, the efficiency remains invariant under this mapping, showing that the catalytic enhancement in the discrete engine carries over to its continuous counterpart. Furthermore, the connection between discrete and continuous Otto-like engines can be captured by a single characteristic time constant—interpreted as the cycle duration—which links the power of a continuous engine (work per unit time) to that of a discrete engine (work per cycle). Overall, this mapping highlights how discrete models provide valuable insights for understanding and optimizing efficiency and power in continuous thermal machines.

\textit{Discrete engine.} We start with a mathematical formulation of the discrete engine. We consider a quantum system defined by a density matrix:  
\begin{equation}
    \rho = \sum_k p_k \dyad{k}
\end{equation}
and Hamiltonian, 
\begin{equation}
    H_0 = H_{0,h} + H_{0,c},
\end{equation}
where we split the degrees of freedom into so-called \emph{hot} and \emph{cold} ones. We consider then a swap process (permutation), such that some of the states $\ket k$  are transposed with each other.  In accordance, we swap states $\ket{u_i}$ with $\ket{d_i}$ (i.e., $\ket{u_i} \leftrightarrow \ket{d_i}$), while remaining states $\ket{r_i}$ remain unchanged. The process is realized via a permutation $S$, such that   
\begin{equation} \label{permutation}
    S\ket{u_i} = \ket{d_i}, \quad S\ket{d_i} = \ket{u_i}, \quad S\ket{r_i} = \ket{r_i}.
\end{equation}
%not current
Then, we define the heat as follows (for $k = h, c$):
\begin{equation}
    Q_k = \Tr[H_{0,k} (\rho - S \rho S^\dag)],
\end{equation}
together with work as a total change in the average energy of the state $\rho$, i.e. 
\begin{equation} \label{work_disc}
    W = \Tr[H_{0} (\rho - S \rho S^\dag)] =\Tr[\left(H_{0}-S^\dag H_0 S\right) \rho ]= Q_h + Q_c,
\end{equation}
which defines the first law of thermodynamics. One can easily show that 
\begin{equation} \label{heat_disc}
    Q_k = \sum_i \Delta \varepsilon_i^k \Delta p_i,
\end{equation}
where 
\begin{align}\label{prob_difference}
    &\Delta p_i = p_{u_i} - p_{d_i}, \quad \Delta \varepsilon_i^k = \varepsilon_{u_i}^k - \varepsilon_{d_i}^k \\
    &\varepsilon_{u_i}^k = \bra{u_i} H_{0,k} \ket{u_i}, \quad \varepsilon_{d_i}^k = \bra{d_i} H_{0,k} \ket{d_i}.
\end{align}
Efficiency of the process is defined as: 
\begin{equation} \label{efficiency_disc}
    \eta_{\text{disc.}} = \frac{W}{Q_h} = 1 + \frac{Q_c}{Q_h} = 1 + \frac{\sum_i \Delta \varepsilon_i^c \Delta p_i}{\sum_i \Delta \varepsilon_i^c \Delta p_i}.
\end{equation}
We refer $\Delta p_i$ as \emph{probability flow}. The following model is purely mathematical, where the \emph{heat} and \emph{work} are not yet physically justified. Nevertheless, the model can function as a discrete thermal machine, provided a dissipative process is executed to restore the system to its initial state and thereby close the thermodynamic cycle. 

In particular, the notion of work and heat naturally emerges, where the engine's state consists of two thermal states at different temperatures, i.e. when
\begin{eqnarray} \label{gibbs_state}
    \varrho_h \otimes \varrho_c = \frac{e^{-\beta_h \omega_h}}{\Tr[e^{-\beta_h \omega_h}]} \otimes \frac{e^{-\beta_c \omega_c}}{\Tr[e^{-\beta_c \omega_c}]}.
\end{eqnarray}
In this case, by thermalizing hot (cold) degrees of freedom, we provide from the corresponding heat bath an energy $Q_h$ ($Q_c$), interpreted commonly as heat. For the non-equilibrium thermal state given by Eq. \eqref{gibbs_state}, one can prove the following Clausius inequality:
\begin{eqnarray}
    \beta_h Q_h + \beta_c Q_c \le 0,
\end{eqnarray}
which ensures the second law and upper bounds efficiency by the Carnot limit. 

\textit{Continuous engine.} Next, we move to the mathematical framework of the continuous engine. Here, we consider a general model with a total Hamiltonian $H_0 + V_t$ given by
\begin{equation}
    H_0 = H_{0,h} + H_{0,c}\quad ;\quad
    V_t = \sum_i g_i\left(\dyad{u_{i}}{d_{i}} e^{- i \Omega_it} + h.c. \right) \label{interaction_term}
\end{equation}
where the states $\ket{u_i}$ and $\ket{d_i}$ are eigenstates of $H_0$, i.e.,
\begin{equation}
    H_{0,k} \ket{u_i} = \varepsilon_{u_i}^k \ket{u_i}, \quad H_{0,k} \ket{d_i} = \varepsilon_{d_i}^k \ket{d_i}.
\end{equation}
for $k = h,c$. We consider a resonant driving such that
\begin{align} \label{resonance_condition}
     \Omega_i = \Delta \varepsilon_{i}^h + \Delta \varepsilon_{i}^c \quad\text{where}\quad  \Delta \varepsilon_{i}^k = \varepsilon_{u_i}^k - \varepsilon_{d_i}^k. 
\end{align}
In the interaction picture, system evolves under the local master equation:
\begin{equation}
    \dot \rho_I(t) = - i [V_0, \rho_I(t)] + \mathcal{D}_h[\rho_I(t)] + \mathcal{D}_c[\rho_I(t)]:=\mathcal{L}[\rho_I(t)],
\end{equation}
where the ``local'' assumption translates to condition \cite{Levy_2014,Dechiara_2018,Hewgill_PRR}:  
\begin{equation}\label{local_condition}
    \mathcal{D}^\dag_h[H_{0,c}] = \mathcal{D}^\dag_c[H_{0,h}] = 0.
\end{equation}
Generally, the local model of the dissipator is justified when the strength of the free Hamiltonian significantly exceeds the coupling strength, i.e., $\| H_0 \| \gg g_i$. The first law is expressed by the equation:
\begin{eqnarray}
 \dot E(t) = P(t) + J_h(t) +J_c(t)
\end{eqnarray}
where the internal energy $E(t)$, power $P(t)$, and heat currents $J_k(t)$ are given by:
\begin{align}
& E(t) = \Tr[(H_0 + e^{i H_0 t} V_t e^{-i H_0 t}) \rho_I(t) ], \\
&P(t)  = -\Tr[\dot V_0 \rho_I(t)], \quad J_{k}(t) = \Tr[\mathcal{D}_k^\dag[H_{0,k}+V_0] \rho_I(t)],
\end{align}
where we used the condition given in Eq. \eqref{local_condition} and $\dot V_0 =  -i \sum_k g_k \Omega_k (\dyad{u_k}{d_k} - \text{H.C.})$. Here $\text{H.C.}$ denotes the Hermitian conjugate. Furthermore, for the Gibbs states, we assume their stationarity with respect to the corresponding dissipators, i.e.,
\begin{eqnarray}
    \mathcal{D}_h [\varrho_h] = \mathcal{D}_c [\varrho_c] = 0,
\end{eqnarray}
one can derive, based on Spohn's inequality \cite{spohn1978entropy,Landi_Review}, the following ``second law inequality": 
\begin{eqnarray} \label{entropy_production}
    \frac{ d S(t)}{dt} \ge \sum_{k=h,c} \beta_k \left(J_k(t) - \Tr[\mathcal{D}_k^\dag [V_0] \rho_I(t)]\right),
\end{eqnarray} 
where $S(t) = - \Tr[\rho_I(t) \ln \rho_I(t)]$ is the von Neumann entropy. It is seen that entropy production is not generally non-negative here, which is a well-known issue with a local master equation \cite{Levy_2014,Dechiara_2018,Hewgill_PRR,Hofer_2017}. However, we will impose a non-negative entropy production in the non-equilibrium steady state (NESS). 

For a time-independent Liouvillian $\mathcal{L}$, we assume a \emph{stationary state} in the long time limit, i.e., $\rho_I(t) \to \rho_I^s$, which implies $\mathcal{L}[\rho_I^s]=0$. Then, the stationary values of the power and heat currents are given by: 
\begin{eqnarray}
    P = -\langle \dot V_0 \rangle, \quad J_k = \langle \mathcal{D}_k^\dag [H_{0,k}] \rangle + \langle \mathcal{D}_k^\dag [V_0] \rangle,
\end{eqnarray}
where $\langle A \rangle = \Tr[A \rho_I^s]$. Then, the non-negative entropy production in the NESS is provided whenever 
\begin{align}\label{int_vanish}
    \langle \mathcal{D}_h^\dag [V_0] \rangle = \langle \mathcal{D}_c^\dag [V_0] \rangle = 0,
\end{align}
from which we get the Clausius inequality:
\begin{eqnarray} \label{clausius_inequality}
    \beta_h J_h + \beta_c J_c \le 0.
\end{eqnarray}
In principle, the condition \eqref{int_vanish} is satisfied for the examples we demonstrate later. Furthermore, with a definition of a stationary \emph{probability current}:
\begin{equation}\label{der:particle_current}
\langle \dot p_i \rangle = ig_i\Big\langle \dyad{u_{i}}{d_{i}} - \dyad{d_{i}}{u_{i}} \Big\rangle
\end{equation}
we get expressions for the power and heat currents in the stationary state: 
\begin{align} \label{power_heat_currents}
& P = \sum_i  \Omega_i \langle \dot p_i \rangle, \quad  J_k = \sum_i \Delta \varepsilon_{i}^k \langle \dot p_i \rangle.
\end{align}
Together with a resonant condition \eqref{resonance_condition}, this reveals the energy conservation:
\begin{equation} \label{work_cont}
    P = J_h + J_c.
\end{equation}
Finally, the efficiency of the introduced continuous engine is given by: 
\begin{equation} \label{efficiency_cont}
    \eta_{\text{cont.}} = 1 + \frac{\sum_i \Delta \varepsilon_{i}^c \langle \dot p_i \rangle}{\sum_i \Delta \varepsilon_{i}^h \langle \dot p_i \rangle}.
\end{equation}

In a nutshell, the general model of the continuous heat engine, that brings the expression for heat currents \eqref{power_heat_currents} together with the first law \eqref{work_cont} and the second law \eqref{clausius_inequality}, is solely based on the resonant interaction (Eqs. \eqref{interaction_term}, \eqref{resonance_condition}), the local master equation model (Eq. \eqref{local_condition}) and the condition given in Eq. \eqref{int_vanish}.

\textit{Catalysis in discrete and continuous engine}.
To complete the thermodynamic cycle of a conventional heat engine, the working medium must be restored to its initial state through a dissipative process. As an alternative, we introduce an ancillary system---the \textit{catalyst}---that satisfies the \textit{catalysis condition}, namely, its marginal state returns to the initial one after the work stroke (see Fig.~\ref{Fig:Discrete_vs_continuous}). In particular, a discrete engine assisted by a catalyst prepared in the state $\rho_s$ may undergo the following two-stroke process:
\begin{align}
    \rho_s\otimes \rho_h \otimes \rho_c 
    &\xrightarrow{\text{work stroke}} 
    S\!\left( \rho_s\otimes \rho_h \otimes \rho_c \right)\!S^\dagger \nonumber\\
    &\xrightarrow{\text{heat stroke}} 
    \Tr_{h,c}\!\left[ S\!\left( \rho_s\otimes \rho_h \otimes \rho_c \right)\!S^\dagger \right] \otimes \rho_h \otimes \rho_c,
\end{align}
where $S$ denotes the global unitary operation implementing the work stroke. 
Under the \emph{catalytic condition}
\begin{equation} \label{Catalyst_condition}
    \rho_s = \Tr_{h,c}\!\left[ S\!\left( \rho_s\otimes \rho_h \otimes \rho_c \right)\!S^\dagger \right],
\end{equation}
the working system returns to its initial state, thereby completing the thermodynamic cycle.
 
Inclusion of a catalyst results in an effective enlargement of the Hilbert space without introducing any additional energy (i.e., $\varepsilon_{u_i}^k$ or $\varepsilon_{d_i}^k$). Thus, without loss of generality, the catalyst may be introduced as a system with a trivial Hamiltonian. %Thus, this extension does not affect any assumptions that we made in the introduced model. , except for the new modified cyclicity condition. 
More generally, the catalysis condition \eqref{Catalyst_condition} is defined as a preservation of the probabilities defined via projectors $\Pi_m$, namely
\begin{equation}\label{Catalyst_defining_Eq}
   \forall m,\quad \Tr[\Pi_m (\rho - S\rho S^\dag)] = \Tr[\left(\Pi_m-S^\dag\Pi_m S\right)\rho]=0,
\end{equation}
where $\Pi_m$ is a projector onto the subspace of the catalytic degrees of freedom. Then, the catalysis condition can be equivalently expressed as
\begin{align}\label{cyclicity_disc}
    \forall m,\quad &\sum_i \Delta \lambda_i^m \Delta p_i = 0 \quad\text{where}\quad\Delta \lambda_i^m = \lambda_{u_i}^m - \lambda_{d_i}^m, \\\text{and}\quad
    \lambda_{u_i}^m &= \bra{u_i} \Pi_m \ket{u_i}, \quad \lambda_{d_i}^m = \bra{d_i} \Pi_m \ket{d_i},
\end{align}
with $\Delta p_i$ introduced in Eq. \eqref{prob_difference}. 

The analogous condition as Eq. \eqref{Catalyst_defining_Eq} for the continuous engine is given by:
\begin{equation}\label{cyclicity_cond_1}
   \forall m,\quad \langle \mathcal{L}^\dag[\Pi_m]\rangle = i \langle [V_0, \Pi_m]\rangle = 0,
\end{equation}
where we assume that $\mathcal{D}^\dag_k [\Pi_m] = 0$, which follows from the fact that the catalyst’s degrees of freedom are not coupled to the heat baths. Similarly to the previous considerations, for the continuous case, we assume that states $\ket{u_i}$ and $\ket{d_i}$ are eigenstates of the projectors $\Pi_m$, such that
\begin{equation} 
    \Pi_m \ket{u_i} = \lambda_{u_i}^m \ket{u_i}, \quad \Pi_m \ket{d_i} = \lambda_{d_i}^m \ket{d_i},
\end{equation}
where for all $m$, we have eigenvalues $\lambda_{b}^{m}\in\{0,1\}$. Using the formulation $\langle\dot p_i\rangle$ from Eq. \eqref{der:particle_current}, one can rewrite Eq. \eqref{cyclicity_cond_1} as  
\begin{equation} \label{cyclicity_cont}
    \forall m,\quad \sum_i \Delta \lambda_i^m \langle \dot p_i \rangle = 0.
\end{equation}
That results in a \emph{continuous catalysis condition}.

We emphasize that for the discrete engine, the catalysis condition is imposed to complete the cycle for degrees of freedom that do not thermalize. In contrast, for the continuous engine the cyclicity condition in Eq.~\eqref{cyclicity_cont} arises naturally from relaxation, provided the system is ergodic. Since the catalyst is not directly coupled to the heat baths, ergodicity must instead be ensured through the interaction $V_0$. 

\textit{Catalytic Otto-like engines.} Within this general framework, we find a one-to-one correspondence between the performance of discrete and continuous engines. Namely, by comparison side by side the expressions for exchanged amounts of heat and heat currents from Eq. \eqref{heat_disc} and Eq. \eqref{power_heat_currents}, efficiency from Eq. \eqref{efficiency_disc} and Eq. \eqref{efficiency_cont}, catalysis conditions in Eq. \eqref{cyclicity_disc} and from Eq. \eqref{cyclicity_cont}, work and power from Eq. \eqref{work_disc} and Eq. \eqref{work_cont}: 
\begin{align}
    &Q_k = \sum_i \Delta \varepsilon_{i}^k \Delta p_i, \quad\quad\quad\;  &J_k = \sum_i \Delta \varepsilon_{i}^k \langle \dot p_i \rangle,\label{Q_k_and_J_k_main}\\
    &\eta_{\text{disc.}}=1 + \frac{\sum_i \Delta \varepsilon_i^c \Delta p_i}{\sum_i \Delta \varepsilon_i^c \Delta p_i}, \quad &\eta_{\text{cont.}}=1 + \frac{\sum_i \Delta \varepsilon_{i}^c \langle \dot p_i \rangle}{\sum_i \Delta \varepsilon_{i}^h \langle \dot p_i \rangle},\label{etadisc_and_etacont_main}\\
     &\forall m,\quad\sum_i \Delta \lambda_i^m \Delta p_i = 0,\quad&\forall m,\quad \sum_i \Delta \lambda_i^m \langle \dot p_i \rangle = 0\label{catalsyis_disc_and_cond},\\
    &W = Q_h + Q_c, \quad\quad\quad\; &P = J_h + J_c, \label{power_disc_and_cond}
\end{align}
and finally with Clausius' inequality:
\begin{gather}
    \beta_h Q_h + \beta_c Q_c \le 0, \quad \beta_h J_h + \beta_c J_c \le 0,
\end{gather}
it is seen that both models are equivalent by replacing 
\begin{equation}\label{Mapping_of_disc_and_cont}
    \Delta p_i \leftrightarrow  \langle \dot p_i \rangle .
\end{equation}
The mapping is then achieved by introducing a set of characteristic times, such that 
\begin{eqnarray} \label{characterstic_time}
   \tau_i  = \Delta p_i / \langle \dot p_i \rangle.
\end{eqnarray}

\begin{figure}[t]
    \centering
    \includegraphics[width=0.43\textwidth]{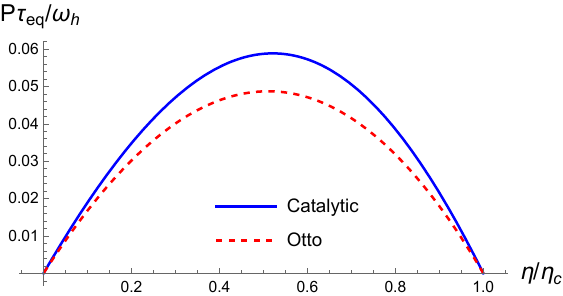}
    \caption{\label{Fig:Power_efficiency_tradeoff}
    The power--efficiency trade-off for the continuous Otto engine and its catalytic extension with a two-dimensional catalyst, showing the full range of catalytic enhancement (cf. Ref. \cite{BiswasPRL}). The engines operate at frequencies $\omega_h$ and $\omega_c$ with inverse temperatures $\beta_h$ and $\beta_c$. Dissipation is characterized by the equilibration time $\tau_{\rm eq}$ (identical for both baths) and driving rate $g$. The characteristic times are $\tau_{\rm Otto} = \tau_{\rm eq}\left(1 + 1/(g\tau_{\rm eq})^2\right)$ and $\tau_{\rm Catalytic} = A\,\tau_{\rm eq}\left(1 + B/(g\tau_{\rm eq})^2\right)$, with $A,B \le 1$ (cf. Eq. \eqref{char_time_ineq}). Parameters: $\beta_h\omega_h=0.1$, $\beta_c/\beta_h=10$, $g\tau_{\rm eq}=10$, and $\eta_c = 1 - \beta_h/\beta_c$.
  }
\end{figure}

In the following, we focus on a special class of catalysis where the work stroke is governed by a particular set of permutations, referred to as \emph{simple permutations}, introduced for discrete Otto-like engines (with a catalyst) in~\cite{BiswasPRL,BiswasPRE}.  It encompasses only those permutations that transfer an equal non-zero amount of probability from one eigenstate of the catalyst to another. Simple permutation is characterized by the unique non-trivial probability flow given by the solution of Eq.~\eqref{cyclicity_disc}:
\begin{eqnarray}\label{simple_per_catalysi}
    \forall\; i,j \quad \Delta p_i = \Delta p_j \equiv \Delta p.
\end{eqnarray}
The primary reason for focusing on such permutations is that, with a suitable catalyst, one can achieve any catalytic enhancement of efficiency, realizing values between the Otto and Carnot limits~\cite{BiswasPRE}. This makes them natural candidates for transferring catalytic advantages to continuous Otto-like engines via the mapping in Eq.~\eqref{Mapping_of_disc_and_cont}.

Analogous to the catalysis condition derived for discrete Otto-like engines in Eq.~\eqref{simple_per_catalysi},  
an equivalent condition for catalysis under simple permutations arises in the continuous setting.  
Specifically, starting from Eq.~\eqref{catalsyis_disc_and_cond}, we obtain  
\begin{align}\label{cat_condition_main}
   \forall\, i,j \quad \langle \dot p_i \rangle = \langle \dot p_j \rangle \equiv \langle \dot p \rangle,
\end{align}
as the equation describing the catalysis condition in the continuous scenario has an identical form to that in the discrete case  (see Eq.~\eqref{catalsyis_disc_and_cond}).

For the class of Otto-like engines defined by Eqs.~\eqref{simple_per_catalysi} and~\eqref{cat_condition_main}, there exists a single characteristic time $\tau$ such that $\langle \dot p \rangle = \Delta p / \tau$. In the discrete case, quantities such as work and heat are evaluated per cycle, while in the continuous case they are expressed as time rates. Hence, $\tau$ represents the cycle duration, linking work per cycle in discrete scenario to power in the continuous scenario via $P = W / \tau$. The corresponding power–efficiency relation for catalytic Otto-like engines is  
\begin{align}\label{Otto_like_defn}
    P = \frac{W}{\tau}, \quad 
    \eta = \frac{W}{Q_h} = \frac{P}{J_h} 
    = 1 + \frac{\sum_i \Delta \varepsilon_i^c}{\sum_i \Delta \varepsilon_i^h}.
\end{align}

We find that the proposed class of heat engines generalizes the Otto formula—expressed as a ratio of energy differences—and therefore refer to them as \emph{catalytic Otto-like engines}.

\textit{Demonstration of the mapping (from discrete to continuous model).}
To illustrate the applicability of our results, we show how the discrete engine can be mapped onto its continuous counterpart. In Refs.~\cite{BiswasPRL,BiswasPRE}, we introduced a class of discrete catalytic Otto-like engines that outperform their non-catalytic analogs. In the discrete setting, these engines are defined by the states $\ket{u_i}$ and $\ket{d_i}$ forming the permutation matrix $S$ in Eq.~\eqref{permutation}. The corresponding continuous model is then obtained by constructing the interaction Hamiltonian $V_t$ in Eq.~\eqref{interaction_term} from these states.

We now map the discrete Otto engine and its simplest catalytic extension with a two-dimensional catalyst~\cite{BiswasPRL,BiswasPRE} onto the continuous regime (see Appendix A and B for details). In the discrete setting, the efficiencies are  
\begin{eqnarray} \label{efficiencies}
\eta_{\rm Otto} = 1-\frac{\omega_c}{\omega_h} < \eta_{\rm Catalytic} = 1-\frac{\omega_c}{2\omega_h}.
\end{eqnarray}  
Furthermore, Ref.~\cite{BiswasPRL} identified a regime where the catalytic engine outperforms the Otto engine in work output for all efficiencies,  
\begin{eqnarray}\label{Work_discrete_ineq}
W_{\rm Otto}(\eta) < W_{\rm Catalytic}(\eta), \quad \eta \in [0, 1-\beta_h/\beta_c],
\end{eqnarray}  
demonstrating an ultimate catalytic enhancement in both work and efficiency. Based on the general result for catalytic Otto-like engines given by Eq. \eqref{Otto_like_defn}, the continuous engine reproduces the same efficiencies \eqref{efficiencies}. Furthermore,  we proved that in this regime the characteristic times satisfy  
\begin{eqnarray}\label{char_time_ineq}
\tau_{\rm Otto} \ge \tau_{\rm Catalytic}.
\end{eqnarray}
(See Sec.~IV~C of the Supplementary Material for detailed proof). Consequently, within the regime where inequality in Eq. \eqref{Work_discrete_ineq} holds, the correspondence between the discrete and continuous engines extends from work to power, yielding  
\begin{eqnarray}
P_{\rm Otto}(\eta) < P_{\rm Catalytic}(\eta).
\end{eqnarray}

This demonstrates how the relatively simple design of the discrete engine can be mapped onto its more experimentally accessible continuous version. The resulting power–efficiency trade-off is presented in Fig.~\ref{Fig:Power_efficiency_tradeoff}. 

\textit{Discussion:}
We establish a mapping between discrete Otto and catalytic Otto-like engines with their continuous counterparts, illustrated for the Otto engine and an Otto-like engine with a two-dimensional catalyst. The key element is the probability flow \(\Delta p\) in the discrete engine, corresponding to the probability current \(\langle \dot p \rangle\) in the continuous case. Notably, for a certain class, the efficiency remains unchanged under this mapping, so any catalytic enhancement in the discrete engine carries over directly. Moreover, all thermodynamically relevant quantities in the discrete setting have analogs in the continuous case, preserving the overall thermodynamic structure.

Our results open several promising directions for future research. An important one is the \emph{experimental demonstration} of the catalytic advantage on platforms such as trapped ions~\cite{Singer} and superconducting qubits~\cite{Aamir2025}. Another is extending the mapping to a \emph{general discrete Otto engine} with a \(d\)-dimensional catalyst operating at efficiency \(1 - (n \omega_c / d \omega_h)\), \(n,d \in \mathbb{N}\), and its continuous analog~\cite{BiswasPRE}. It would also be valuable to quantify the cost of catalytic enhancement in terms of cycle duration and control complexity, following the thermodynamic cost analysis in~\cite{Landauer_vs_Nernst}. Finally, exploring how quantum resources such as entanglement and coherence transform under this mapping could clarify their role in enhancing the performance of thermal machines \cite{Levy2018,Latune2021,Manzano,BiswasDattaPRL}.

\textit{Acknowledgment:} T.B acknowledges Luis Pedro Garcia-Pintos and Gabriel T. Landi for insightful discussions and comments.  The part of the work done at Los Alamos National Laboratory (LANL) was carried out under the auspices of the U.S. Department
of Energy and National Nuclear Security Administration under Contract No.~DEAC52-06NA25396. TB also acknowledges support by the Department of Energy Office of Science, Office of Advanced Scientific Computing Research, Accelerated Research for Quantum Computing program, Fundamental Algorithmic Research for Quantum Computing (FAR-QC) project. M.H. acknowledges the support by the Polish National Science Centre grant OPUS-21 (No: 2021/41/B/ST2/03207).

\bibliography{main}

\begin{thebibliography}{66}%
\makeatletter
\providecommand \@ifxundefined [1]{%
 \@ifx{#1\undefined}
}%
\providecommand \@ifnum [1]{%
 \ifnum #1\expandafter \@firstoftwo
 \else \expandafter \@secondoftwo
 \fi
}%
\providecommand \@ifx [1]{%
 \ifx #1\expandafter \@firstoftwo
 \else \expandafter \@secondoftwo
 \fi
}%
\providecommand \natexlab [1]{#1}%
\providecommand \enquote  [1]{``#1''}%
\providecommand \bibnamefont  [1]{#1}%
\providecommand \bibfnamefont [1]{#1}%
\providecommand \citenamefont [1]{#1}%
\providecommand \href@noop [0]{\@secondoftwo}%
\providecommand \href [0]{\begingroup \@sanitize@url \@href}%
\providecommand \@href[1]{\@@startlink{#1}\@@href}%
\providecommand \@@href[1]{\endgroup#1\@@endlink}%
\providecommand \@sanitize@url [0]{\catcode `\\12\catcode `\$12\catcode
  `\&12\catcode `\#12\catcode `\^12\catcode `\_12\catcode `\%12\relax}%
\providecommand \@@startlink[1]{}%
\providecommand \@@endlink[0]{}%
\providecommand \url  [0]{\begingroup\@sanitize@url \@url }%
\providecommand \@url [1]{\endgroup\@href {#1}{\urlprefix }}%
\providecommand \urlprefix  [0]{URL }%
\providecommand \Eprint [0]{\href }%
\providecommand \doibase [0]{https://doi.org/}%
\providecommand \selectlanguage [0]{\@gobble}%
\providecommand \bibinfo  [0]{\@secondoftwo}%
\providecommand \bibfield  [0]{\@secondoftwo}%
\providecommand \translation [1]{[#1]}%
\providecommand \BibitemOpen [0]{}%
\providecommand \bibitemStop [0]{}%
\providecommand \bibitemNoStop [0]{.\EOS\space}%
\providecommand \EOS [0]{\spacefactor3000\relax}%
\providecommand \BibitemShut  [1]{\csname bibitem#1\endcsname}%
\let\auto@bib@innerbib\@empty
%</preamble>
\bibitem [{\citenamefont {Scovil}\ and\ \citenamefont
  {Schulz-DuBois}(1959)}]{Scovil1959}%
  \BibitemOpen
  \bibfield  {author} {\bibinfo {author} {\bibfnamefont {H.~E.~D.}\
  \bibnamefont {Scovil}}\ and\ \bibinfo {author} {\bibfnamefont {E.~O.}\
  \bibnamefont {Schulz-DuBois}},\ }\bibfield  {title} {\bibinfo {title}
  {Three-level masers as heat engines},\ }\href
  {https://doi.org/10.1103/PhysRevLett.2.262} {\bibfield  {journal} {\bibinfo
  {journal} {Phys. Rev. Lett.}\ }\textbf {\bibinfo {volume} {2}},\ \bibinfo
  {pages} {262} (\bibinfo {year} {1959})}\BibitemShut {NoStop}%
\bibitem [{\citenamefont {Geusic}\ \emph {et~al.}(1967)\citenamefont {Geusic},
  \citenamefont {Schulz-DuBios},\ and\ \citenamefont {Scovil}}]{Geusic1967}%
  \BibitemOpen
  \bibfield  {author} {\bibinfo {author} {\bibfnamefont {J.~E.}\ \bibnamefont
  {Geusic}}, \bibinfo {author} {\bibfnamefont {E.~O.}\ \bibnamefont
  {Schulz-DuBios}},\ and\ \bibinfo {author} {\bibfnamefont {H.~E.~D.}\
  \bibnamefont {Scovil}},\ }\bibfield  {title} {\bibinfo {title} {Quantum
  equivalent of the carnot cycle},\ }\href
  {https://doi.org/10.1103/PhysRev.156.343} {\bibfield  {journal} {\bibinfo
  {journal} {Phys. Rev.}\ }\textbf {\bibinfo {volume} {156}},\ \bibinfo {pages}
  {343} (\bibinfo {year} {1967})}\BibitemShut {NoStop}%
\bibitem [{\citenamefont {Alicki}(1979)}]{Alicki_1979}%
  \BibitemOpen
  \bibfield  {author} {\bibinfo {author} {\bibfnamefont {R.}~\bibnamefont
  {Alicki}},\ }\bibfield  {title} {\bibinfo {title} {The quantum open system as
  a model of the heat engine},\ }\href
  {https://doi.org/10.1088/0305-4470/12/5/007} {\bibfield  {journal} {\bibinfo
  {journal} {J. Phys. A: Math. Gen.}\ }\textbf {\bibinfo {volume} {12}},\
  \bibinfo {pages} {L103} (\bibinfo {year} {1979})}\BibitemShut {NoStop}%
\bibitem [{\citenamefont {Kosloff}\ and\ \citenamefont
  {Rezek}(2017)}]{KosloffRezek}%
  \BibitemOpen
  \bibfield  {author} {\bibinfo {author} {\bibfnamefont {R.}~\bibnamefont
  {Kosloff}}\ and\ \bibinfo {author} {\bibfnamefont {Y.}~\bibnamefont
  {Rezek}},\ }\bibfield  {title} {\bibinfo {title} {The quantum harmonic otto
  cycle},\ }\bibfield  {journal} {\bibinfo  {journal} {Entropy}\ }\textbf
  {\bibinfo {volume} {19}},\ \href {https://doi.org/10.3390/e19040136}
  {10.3390/e19040136} (\bibinfo {year} {2017})\BibitemShut {NoStop}%
\bibitem [{\citenamefont {Zhang}\ \emph
  {et~al.}(2014{\natexlab{a}})\citenamefont {Zhang}, \citenamefont {Bariani},\
  and\ \citenamefont {Meystre}}]{Zhang_meystre_PRL}%
  \BibitemOpen
  \bibfield  {author} {\bibinfo {author} {\bibfnamefont {K.}~\bibnamefont
  {Zhang}}, \bibinfo {author} {\bibfnamefont {F.}~\bibnamefont {Bariani}},\
  and\ \bibinfo {author} {\bibfnamefont {P.}~\bibnamefont {Meystre}},\
  }\bibfield  {title} {\bibinfo {title} {Quantum optomechanical heat engine},\
  }\href {https://doi.org/10.1103/PhysRevLett.112.150602} {\bibfield  {journal}
  {\bibinfo  {journal} {Phys. Rev. Lett.}\ }\textbf {\bibinfo {volume} {112}},\
  \bibinfo {pages} {150602} (\bibinfo {year} {2014}{\natexlab{a}})}\BibitemShut
  {NoStop}%
\bibitem [{\citenamefont {Zhang}\ \emph
  {et~al.}(2014{\natexlab{b}})\citenamefont {Zhang}, \citenamefont {Bariani},\
  and\ \citenamefont {Meystre}}]{Zhang_PRA}%
  \BibitemOpen
  \bibfield  {author} {\bibinfo {author} {\bibfnamefont {K.}~\bibnamefont
  {Zhang}}, \bibinfo {author} {\bibfnamefont {F.}~\bibnamefont {Bariani}},\
  and\ \bibinfo {author} {\bibfnamefont {P.}~\bibnamefont {Meystre}},\
  }\bibfield  {title} {\bibinfo {title} {Theory of an optomechanical quantum
  heat engine},\ }\href {https://doi.org/10.1103/PhysRevA.90.023819} {\bibfield
   {journal} {\bibinfo  {journal} {Phys. Rev. A}\ }\textbf {\bibinfo {volume}
  {90}},\ \bibinfo {pages} {023819} (\bibinfo {year}
  {2014}{\natexlab{b}})}\BibitemShut {NoStop}%
\bibitem [{\citenamefont {Xiao}\ \emph {et~al.}(2023)\citenamefont {Xiao},
  \citenamefont {Liu}, \citenamefont {He}, \citenamefont {Ma}, \citenamefont
  {Wu},\ and\ \citenamefont {Wang}}]{XiaoOtto}%
  \BibitemOpen
  \bibfield  {author} {\bibinfo {author} {\bibfnamefont {Y.}~\bibnamefont
  {Xiao}}, \bibinfo {author} {\bibfnamefont {D.}~\bibnamefont {Liu}}, \bibinfo
  {author} {\bibfnamefont {J.}~\bibnamefont {He}}, \bibinfo {author}
  {\bibfnamefont {Y.}~\bibnamefont {Ma}}, \bibinfo {author} {\bibfnamefont
  {Z.}~\bibnamefont {Wu}},\ and\ \bibinfo {author} {\bibfnamefont
  {J.}~\bibnamefont {Wang}},\ }\bibfield  {title} {\bibinfo {title} {Quantum
  otto engine with quantum correlations},\ }\href
  {https://doi.org/10.1103/PhysRevA.108.042614} {\bibfield  {journal} {\bibinfo
   {journal} {Phys. Rev. A}\ }\textbf {\bibinfo {volume} {108}},\ \bibinfo
  {pages} {042614} (\bibinfo {year} {2023})}\BibitemShut {NoStop}%
\bibitem [{\citenamefont {Quan}\ \emph {et~al.}(2007)\citenamefont {Quan},
  \citenamefont {Liu}, \citenamefont {Sun},\ and\ \citenamefont
  {Nori}}]{QuanPRE}%
  \BibitemOpen
  \bibfield  {author} {\bibinfo {author} {\bibfnamefont {H.~T.}\ \bibnamefont
  {Quan}}, \bibinfo {author} {\bibfnamefont {Y.-x.}\ \bibnamefont {Liu}},
  \bibinfo {author} {\bibfnamefont {C.~P.}\ \bibnamefont {Sun}},\ and\ \bibinfo
  {author} {\bibfnamefont {F.}~\bibnamefont {Nori}},\ }\bibfield  {title}
  {\bibinfo {title} {Quantum thermodynamic cycles and quantum heat engines},\
  }\href {https://doi.org/10.1103/PhysRevE.76.031105} {\bibfield  {journal}
  {\bibinfo  {journal} {Phys. Rev. E}\ }\textbf {\bibinfo {volume} {76}},\
  \bibinfo {pages} {031105} (\bibinfo {year} {2007})}\BibitemShut {NoStop}%
\bibitem [{\citenamefont {Quan}(2009)}]{Quan2_PRE}%
  \BibitemOpen
  \bibfield  {author} {\bibinfo {author} {\bibfnamefont {H.~T.}\ \bibnamefont
  {Quan}},\ }\bibfield  {title} {\bibinfo {title} {Quantum thermodynamic cycles
  and quantum heat engines. ii.},\ }\href
  {https://doi.org/10.1103/PhysRevE.79.041129} {\bibfield  {journal} {\bibinfo
  {journal} {Phys. Rev. E}\ }\textbf {\bibinfo {volume} {79}},\ \bibinfo
  {pages} {041129} (\bibinfo {year} {2009})}\BibitemShut {NoStop}%
\bibitem [{\citenamefont {Deffner}(2018)}]{deffner_entropy}%
  \BibitemOpen
  \bibfield  {author} {\bibinfo {author} {\bibfnamefont {S.}~\bibnamefont
  {Deffner}},\ }\bibfield  {title} {\bibinfo {title} {Efficiency of harmonic
  quantum otto engines at maximal power},\ }\bibfield  {journal} {\bibinfo
  {journal} {Entropy}\ }\textbf {\bibinfo {volume} {20}},\ \href
  {https://doi.org/10.3390/e20110875} {10.3390/e20110875} (\bibinfo {year}
  {2018})\BibitemShut {NoStop}%
\bibitem [{\citenamefont {Kosloff}(2013)}]{Kosloff_entropy}%
  \BibitemOpen
  \bibfield  {author} {\bibinfo {author} {\bibfnamefont {R.}~\bibnamefont
  {Kosloff}},\ }\bibfield  {title} {\bibinfo {title} {Quantum thermodynamics: A
  dynamical viewpoint},\ }\href {https://doi.org/10.3390/e15062100} {\bibfield
  {journal} {\bibinfo  {journal} {Entropy}\ }\textbf {\bibinfo {volume} {15}},\
  \bibinfo {pages} {2100} (\bibinfo {year} {2013})}\BibitemShut {NoStop}%
\bibitem [{\citenamefont {Geva}\ and\ \citenamefont
  {Kosloff}(1992)}]{GevaKosloff1992}%
  \BibitemOpen
  \bibfield  {author} {\bibinfo {author} {\bibfnamefont {E.}~\bibnamefont
  {Geva}}\ and\ \bibinfo {author} {\bibfnamefont {R.}~\bibnamefont {Kosloff}},\
  }\bibfield  {title} {\bibinfo {title} {On the classical limit of quantum
  thermodynamics in finite time},\ }\href {https://doi.org/10.1063/1.463909}
  {\bibfield  {journal} {\bibinfo  {journal} {J. Chem. Phys.}\ }\textbf
  {\bibinfo {volume} {97}},\ \bibinfo {pages} {4398} (\bibinfo {year}
  {1992})}\BibitemShut {NoStop}%
\bibitem [{\citenamefont {Feldmann}\ and\ \citenamefont
  {Kosloff}(2004)}]{Feldman_Kosloff}%
  \BibitemOpen
  \bibfield  {author} {\bibinfo {author} {\bibfnamefont {T.}~\bibnamefont
  {Feldmann}}\ and\ \bibinfo {author} {\bibfnamefont {R.}~\bibnamefont
  {Kosloff}},\ }\bibfield  {title} {\bibinfo {title} {Characteristics of the
  limit cycle of a reciprocating quantum heat engine},\ }\href
  {https://doi.org/10.1103/PhysRevE.70.046110} {\bibfield  {journal} {\bibinfo
  {journal} {Phys. Rev. E}\ }\textbf {\bibinfo {volume} {70}},\ \bibinfo
  {pages} {046110} (\bibinfo {year} {2004})}\BibitemShut {NoStop}%
\bibitem [{\citenamefont {Feldmann}\ and\ \citenamefont
  {Kosloff}(2003)}]{Feldman_Kosloff2003}%
  \BibitemOpen
  \bibfield  {author} {\bibinfo {author} {\bibfnamefont {T.}~\bibnamefont
  {Feldmann}}\ and\ \bibinfo {author} {\bibfnamefont {R.}~\bibnamefont
  {Kosloff}},\ }\bibfield  {title} {\bibinfo {title} {Quantum four-stroke heat
  engine: Thermodynamic observables in a model with intrinsic friction},\
  }\href {https://doi.org/10.1103/PhysRevE.68.016101} {\bibfield  {journal}
  {\bibinfo  {journal} {Phys. Rev. E}\ }\textbf {\bibinfo {volume} {68}},\
  \bibinfo {pages} {016101} (\bibinfo {year} {2003})}\BibitemShut {NoStop}%
\bibitem [{\citenamefont {Solfanelli}\ \emph {et~al.}(2020)\citenamefont
  {Solfanelli}, \citenamefont {Falsetti},\ and\ \citenamefont
  {Campisi}}]{CampisiPRB}%
  \BibitemOpen
  \bibfield  {author} {\bibinfo {author} {\bibfnamefont {A.}~\bibnamefont
  {Solfanelli}}, \bibinfo {author} {\bibfnamefont {M.}~\bibnamefont
  {Falsetti}},\ and\ \bibinfo {author} {\bibfnamefont {M.}~\bibnamefont
  {Campisi}},\ }\bibfield  {title} {\bibinfo {title} {Nonadiabatic single-qubit
  quantum otto engine},\ }\href {https://doi.org/10.1103/PhysRevB.101.054513}
  {\bibfield  {journal} {\bibinfo  {journal} {Phys. Rev. B}\ }\textbf {\bibinfo
  {volume} {101}},\ \bibinfo {pages} {054513} (\bibinfo {year}
  {2020})}\BibitemShut {NoStop}%
\bibitem [{\citenamefont {Campisi}\ \emph {et~al.}(2015)\citenamefont
  {Campisi}, \citenamefont {Pekola},\ and\ \citenamefont {Fazio}}]{Campisi4}%
  \BibitemOpen
  \bibfield  {author} {\bibinfo {author} {\bibfnamefont {M.}~\bibnamefont
  {Campisi}}, \bibinfo {author} {\bibfnamefont {J.}~\bibnamefont {Pekola}},\
  and\ \bibinfo {author} {\bibfnamefont {R.}~\bibnamefont {Fazio}},\ }\bibfield
   {title} {\bibinfo {title} {Nonequilibrium fluctuations in quantum heat
  engines: theory, example, and possible solid state experiments},\ }\href
  {https://doi.org/10.1088/1367-2630/17/3/035012} {\bibfield  {journal}
  {\bibinfo  {journal} {New J. Phys.}\ }\textbf {\bibinfo {volume} {17}},\
  \bibinfo {pages} {035012} (\bibinfo {year} {2015})}\BibitemShut {NoStop}%
\bibitem [{\citenamefont {Biswas}\ \emph {et~al.}(2024)\citenamefont {Biswas},
  \citenamefont {\L{}obejko}, \citenamefont {Mazurek},\ and\ \citenamefont
  {Horodecki}}]{BiswasPRE}%
  \BibitemOpen
  \bibfield  {author} {\bibinfo {author} {\bibfnamefont {T.}~\bibnamefont
  {Biswas}}, \bibinfo {author} {\bibfnamefont {M.}~\bibnamefont {\L{}obejko}},
  \bibinfo {author} {\bibfnamefont {P.}~\bibnamefont {Mazurek}},\ and\ \bibinfo
  {author} {\bibfnamefont {M.}~\bibnamefont {Horodecki}},\ }\bibfield  {title}
  {\bibinfo {title} {Catalytic enhancement in the performance of the
  microscopic two-stroke heat engine},\ }\href
  {https://doi.org/10.1103/PhysRevE.110.044120} {\bibfield  {journal} {\bibinfo
   {journal} {Phys. Rev. E}\ }\textbf {\bibinfo {volume} {110}},\ \bibinfo
  {pages} {044120} (\bibinfo {year} {2024})}\BibitemShut {NoStop}%
\bibitem [{\citenamefont {\L{}obejko}\ \emph {et~al.}(2024)\citenamefont
  {\L{}obejko}, \citenamefont {Biswas}, \citenamefont {Mazurek},\ and\
  \citenamefont {Horodecki}}]{BiswasPRL}%
  \BibitemOpen
  \bibfield  {author} {\bibinfo {author} {\bibfnamefont {M.}~\bibnamefont
  {\L{}obejko}}, \bibinfo {author} {\bibfnamefont {T.}~\bibnamefont {Biswas}},
  \bibinfo {author} {\bibfnamefont {P.}~\bibnamefont {Mazurek}},\ and\ \bibinfo
  {author} {\bibfnamefont {M.}~\bibnamefont {Horodecki}},\ }\bibfield  {title}
  {\bibinfo {title} {Catalytic advantage in otto-like two-stroke quantum
  engines},\ }\href {https://doi.org/10.1103/PhysRevLett.132.260403} {\bibfield
   {journal} {\bibinfo  {journal} {Phys. Rev. Lett.}\ }\textbf {\bibinfo
  {volume} {132}},\ \bibinfo {pages} {260403} (\bibinfo {year}
  {2024})}\BibitemShut {NoStop}%
\bibitem [{\citenamefont {Fu}\ \emph {et~al.}(2025)\citenamefont {Fu},
  \citenamefont {Pan}, \citenamefont {Fan}, \citenamefont {Tang}, \citenamefont
  {Su}, \citenamefont {Lin},\ and\ \citenamefont
  {Chen}}]{fu2025extendinglimitedperformancequantum}%
  \BibitemOpen
  \bibfield  {author} {\bibinfo {author} {\bibfnamefont {C.}~\bibnamefont
  {Fu}}, \bibinfo {author} {\bibfnamefont {O.}~\bibnamefont {Pan}}, \bibinfo
  {author} {\bibfnamefont {Z.}~\bibnamefont {Fan}}, \bibinfo {author}
  {\bibfnamefont {Y.}~\bibnamefont {Tang}}, \bibinfo {author} {\bibfnamefont
  {S.}~\bibnamefont {Su}}, \bibinfo {author} {\bibfnamefont {Y.}~\bibnamefont
  {Lin}},\ and\ \bibinfo {author} {\bibfnamefont {J.}~\bibnamefont {Chen}},\
  }\href {https://arxiv.org/abs/2507.12016} {\bibinfo {title} {Extending the
  limited performance of the quantum refrigerator with catalysts}} (\bibinfo
  {year} {2025}),\ \Eprint {https://arxiv.org/abs/2507.12016} {arXiv:2507.12016
  [quant-ph]} \BibitemShut {NoStop}%
\bibitem [{\citenamefont {Henao}\ and\ \citenamefont
  {Uzdin}(2021)}]{Henao2021catalytic}%
  \BibitemOpen
  \bibfield  {author} {\bibinfo {author} {\bibfnamefont {I.}~\bibnamefont
  {Henao}}\ and\ \bibinfo {author} {\bibfnamefont {R.}~\bibnamefont {Uzdin}},\
  }\bibfield  {title} {\bibinfo {title} {Catalytic transformations with
  finite-size environments: applications to cooling and thermometry},\ }\href
  {https://doi.org/10.22331/q-2021-09-21-547} {\bibfield  {journal} {\bibinfo
  {journal} {{Quantum}}\ }\textbf {\bibinfo {volume} {5}},\ \bibinfo {pages}
  {547} (\bibinfo {year} {2021})}\BibitemShut {NoStop}%
\bibitem [{\citenamefont {Henao}\ and\ \citenamefont
  {Uzdin}(2023)}]{HenaoUzdin}%
  \BibitemOpen
  \bibfield  {author} {\bibinfo {author} {\bibfnamefont {I.}~\bibnamefont
  {Henao}}\ and\ \bibinfo {author} {\bibfnamefont {R.}~\bibnamefont {Uzdin}},\
  }\bibfield  {title} {\bibinfo {title} {Catalytic leverage of correlations and
  mitigation of dissipation in information erasure},\ }\href
  {https://doi.org/10.1103/PhysRevLett.130.020403} {\bibfield  {journal}
  {\bibinfo  {journal} {Phys. Rev. Lett.}\ }\textbf {\bibinfo {volume} {130}},\
  \bibinfo {pages} {020403} (\bibinfo {year} {2023})}\BibitemShut {NoStop}%
\bibitem [{\citenamefont {{Brand\~ao}}\ \emph {et~al.}(2015)\citenamefont
  {{Brand\~ao}}, \citenamefont {{Horodecki}}, \citenamefont {{Ng}},
  \citenamefont {{Oppenheim}},\ and\ \citenamefont
  {{Wehner}}}]{brandao2015second}%
  \BibitemOpen
  \bibfield  {author} {\bibinfo {author} {\bibfnamefont {F.~G.~S.~L.}\
  \bibnamefont {{Brand\~ao}}}, \bibinfo {author} {\bibfnamefont
  {M.}~\bibnamefont {{Horodecki}}}, \bibinfo {author} {\bibfnamefont
  {N.~H.~Y.}\ \bibnamefont {{Ng}}}, \bibinfo {author} {\bibfnamefont
  {J.}~\bibnamefont {{Oppenheim}}},\ and\ \bibinfo {author} {\bibfnamefont
  {S.}~\bibnamefont {{Wehner}}},\ }\bibfield  {title} {\bibinfo {title} {{The
  second laws of quantum thermodynamics}},\ }\href
  {https://doi.org/10.1073/pnas.1411728112} {\bibfield  {journal} {\bibinfo
  {journal} {Proc. Natl. Acad. Sci. U.S.A.}\ }\textbf {\bibinfo {volume}
  {112}},\ \bibinfo {pages} {3275} (\bibinfo {year} {2015})}\BibitemShut
  {NoStop}%
\bibitem [{\citenamefont {Wilming}(2021)}]{Wilming_catalsysis}%
  \BibitemOpen
  \bibfield  {author} {\bibinfo {author} {\bibfnamefont {H.}~\bibnamefont
  {Wilming}},\ }\bibfield  {title} {\bibinfo {title} {Entropy and reversible
  catalysis},\ }\href {https://doi.org/10.1103/PhysRevLett.127.260402}
  {\bibfield  {journal} {\bibinfo  {journal} {Phys. Rev. Lett.}\ }\textbf
  {\bibinfo {volume} {127}},\ \bibinfo {pages} {260402} (\bibinfo {year}
  {2021})}\BibitemShut {NoStop}%
\bibitem [{\citenamefont {Shiraishi}\ and\ \citenamefont
  {Sagawa}(2021)}]{Shiraishi_sagawa_PRL}%
  \BibitemOpen
  \bibfield  {author} {\bibinfo {author} {\bibfnamefont {N.}~\bibnamefont
  {Shiraishi}}\ and\ \bibinfo {author} {\bibfnamefont {T.}~\bibnamefont
  {Sagawa}},\ }\bibfield  {title} {\bibinfo {title} {Quantum thermodynamics of
  correlated-catalytic state conversion at small scale},\ }\href
  {https://doi.org/10.1103/PhysRevLett.126.150502} {\bibfield  {journal}
  {\bibinfo  {journal} {Phys. Rev. Lett.}\ }\textbf {\bibinfo {volume} {126}},\
  \bibinfo {pages} {150502} (\bibinfo {year} {2021})}\BibitemShut {NoStop}%
\bibitem [{\citenamefont {Datta}\ \emph {et~al.}(2023)\citenamefont {Datta},
  \citenamefont {Kondra}, \citenamefont {Miller},\ and\ \citenamefont
  {Streltsov}}]{CDatta_Review}%
  \BibitemOpen
  \bibfield  {author} {\bibinfo {author} {\bibfnamefont {C.}~\bibnamefont
  {Datta}}, \bibinfo {author} {\bibfnamefont {T.~V.}\ \bibnamefont {Kondra}},
  \bibinfo {author} {\bibfnamefont {M.}~\bibnamefont {Miller}},\ and\ \bibinfo
  {author} {\bibfnamefont {A.}~\bibnamefont {Streltsov}},\ }\bibfield  {title}
  {\bibinfo {title} {Catalysis of entanglement and other quantum resources},\
  }\href {https://doi.org/10.1088/1361-6633/acfbec} {\bibfield  {journal}
  {\bibinfo  {journal} {Rep. Prog. Phys}\ }\textbf {\bibinfo {volume} {86}},\
  \bibinfo {pages} {116002} (\bibinfo {year} {2023})}\BibitemShut {NoStop}%
\bibitem [{\citenamefont {Lipka-Bartosik}\ \emph {et~al.}(2024)\citenamefont
  {Lipka-Bartosik}, \citenamefont {Wilming},\ and\ \citenamefont
  {Ng}}]{Bartosik_review}%
  \BibitemOpen
  \bibfield  {author} {\bibinfo {author} {\bibfnamefont {P.}~\bibnamefont
  {Lipka-Bartosik}}, \bibinfo {author} {\bibfnamefont {H.}~\bibnamefont
  {Wilming}},\ and\ \bibinfo {author} {\bibfnamefont {N.~H.~Y.}\ \bibnamefont
  {Ng}},\ }\bibfield  {title} {\bibinfo {title} {Catalysis in quantum
  information theory},\ }\href {https://doi.org/10.1103/RevModPhys.96.025005}
  {\bibfield  {journal} {\bibinfo  {journal} {Rev. Mod. Phys.}\ }\textbf
  {\bibinfo {volume} {96}},\ \bibinfo {pages} {025005} (\bibinfo {year}
  {2024})}\BibitemShut {NoStop}%
\bibitem [{\citenamefont {Kosloff}(1984)}]{Kosloff1984}%
  \BibitemOpen
  \bibfield  {author} {\bibinfo {author} {\bibfnamefont {R.}~\bibnamefont
  {Kosloff}},\ }\bibfield  {title} {\bibinfo {title} {A quantum mechanical open
  system as a model of a heat engine},\ }\bibfield  {journal} {\bibinfo
  {journal} {J. Chem. Phys.}\ }\textbf {\bibinfo {volume} {80}},\ \href
  {https://doi.org/10.1063/1.446862} {10.1063/1.446862} (\bibinfo {year}
  {1984})\BibitemShut {NoStop}%
\bibitem [{\citenamefont {Kosloff}\ and\ \citenamefont
  {Levy}(2014)}]{KosloffLevy2014}%
  \BibitemOpen
  \bibfield  {author} {\bibinfo {author} {\bibfnamefont {R.}~\bibnamefont
  {Kosloff}}\ and\ \bibinfo {author} {\bibfnamefont {A.}~\bibnamefont {Levy}},\
  }\bibfield  {title} {\bibinfo {title} {Quantum heat engines and
  refrigerators: Continuous devices},\ }\href
  {https://doi.org/10.1146/annurev-physchem-040513-103724} {\bibfield
  {journal} {\bibinfo  {journal} {Annu. Rev. Phys. Chem.}\ }\textbf {\bibinfo
  {volume} {65}},\ \bibinfo {pages} {365} (\bibinfo {year} {2014})}\BibitemShut
  {NoStop}%
\bibitem [{\citenamefont {Mitchison}(2019)}]{MitchisonContemp}%
  \BibitemOpen
  \bibfield  {author} {\bibinfo {author} {\bibfnamefont {M.~T.}\ \bibnamefont
  {Mitchison}},\ }\bibfield  {title} {\bibinfo {title} {Quantum thermal
  absorption machines: refrigerators, engines and clocks},\ }\href
  {https://doi.org/10.1080/00107514.2019.1631555} {\bibfield  {journal}
  {\bibinfo  {journal} {Contemp. Phys.}\ }\textbf {\bibinfo {volume} {60}},\
  \bibinfo {pages} {164} (\bibinfo {year} {2019})}\BibitemShut {NoStop}%
\bibitem [{\citenamefont {Allahverdyan}\ \emph {et~al.}(2010)\citenamefont
  {Allahverdyan}, \citenamefont {Hovhannisyan},\ and\ \citenamefont
  {Mahler}}]{two_stroke_Allahverdyan}%
  \BibitemOpen
  \bibfield  {author} {\bibinfo {author} {\bibfnamefont {A.~E.}\ \bibnamefont
  {Allahverdyan}}, \bibinfo {author} {\bibfnamefont {K.}~\bibnamefont
  {Hovhannisyan}},\ and\ \bibinfo {author} {\bibfnamefont {G.}~\bibnamefont
  {Mahler}},\ }\bibfield  {title} {\bibinfo {title} {Optimal refrigerator},\
  }\href {https://doi.org/10.1103/PhysRevE.81.051129} {\bibfield  {journal}
  {\bibinfo  {journal} {Phys. Rev. E}\ }\textbf {\bibinfo {volume} {81}},\
  \bibinfo {pages} {051129} (\bibinfo {year} {2010})}\BibitemShut {NoStop}%
\bibitem [{\citenamefont {Antonio Marín~Guzmán}\ \emph
  {et~al.}(2024)\citenamefont {Antonio Marín~Guzmán}, \citenamefont {Erker},
  \citenamefont {Gasparinetti}, \citenamefont {Huber},\ and\ \citenamefont
  {Yunger~Halpern}}]{AntonioGuzmán_2024}%
  \BibitemOpen
  \bibfield  {author} {\bibinfo {author} {\bibfnamefont {J.}~\bibnamefont
  {Antonio Marín~Guzmán}}, \bibinfo {author} {\bibfnamefont {P.}~\bibnamefont
  {Erker}}, \bibinfo {author} {\bibfnamefont {S.}~\bibnamefont {Gasparinetti}},
  \bibinfo {author} {\bibfnamefont {M.}~\bibnamefont {Huber}},\ and\ \bibinfo
  {author} {\bibfnamefont {N.}~\bibnamefont {Yunger~Halpern}},\ }\bibfield
  {title} {\bibinfo {title} {Key issues review: useful autonomous quantum
  machines},\ }\href {https://doi.org/10.1088/1361-6633/ad8803} {\bibfield
  {journal} {\bibinfo  {journal} {Rep. Prog. Phys.}\ }\textbf {\bibinfo
  {volume} {87}},\ \bibinfo {pages} {122001} (\bibinfo {year}
  {2024})}\BibitemShut {NoStop}%
\bibitem [{\citenamefont {Brunner}\ \emph {et~al.}(2014)\citenamefont
  {Brunner}, \citenamefont {Huber}, \citenamefont {Linden}, \citenamefont
  {Popescu}, \citenamefont {Silva},\ and\ \citenamefont
  {Skrzypczyk}}]{SilvaPRE1}%
  \BibitemOpen
  \bibfield  {author} {\bibinfo {author} {\bibfnamefont {N.}~\bibnamefont
  {Brunner}}, \bibinfo {author} {\bibfnamefont {M.}~\bibnamefont {Huber}},
  \bibinfo {author} {\bibfnamefont {N.}~\bibnamefont {Linden}}, \bibinfo
  {author} {\bibfnamefont {S.}~\bibnamefont {Popescu}}, \bibinfo {author}
  {\bibfnamefont {R.}~\bibnamefont {Silva}},\ and\ \bibinfo {author}
  {\bibfnamefont {P.}~\bibnamefont {Skrzypczyk}},\ }\bibfield  {title}
  {\bibinfo {title} {Entanglement enhances cooling in microscopic quantum
  refrigerators},\ }\href {https://doi.org/10.1103/PhysRevE.89.032115}
  {\bibfield  {journal} {\bibinfo  {journal} {Phys. Rev. E}\ }\textbf {\bibinfo
  {volume} {89}},\ \bibinfo {pages} {032115} (\bibinfo {year}
  {2014})}\BibitemShut {NoStop}%
\bibitem [{\citenamefont {Clivaz}\ \emph
  {et~al.}(2019{\natexlab{a}})\citenamefont {Clivaz}, \citenamefont {Silva},
  \citenamefont {Haack}, \citenamefont {Brask}, \citenamefont {Brunner},\ and\
  \citenamefont {Huber}}]{SilvaPRL}%
  \BibitemOpen
  \bibfield  {author} {\bibinfo {author} {\bibfnamefont {F.}~\bibnamefont
  {Clivaz}}, \bibinfo {author} {\bibfnamefont {R.}~\bibnamefont {Silva}},
  \bibinfo {author} {\bibfnamefont {G.}~\bibnamefont {Haack}}, \bibinfo
  {author} {\bibfnamefont {J.~B.}\ \bibnamefont {Brask}}, \bibinfo {author}
  {\bibfnamefont {N.}~\bibnamefont {Brunner}},\ and\ \bibinfo {author}
  {\bibfnamefont {M.}~\bibnamefont {Huber}},\ }\bibfield  {title} {\bibinfo
  {title} {Unifying paradigms of quantum refrigeration: A universal and
  attainable bound on cooling},\ }\href
  {https://doi.org/10.1103/PhysRevLett.123.170605} {\bibfield  {journal}
  {\bibinfo  {journal} {Phys. Rev. Lett.}\ }\textbf {\bibinfo {volume} {123}},\
  \bibinfo {pages} {170605} (\bibinfo {year} {2019}{\natexlab{a}})}\BibitemShut
  {NoStop}%
\bibitem [{\citenamefont {Clivaz}\ \emph
  {et~al.}(2019{\natexlab{b}})\citenamefont {Clivaz}, \citenamefont {Silva},
  \citenamefont {Haack}, \citenamefont {Brask}, \citenamefont {Brunner},\ and\
  \citenamefont {Huber}}]{SilvaPRE3}%
  \BibitemOpen
  \bibfield  {author} {\bibinfo {author} {\bibfnamefont {F.}~\bibnamefont
  {Clivaz}}, \bibinfo {author} {\bibfnamefont {R.}~\bibnamefont {Silva}},
  \bibinfo {author} {\bibfnamefont {G.}~\bibnamefont {Haack}}, \bibinfo
  {author} {\bibfnamefont {J.~B.}\ \bibnamefont {Brask}}, \bibinfo {author}
  {\bibfnamefont {N.}~\bibnamefont {Brunner}},\ and\ \bibinfo {author}
  {\bibfnamefont {M.}~\bibnamefont {Huber}},\ }\bibfield  {title} {\bibinfo
  {title} {Unifying paradigms of quantum refrigeration: Fundamental limits of
  cooling and associated work costs},\ }\href
  {https://doi.org/10.1103/PhysRevE.100.042130} {\bibfield  {journal} {\bibinfo
   {journal} {Phys. Rev. E}\ }\textbf {\bibinfo {volume} {100}},\ \bibinfo
  {pages} {042130} (\bibinfo {year} {2019}{\natexlab{b}})}\BibitemShut
  {NoStop}%
\bibitem [{\citenamefont {{\L{}}obejko}\ \emph {et~al.}(2020)\citenamefont
  {{\L{}}obejko}, \citenamefont {Mazurek},\ and\ \citenamefont
  {Horodecki}}]{Lobejko2020thermodynamicsof}%
  \BibitemOpen
  \bibfield  {author} {\bibinfo {author} {\bibfnamefont {M.}~\bibnamefont
  {{\L{}}obejko}}, \bibinfo {author} {\bibfnamefont {P.}~\bibnamefont
  {Mazurek}},\ and\ \bibinfo {author} {\bibfnamefont {M.}~\bibnamefont
  {Horodecki}},\ }\bibfield  {title} {\bibinfo {title} {Thermodynamics of
  {M}inimal {C}oupling {Q}uantum {H}eat {E}ngines},\ }\href
  {https://doi.org/10.22331/q-2020-12-23-375} {\bibfield  {journal} {\bibinfo
  {journal} {{Quantum}}\ }\textbf {\bibinfo {volume} {4}},\ \bibinfo {pages}
  {375} (\bibinfo {year} {2020})}\BibitemShut {NoStop}%
\bibitem [{\citenamefont {Biswas}\ \emph {et~al.}(2022)\citenamefont {Biswas},
  \citenamefont {{\L{}}obejko}, \citenamefont {Mazurek}, \citenamefont
  {Ja{\l{}}owiecki},\ and\ \citenamefont {Horodecki}}]{Biswas2022extractionof}%
  \BibitemOpen
  \bibfield  {author} {\bibinfo {author} {\bibfnamefont {T.}~\bibnamefont
  {Biswas}}, \bibinfo {author} {\bibfnamefont {M.}~\bibnamefont
  {{\L{}}obejko}}, \bibinfo {author} {\bibfnamefont {P.}~\bibnamefont
  {Mazurek}}, \bibinfo {author} {\bibfnamefont {K.}~\bibnamefont
  {Ja{\l{}}owiecki}},\ and\ \bibinfo {author} {\bibfnamefont {M.}~\bibnamefont
  {Horodecki}},\ }\bibfield  {title} {\bibinfo {title} {Extraction of
  ergotropy: free energy bound and application to open cycle engines},\ }\href
  {https://doi.org/10.22331/q-2022-10-17-841} {\bibfield  {journal} {\bibinfo
  {journal} {{Quantum}}\ }\textbf {\bibinfo {volume} {6}},\ \bibinfo {pages}
  {841} (\bibinfo {year} {2022})}\BibitemShut {NoStop}%
\bibitem [{\citenamefont {Biswas}\ and\ \citenamefont
  {Datta}(2024)}]{BiswasDatta}%
  \BibitemOpen
  \bibfield  {author} {\bibinfo {author} {\bibfnamefont {T.}~\bibnamefont
  {Biswas}}\ and\ \bibinfo {author} {\bibfnamefont {C.}~\bibnamefont {Datta}},\
  }\bibfield  {title} {\bibinfo {title} {Optimal performance of a three-stroke
  heat engine in the microscopic regime},\ }\href
  {https://doi.org/10.1103/PhysRevA.110.032218} {\bibfield  {journal} {\bibinfo
   {journal} {Phys. Rev. A}\ }\textbf {\bibinfo {volume} {110}},\ \bibinfo
  {pages} {032218} (\bibinfo {year} {2024})}\BibitemShut {NoStop}%
\bibitem [{\citenamefont {Dann}\ and\ \citenamefont
  {Kosloff}(2023)}]{DannKosloff_2023}%
  \BibitemOpen
  \bibfield  {author} {\bibinfo {author} {\bibfnamefont {R.}~\bibnamefont
  {Dann}}\ and\ \bibinfo {author} {\bibfnamefont {R.}~\bibnamefont {Kosloff}},\
  }\bibfield  {title} {\bibinfo {title} {Unification of the first law of
  quantum thermodynamics},\ }\href {https://doi.org/10.1088/1367-2630/acc967}
  {\bibfield  {journal} {\bibinfo  {journal} {New J. Phys.}\ }\textbf {\bibinfo
  {volume} {25}},\ \bibinfo {pages} {043019} (\bibinfo {year}
  {2023})}\BibitemShut {NoStop}%
\bibitem [{\citenamefont {Linden}\ \emph
  {et~al.}(2010{\natexlab{a}})\citenamefont {Linden}, \citenamefont {Popescu},\
  and\ \citenamefont {Skrzypczyk}}]{PopescuSmallestHeatEngine2010}%
  \BibitemOpen
  \bibfield  {author} {\bibinfo {author} {\bibfnamefont {N.}~\bibnamefont
  {Linden}}, \bibinfo {author} {\bibfnamefont {S.}~\bibnamefont {Popescu}},\
  and\ \bibinfo {author} {\bibfnamefont {P.}~\bibnamefont {Skrzypczyk}},\
  }\bibfield  {title} {\bibinfo {title} {The smallest possible heat engines},\
  }\href {https://doi.org/10.48550/arXiv.1010.6029} {\bibfield  {journal}
  {\bibinfo  {journal} {arXiv:1010.6029}\ } (\bibinfo {year}
  {2010}{\natexlab{a}})}\BibitemShut {NoStop}%
\bibitem [{\citenamefont {Linden}\ \emph
  {et~al.}(2010{\natexlab{b}})\citenamefont {Linden}, \citenamefont {Popescu},\
  and\ \citenamefont {Skrzypczyk}}]{Popescu_PRL}%
  \BibitemOpen
  \bibfield  {author} {\bibinfo {author} {\bibfnamefont {N.}~\bibnamefont
  {Linden}}, \bibinfo {author} {\bibfnamefont {S.}~\bibnamefont {Popescu}},\
  and\ \bibinfo {author} {\bibfnamefont {P.}~\bibnamefont {Skrzypczyk}},\
  }\bibfield  {title} {\bibinfo {title} {How small can thermal machines be? the
  smallest possible refrigerator},\ }\href
  {https://doi.org/10.1103/PhysRevLett.105.130401} {\bibfield  {journal}
  {\bibinfo  {journal} {Phys. Rev. Lett.}\ }\textbf {\bibinfo {volume} {105}},\
  \bibinfo {pages} {130401} (\bibinfo {year} {2010}{\natexlab{b}})}\BibitemShut
  {NoStop}%
\bibitem [{\citenamefont {{Gelbwaser-Klimovsky, D.}}\ \emph
  {et~al.}(2013)\citenamefont {{Gelbwaser-Klimovsky, D.}}, \citenamefont
  {{Alicki, R.}},\ and\ \citenamefont {{Kurizki, G.}}}]{Klimkovsky2013}%
  \BibitemOpen
  \bibfield  {author} {\bibinfo {author} {\bibnamefont {{Gelbwaser-Klimovsky,
  D.}}}, \bibinfo {author} {\bibnamefont {{Alicki, R.}}},\ and\ \bibinfo
  {author} {\bibnamefont {{Kurizki, G.}}},\ }\bibfield  {title} {\bibinfo
  {title} {Work and energy gain of heat-pumped quantized amplifiers},\ }\href
  {https://doi.org/10.1209/0295-5075/103/60005} {\bibfield  {journal} {\bibinfo
   {journal} {Euro. Phys. Lett.}\ }\textbf {\bibinfo {volume} {103}},\ \bibinfo
  {pages} {60005} (\bibinfo {year} {2013})}\BibitemShut {NoStop}%
\bibitem [{\citenamefont {Cangemi}\ \emph {et~al.}(2024)\citenamefont
  {Cangemi}, \citenamefont {Bhadra},\ and\ \citenamefont
  {Levy}}]{cangemni_Levy_engines}%
  \BibitemOpen
  \bibfield  {author} {\bibinfo {author} {\bibfnamefont {L.~M.}\ \bibnamefont
  {Cangemi}}, \bibinfo {author} {\bibfnamefont {C.}~\bibnamefont {Bhadra}},\
  and\ \bibinfo {author} {\bibfnamefont {A.}~\bibnamefont {Levy}},\ }\bibfield
  {title} {\bibinfo {title} {Quantum engines and refrigerators},\ }\href
  {https://doi.org/https://doi.org/10.1016/j.physrep.2024.07.001} {\bibfield
  {journal} {\bibinfo  {journal} {Phys. Rep.}\ }\textbf {\bibinfo {volume}
  {1087}},\ \bibinfo {pages} {1} (\bibinfo {year} {2024})}\BibitemShut
  {NoStop}%
\bibitem [{\citenamefont
  {Popescu}(2010)}]{PopescuSmallestHeatEngineprinciple2010}%
  \BibitemOpen
  \bibfield  {author} {\bibinfo {author} {\bibfnamefont {S.}~\bibnamefont
  {Popescu}},\ }\bibfield  {title} {\bibinfo {title} {Maximally efficient
  quantum thermal machines: The basic principles},\ }\href
  {https://arxiv.org/pdf/1009.2536.pdf} {\bibfield  {journal} {\bibinfo
  {journal} {arXiv:1009.2536}\ } (\bibinfo {year} {2010})}\BibitemShut
  {NoStop}%
\bibitem [{\citenamefont {Uzdin}\ \emph {et~al.}(2015)\citenamefont {Uzdin},
  \citenamefont {Levy},\ and\ \citenamefont {Kosloff}}]{UzdinLevyKosloff2015}%
  \BibitemOpen
  \bibfield  {author} {\bibinfo {author} {\bibfnamefont {R.}~\bibnamefont
  {Uzdin}}, \bibinfo {author} {\bibfnamefont {A.}~\bibnamefont {Levy}},\ and\
  \bibinfo {author} {\bibfnamefont {R.}~\bibnamefont {Kosloff}},\ }\bibfield
  {title} {\bibinfo {title} {Equivalence of quantum heat machines, and
  quantum-thermodynamic signatures},\ }\href
  {https://doi.org/10.1103/PhysRevX.5.031044} {\bibfield  {journal} {\bibinfo
  {journal} {Phys. Rev. X}\ }\textbf {\bibinfo {volume} {5}},\ \bibinfo {pages}
  {031044} (\bibinfo {year} {2015})}\BibitemShut {NoStop}%
\bibitem [{\citenamefont {Chiara}\ \emph {et~al.}(2018)\citenamefont {Chiara},
  \citenamefont {Landi}, \citenamefont {Hewgill}, \citenamefont {Reid},
  \citenamefont {Ferraro}, \citenamefont {Roncaglia},\ and\ \citenamefont
  {Antezza}}]{Dechiara_2018}%
  \BibitemOpen
  \bibfield  {author} {\bibinfo {author} {\bibfnamefont {G.~D.}\ \bibnamefont
  {Chiara}}, \bibinfo {author} {\bibfnamefont {G.}~\bibnamefont {Landi}},
  \bibinfo {author} {\bibfnamefont {A.}~\bibnamefont {Hewgill}}, \bibinfo
  {author} {\bibfnamefont {B.}~\bibnamefont {Reid}}, \bibinfo {author}
  {\bibfnamefont {A.}~\bibnamefont {Ferraro}}, \bibinfo {author} {\bibfnamefont
  {A.~J.}\ \bibnamefont {Roncaglia}},\ and\ \bibinfo {author} {\bibfnamefont
  {M.}~\bibnamefont {Antezza}},\ }\bibfield  {title} {\bibinfo {title}
  {Reconciliation of quantum local master equations with thermodynamics},\
  }\href {https://doi.org/10.1088/1367-2630/aaecee} {\bibfield  {journal}
  {\bibinfo  {journal} {New J. Phys.}\ }\textbf {\bibinfo {volume} {20}},\
  \bibinfo {pages} {113024} (\bibinfo {year} {2018})}\BibitemShut {NoStop}%
\bibitem [{\citenamefont {Molitor}\ and\ \citenamefont
  {Landi}(2020)}]{Molitor_Landi}%
  \BibitemOpen
  \bibfield  {author} {\bibinfo {author} {\bibfnamefont {O.~A.~D.}\
  \bibnamefont {Molitor}}\ and\ \bibinfo {author} {\bibfnamefont {G.~T.}\
  \bibnamefont {Landi}},\ }\bibfield  {title} {\bibinfo {title} {Stroboscopic
  two-stroke quantum heat engines},\ }\href
  {https://doi.org/10.1103/PhysRevA.102.042217} {\bibfield  {journal} {\bibinfo
   {journal} {Phys. Rev. A}\ }\textbf {\bibinfo {volume} {102}},\ \bibinfo
  {pages} {042217} (\bibinfo {year} {2020})}\BibitemShut {NoStop}%
\bibitem [{\citenamefont {Melo}\ \emph {et~al.}(2022)\citenamefont {Melo},
  \citenamefont {S\'a}, \citenamefont {Roditi}, \citenamefont {Souza},
  \citenamefont {Oliveira}, \citenamefont {Sarthour},\ and\ \citenamefont
  {Landi}}]{Melo_Landi}%
  \BibitemOpen
  \bibfield  {author} {\bibinfo {author} {\bibfnamefont {F.~V.}\ \bibnamefont
  {Melo}}, \bibinfo {author} {\bibfnamefont {N.}~\bibnamefont {S\'a}}, \bibinfo
  {author} {\bibfnamefont {I.}~\bibnamefont {Roditi}}, \bibinfo {author}
  {\bibfnamefont {A.~M.}\ \bibnamefont {Souza}}, \bibinfo {author}
  {\bibfnamefont {I.~S.}\ \bibnamefont {Oliveira}}, \bibinfo {author}
  {\bibfnamefont {R.~S.}\ \bibnamefont {Sarthour}},\ and\ \bibinfo {author}
  {\bibfnamefont {G.~T.}\ \bibnamefont {Landi}},\ }\bibfield  {title} {\bibinfo
  {title} {Implementation of a two-stroke quantum heat engine with a
  collisional model},\ }\href {https://doi.org/10.1103/PhysRevA.106.032410}
  {\bibfield  {journal} {\bibinfo  {journal} {Phys. Rev. A}\ }\textbf {\bibinfo
  {volume} {106}},\ \bibinfo {pages} {032410} (\bibinfo {year}
  {2022})}\BibitemShut {NoStop}%
\bibitem [{\citenamefont {Campbell}\ and\ \citenamefont
  {Vacchini}(2021)}]{Campbell_2021}%
  \BibitemOpen
  \bibfield  {author} {\bibinfo {author} {\bibfnamefont {S.}~\bibnamefont
  {Campbell}}\ and\ \bibinfo {author} {\bibfnamefont {B.}~\bibnamefont
  {Vacchini}},\ }\bibfield  {title} {\bibinfo {title} {Collision models in open
  system dynamics: A versatile tool for deeper insights?},\ }\href
  {https://doi.org/10.1209/0295-5075/133/60001} {\bibfield  {journal} {\bibinfo
   {journal} {Europhys. Lett.}\ }\textbf {\bibinfo {volume} {133}},\ \bibinfo
  {pages} {60001} (\bibinfo {year} {2021})}\BibitemShut {NoStop}%
\bibitem [{\citenamefont {Strasberg}(2019)}]{Strasberg}%
  \BibitemOpen
  \bibfield  {author} {\bibinfo {author} {\bibfnamefont {P.}~\bibnamefont
  {Strasberg}},\ }\bibfield  {title} {\bibinfo {title} {Repeated interactions
  and quantum stochastic thermodynamics at strong coupling},\ }\href
  {https://doi.org/10.1103/PhysRevLett.123.180604} {\bibfield  {journal}
  {\bibinfo  {journal} {Phys. Rev. Lett.}\ }\textbf {\bibinfo {volume} {123}},\
  \bibinfo {pages} {180604} (\bibinfo {year} {2019})}\BibitemShut {NoStop}%
\bibitem [{\citenamefont {Bettmann}\ \emph {et~al.}(2023)\citenamefont
  {Bettmann}, \citenamefont {Kewming},\ and\ \citenamefont {Goold}}]{Bettman}%
  \BibitemOpen
  \bibfield  {author} {\bibinfo {author} {\bibfnamefont {L.~P.}\ \bibnamefont
  {Bettmann}}, \bibinfo {author} {\bibfnamefont {M.~J.}\ \bibnamefont
  {Kewming}},\ and\ \bibinfo {author} {\bibfnamefont {J.}~\bibnamefont
  {Goold}},\ }\bibfield  {title} {\bibinfo {title} {Thermodynamics of a
  continuously monitored double-quantum-dot heat engine in the repeated
  interactions framework},\ }\href
  {https://doi.org/10.1103/PhysRevE.107.044102} {\bibfield  {journal} {\bibinfo
   {journal} {Phys. Rev. E}\ }\textbf {\bibinfo {volume} {107}},\ \bibinfo
  {pages} {044102} (\bibinfo {year} {2023})}\BibitemShut {NoStop}%
\bibitem [{\citenamefont {Piccione}\ \emph {et~al.}(2021)\citenamefont
  {Piccione}, \citenamefont {De~Chiara},\ and\ \citenamefont
  {Bellomo}}]{Piccione2021}%
  \BibitemOpen
  \bibfield  {author} {\bibinfo {author} {\bibfnamefont {N.}~\bibnamefont
  {Piccione}}, \bibinfo {author} {\bibfnamefont {G.}~\bibnamefont
  {De~Chiara}},\ and\ \bibinfo {author} {\bibfnamefont {B.}~\bibnamefont
  {Bellomo}},\ }\bibfield  {title} {\bibinfo {title} {Power maximization of
  two-stroke quantum thermal machines},\ }\href
  {https://doi.org/10.1103/PhysRevA.103.032211} {\bibfield  {journal} {\bibinfo
   {journal} {Phys. Rev. A}\ }\textbf {\bibinfo {volume} {103}},\ \bibinfo
  {pages} {032211} (\bibinfo {year} {2021})}\BibitemShut {NoStop}%
\bibitem [{\citenamefont {Saha}\ \emph {et~al.}(2024)\citenamefont {Saha},
  \citenamefont {Das},\ and\ \citenamefont {Ghosh}}]{Saha_2024}%
  \BibitemOpen
  \bibfield  {author} {\bibinfo {author} {\bibfnamefont {T.}~\bibnamefont
  {Saha}}, \bibinfo {author} {\bibfnamefont {A.}~\bibnamefont {Das}},\ and\
  \bibinfo {author} {\bibfnamefont {S.}~\bibnamefont {Ghosh}},\ }\bibfield
  {title} {\bibinfo {title} {Quantum homogenization in non-markovian
  collisional model},\ }\href {https://doi.org/10.1088/1367-2630/ad212f}
  {\bibfield  {journal} {\bibinfo  {journal} {New J. Phys.}\ }\textbf {\bibinfo
  {volume} {26}},\ \bibinfo {pages} {023011} (\bibinfo {year}
  {2024})}\BibitemShut {NoStop}%
\bibitem [{\citenamefont {Cusumano}\ and\ \citenamefont
  {De~Chiara}(2024)}]{Cusumano_2024}%
  \BibitemOpen
  \bibfield  {author} {\bibinfo {author} {\bibfnamefont {S.}~\bibnamefont
  {Cusumano}}\ and\ \bibinfo {author} {\bibfnamefont {G.}~\bibnamefont
  {De~Chiara}},\ }\bibfield  {title} {\bibinfo {title} {Structured quantum
  collision models: generating coherence with thermal resources},\ }\href
  {https://doi.org/10.1088/1367-2630/ad202a} {\bibfield  {journal} {\bibinfo
  {journal} {New J. Phys.}\ }\textbf {\bibinfo {volume} {26}},\ \bibinfo
  {pages} {023001} (\bibinfo {year} {2024})}\BibitemShut {NoStop}%
\bibitem [{\citenamefont {Ciccarello}\ \emph {et~al.}(2022)\citenamefont
  {Ciccarello}, \citenamefont {Lorenzo}, \citenamefont {Giovannetti},\ and\
  \citenamefont {Palma}}]{CICCARELLO20221}%
  \BibitemOpen
  \bibfield  {author} {\bibinfo {author} {\bibfnamefont {F.}~\bibnamefont
  {Ciccarello}}, \bibinfo {author} {\bibfnamefont {S.}~\bibnamefont {Lorenzo}},
  \bibinfo {author} {\bibfnamefont {V.}~\bibnamefont {Giovannetti}},\ and\
  \bibinfo {author} {\bibfnamefont {G.~M.}\ \bibnamefont {Palma}},\ }\bibfield
  {title} {\bibinfo {title} {Quantum collision models: Open system dynamics
  from repeated interactions},\ }\href
  {https://doi.org/https://doi.org/10.1016/j.physrep.2022.01.001} {\bibfield
  {journal} {\bibinfo  {journal} {Phys. Rep.}\ }\textbf {\bibinfo {volume}
  {954}},\ \bibinfo {pages} {1} (\bibinfo {year} {2022})}\BibitemShut {NoStop}%
\bibitem [{\citenamefont {Levy}\ and\ \citenamefont
  {Kosloff}(2014)}]{Levy_2014}%
  \BibitemOpen
  \bibfield  {author} {\bibinfo {author} {\bibfnamefont {A.}~\bibnamefont
  {Levy}}\ and\ \bibinfo {author} {\bibfnamefont {R.}~\bibnamefont {Kosloff}},\
  }\bibfield  {title} {\bibinfo {title} {The local approach to quantum
  transport may violate the second law of thermodynamics},\ }\href
  {https://doi.org/10.1209/0295-5075/107/20004} {\bibfield  {journal} {\bibinfo
   {journal} {Europhysics Letters}\ }\textbf {\bibinfo {volume} {107}},\
  \bibinfo {pages} {20004} (\bibinfo {year} {2014})}\BibitemShut {NoStop}%
\bibitem [{\citenamefont {Hewgill}\ \emph {et~al.}(2021)\citenamefont
  {Hewgill}, \citenamefont {De~Chiara},\ and\ \citenamefont
  {Imparato}}]{Hewgill_PRR}%
  \BibitemOpen
  \bibfield  {author} {\bibinfo {author} {\bibfnamefont {A.}~\bibnamefont
  {Hewgill}}, \bibinfo {author} {\bibfnamefont {G.}~\bibnamefont {De~Chiara}},\
  and\ \bibinfo {author} {\bibfnamefont {A.}~\bibnamefont {Imparato}},\
  }\bibfield  {title} {\bibinfo {title} {Quantum thermodynamically consistent
  local master equations},\ }\href
  {https://doi.org/10.1103/PhysRevResearch.3.013165} {\bibfield  {journal}
  {\bibinfo  {journal} {Phys. Rev. Res.}\ }\textbf {\bibinfo {volume} {3}},\
  \bibinfo {pages} {013165} (\bibinfo {year} {2021})}\BibitemShut {NoStop}%
\bibitem [{\citenamefont {Spohn}(1978)}]{spohn1978entropy}%
  \BibitemOpen
  \bibfield  {author} {\bibinfo {author} {\bibfnamefont {H.}~\bibnamefont
  {Spohn}},\ }\bibfield  {title} {\bibinfo {title} {Entropy production for
  quantum dynamical semigroups},\ }\href
  {https://pubs.aip.org/aip/jmp/article/19/5/1227/460049/Entropy-production-for-quantum-dynamical}
  {\bibfield  {journal} {\bibinfo  {journal} {J. Math. Phys.}\ }\textbf
  {\bibinfo {volume} {19}},\ \bibinfo {pages} {1227} (\bibinfo {year}
  {1978})}\BibitemShut {NoStop}%
\bibitem [{\citenamefont {Landi}\ and\ \citenamefont
  {Paternostro}(2021)}]{Landi_Review}%
  \BibitemOpen
  \bibfield  {author} {\bibinfo {author} {\bibfnamefont {G.~T.}\ \bibnamefont
  {Landi}}\ and\ \bibinfo {author} {\bibfnamefont {M.}~\bibnamefont
  {Paternostro}},\ }\bibfield  {title} {\bibinfo {title} {Irreversible entropy
  production: From classical to quantum},\ }\href
  {https://doi.org/10.1103/RevModPhys.93.035008} {\bibfield  {journal}
  {\bibinfo  {journal} {Rev. Mod. Phys.}\ }\textbf {\bibinfo {volume} {93}},\
  \bibinfo {pages} {035008} (\bibinfo {year} {2021})}\BibitemShut {NoStop}%
\bibitem [{\citenamefont {Hofer}\ \emph {et~al.}(2017)\citenamefont {Hofer},
  \citenamefont {Perarnau-Llobet}, \citenamefont {Miranda}, \citenamefont
  {Haack}, \citenamefont {Silva}, \citenamefont {Brask},\ and\ \citenamefont
  {Brunner}}]{Hofer_2017}%
  \BibitemOpen
  \bibfield  {author} {\bibinfo {author} {\bibfnamefont {P.~P.}\ \bibnamefont
  {Hofer}}, \bibinfo {author} {\bibfnamefont {M.}~\bibnamefont
  {Perarnau-Llobet}}, \bibinfo {author} {\bibfnamefont {L.~D.~M.}\ \bibnamefont
  {Miranda}}, \bibinfo {author} {\bibfnamefont {G.}~\bibnamefont {Haack}},
  \bibinfo {author} {\bibfnamefont {R.}~\bibnamefont {Silva}}, \bibinfo
  {author} {\bibfnamefont {J.~B.}\ \bibnamefont {Brask}},\ and\ \bibinfo
  {author} {\bibfnamefont {N.}~\bibnamefont {Brunner}},\ }\bibfield  {title}
  {\bibinfo {title} {Markovian master equations for quantum thermal machines:
  local versus global approach},\ }\href
  {https://doi.org/10.1088/1367-2630/aa964f} {\bibfield  {journal} {\bibinfo
  {journal} {New J. Phys.}\ }\textbf {\bibinfo {volume} {19}},\ \bibinfo
  {pages} {123037} (\bibinfo {year} {2017})}\BibitemShut {NoStop}%
\bibitem [{\citenamefont {Roßnagel}\ \emph {et~al.}(2016)\citenamefont
  {Roßnagel}, \citenamefont {Dawkins}, \citenamefont {Tolazzi}, \citenamefont
  {Abah}, \citenamefont {Lutz}, \citenamefont {Schmidt-Kaler},\ and\
  \citenamefont {Singer}}]{Singer}%
  \BibitemOpen
  \bibfield  {author} {\bibinfo {author} {\bibfnamefont {J.}~\bibnamefont
  {Roßnagel}}, \bibinfo {author} {\bibfnamefont {S.~T.}\ \bibnamefont
  {Dawkins}}, \bibinfo {author} {\bibfnamefont {K.~N.}\ \bibnamefont
  {Tolazzi}}, \bibinfo {author} {\bibfnamefont {O.}~\bibnamefont {Abah}},
  \bibinfo {author} {\bibfnamefont {E.}~\bibnamefont {Lutz}}, \bibinfo {author}
  {\bibfnamefont {F.}~\bibnamefont {Schmidt-Kaler}},\ and\ \bibinfo {author}
  {\bibfnamefont {K.}~\bibnamefont {Singer}},\ }\bibfield  {title} {\bibinfo
  {title} {A single-atom heat engine},\ }\href
  {https://doi.org/10.1126/science.aad6320} {\bibfield  {journal} {\bibinfo
  {journal} {Science}\ }\textbf {\bibinfo {volume} {352}},\ \bibinfo {pages}
  {325} (\bibinfo {year} {2016})}\BibitemShut {NoStop}%
\bibitem [{\citenamefont {Aamir}\ \emph {et~al.}(2025)\citenamefont {Aamir},
  \citenamefont {Jamet~Suria}, \citenamefont {Mar{\'i}n~Guzm{\'a}n},
  \citenamefont {Castillo-Moreno}, \citenamefont {Epstein}, \citenamefont
  {Yunger~Halpern},\ and\ \citenamefont {Gasparinetti}}]{Aamir2025}%
  \BibitemOpen
  \bibfield  {author} {\bibinfo {author} {\bibfnamefont {M.~A.}\ \bibnamefont
  {Aamir}}, \bibinfo {author} {\bibfnamefont {P.}~\bibnamefont {Jamet~Suria}},
  \bibinfo {author} {\bibfnamefont {J.~A.}\ \bibnamefont
  {Mar{\'i}n~Guzm{\'a}n}}, \bibinfo {author} {\bibfnamefont {C.}~\bibnamefont
  {Castillo-Moreno}}, \bibinfo {author} {\bibfnamefont {J.~M.}\ \bibnamefont
  {Epstein}}, \bibinfo {author} {\bibfnamefont {N.}~\bibnamefont
  {Yunger~Halpern}},\ and\ \bibinfo {author} {\bibfnamefont {S.}~\bibnamefont
  {Gasparinetti}},\ }\bibfield  {title} {\bibinfo {title} {Thermally driven
  quantum refrigerator autonomously resets a superconducting qubit},\ }\href
  {https://doi.org/10.1038/s41567-024-02708-5} {\bibfield  {journal} {\bibinfo
  {journal} {Nat. Phys.}\ }\textbf {\bibinfo {volume} {21}},\ \bibinfo {pages}
  {318} (\bibinfo {year} {2025})}\BibitemShut {NoStop}%
\bibitem [{\citenamefont {Taranto}\ \emph {et~al.}(2023)\citenamefont
  {Taranto}, \citenamefont {Bakhshinezhad}, \citenamefont {Bluhm},
  \citenamefont {Silva}, \citenamefont {Friis}, \citenamefont {Lock},
  \citenamefont {Vitagliano}, \citenamefont {Binder}, \citenamefont {Debarba},
  \citenamefont {Schwarzhans}, \citenamefont {Clivaz},\ and\ \citenamefont
  {Huber}}]{Landauer_vs_Nernst}%
  \BibitemOpen
  \bibfield  {author} {\bibinfo {author} {\bibfnamefont {P.}~\bibnamefont
  {Taranto}}, \bibinfo {author} {\bibfnamefont {F.}~\bibnamefont
  {Bakhshinezhad}}, \bibinfo {author} {\bibfnamefont {A.}~\bibnamefont
  {Bluhm}}, \bibinfo {author} {\bibfnamefont {R.}~\bibnamefont {Silva}},
  \bibinfo {author} {\bibfnamefont {N.}~\bibnamefont {Friis}}, \bibinfo
  {author} {\bibfnamefont {M.~P.}\ \bibnamefont {Lock}}, \bibinfo {author}
  {\bibfnamefont {G.}~\bibnamefont {Vitagliano}}, \bibinfo {author}
  {\bibfnamefont {F.~C.}\ \bibnamefont {Binder}}, \bibinfo {author}
  {\bibfnamefont {T.}~\bibnamefont {Debarba}}, \bibinfo {author} {\bibfnamefont
  {E.}~\bibnamefont {Schwarzhans}}, \bibinfo {author} {\bibfnamefont
  {F.}~\bibnamefont {Clivaz}},\ and\ \bibinfo {author} {\bibfnamefont
  {M.}~\bibnamefont {Huber}},\ }\bibfield  {title} {\bibinfo {title} {Landauer
  versus nernst: What is the true cost of cooling a quantum system?},\ }\href
  {https://doi.org/10.1103/PRXQuantum.4.010332} {\bibfield  {journal} {\bibinfo
   {journal} {PRX Quantum}\ }\textbf {\bibinfo {volume} {4}},\ \bibinfo {pages}
  {010332} (\bibinfo {year} {2023})}\BibitemShut {NoStop}%
\bibitem [{\citenamefont {Levy}\ and\ \citenamefont
  {Gelbwaser-Klimovsky}(2018)}]{Levy2018}%
  \BibitemOpen
  \bibfield  {author} {\bibinfo {author} {\bibfnamefont {A.}~\bibnamefont
  {Levy}}\ and\ \bibinfo {author} {\bibfnamefont {D.}~\bibnamefont
  {Gelbwaser-Klimovsky}},\ }\bibinfo {title} {Quantum features and signatures
  of quantum thermal machines},\ in\ \href
  {https://doi.org/10.1007/978-3-319-99046-0_4} {\emph {\bibinfo {booktitle}
  {Thermodynamics in the Quantum Regime: Fundamental Aspects and New
  Directions}}}\ (\bibinfo  {publisher} {Springer International Publishing},\
  \bibinfo {address} {Cham},\ \bibinfo {year} {2018})\ pp.\ \bibinfo {pages}
  {87--126}\BibitemShut {NoStop}%
\bibitem [{\citenamefont {Latune}\ \emph {et~al.}(2021)\citenamefont {Latune},
  \citenamefont {Sinayskiy},\ and\ \citenamefont {Petruccione}}]{Latune2021}%
  \BibitemOpen
  \bibfield  {author} {\bibinfo {author} {\bibfnamefont {C.~L.}\ \bibnamefont
  {Latune}}, \bibinfo {author} {\bibfnamefont {I.}~\bibnamefont {Sinayskiy}},\
  and\ \bibinfo {author} {\bibfnamefont {F.}~\bibnamefont {Petruccione}},\
  }\bibfield  {title} {\bibinfo {title} {Roles of quantum coherences in thermal
  machines},\ }\href {https://doi.org/10.1140/epjs/s11734-021-00085-1}
  {\bibfield  {journal} {\bibinfo  {journal} {The European Physical Journal
  Special Topics}\ }\textbf {\bibinfo {volume} {230}},\ \bibinfo {pages} {841}
  (\bibinfo {year} {2021})}\BibitemShut {NoStop}%
\bibitem [{\citenamefont {Manzano}\ \emph {et~al.}(2019)\citenamefont
  {Manzano}, \citenamefont {Silva},\ and\ \citenamefont {Parrondo}}]{Manzano}%
  \BibitemOpen
  \bibfield  {author} {\bibinfo {author} {\bibfnamefont {G.}~\bibnamefont
  {Manzano}}, \bibinfo {author} {\bibfnamefont {R.}~\bibnamefont {Silva}},\
  and\ \bibinfo {author} {\bibfnamefont {J.~M.~R.}\ \bibnamefont {Parrondo}},\
  }\bibfield  {title} {\bibinfo {title} {Autonomous thermal machine for
  amplification and control of energetic coherence},\ }\href
  {https://doi.org/10.1103/PhysRevE.99.042135} {\bibfield  {journal} {\bibinfo
  {journal} {Phys. Rev. E}\ }\textbf {\bibinfo {volume} {99}},\ \bibinfo
  {pages} {042135} (\bibinfo {year} {2019})}\BibitemShut {NoStop}%
\bibitem [{\citenamefont {Biswas}\ \emph {et~al.}(2025)\citenamefont {Biswas},
  \citenamefont {Datta},\ and\ \citenamefont
  {Garc\'{\i}a-Pintos}}]{BiswasDattaPRL}%
  \BibitemOpen
  \bibfield  {author} {\bibinfo {author} {\bibfnamefont {T.}~\bibnamefont
  {Biswas}}, \bibinfo {author} {\bibfnamefont {C.}~\bibnamefont {Datta}},\ and\
  \bibinfo {author} {\bibfnamefont {L.~P.}\ \bibnamefont
  {Garc\'{\i}a-Pintos}},\ }\bibfield  {title} {\bibinfo {title} {Quantum
  thermodynamic advantage in work extraction from steerable quantum
  correlations},\ }\href {https://doi.org/10.1103/9qcc-7lq5} {\bibfield
  {journal} {\bibinfo  {journal} {Phys. Rev. Lett.}\ }\textbf {\bibinfo
  {volume} {135}},\ \bibinfo {pages} {110402} (\bibinfo {year}
  {2025})}\BibitemShut {NoStop}%
\end{thebibliography}%

\appendix
\newpage
%\onecolumngrid
%\section*{End Matter}

%\section{Examples}
%\section{Example 1: Otto engines (without a catalyst)}\label{Examples}
\textit{Appendix A: Example 1- Otto engines (without a catalyst)}
%\subsection{Discrete Otto engine}
Discrete Otto engine: The initial density matrix describing the engine is taken as a product of the Gibbs states of the hot and cold systems,  
\begin{equation}\label{App:Qubit_engine_state_without1}
    \rho = \frac{e^{-\beta_h H_{0,h}}}{\Tr[e^{-\beta_h H_{0,h}}]} \otimes \frac{e^{-\beta_c H_{0,c}}}{\Tr[e^{-\beta_c H_{0,c}}]},
\end{equation}
where each subsystem is modeled as a two-level system with Hamiltonian
\begin{equation}
    H_{0,k} = \omega_k \dyad{1}_k, \quad k \in \{h,c\}.
\end{equation}
We consider a single swap operation $S$ defined by  
\begin{equation}\label{app_u_and_d_Otto}
    \ket{u} := \ket{10}_{h,c} \xleftrightarrow{S} \ket{01}_{h,c} =: \ket{d},
\end{equation}
which acts on the computational basis states as
\begin{equation}
    S \ket{01}= \ket{10}, \;\; S \ket{10} = \ket{01}, \;\;
    S \ket{11} = \ket{11}, \;\; S \ket{00} = \ket{00}.
\end{equation} 
For brevity, we omit the subscript and follow this convention throughout the section. We can obtain \( Q_k \) from the definition given in Eq.~\eqref{Q_k_and_J_k_main} by calculating the corresponding probabilities and energy differences, which gives
\begin{align}
   \Delta p&= p_u-p_d = \bra{u}\rho\ket{u}-\bra{d}\rho\ket{d}, 
    = \frac{\left(e^{-\beta_h\omega_h}-e^{-\beta_c\omega_c}\right)}{(1+e^{-\beta_h\omega_h})(1+e^{-\beta_c\omega_c})}\label{Eq_deltap_main},\\
     \Delta \varepsilon^h&= \varepsilon^h_u-\varepsilon^h_d=\left(\langle u|H_{0,h}|u\rangle-\langle d|H_{0,h}|d\rangle\right)=\omega_h,\label{app_delta_varepsilon_next0}\\
    \Delta \varepsilon^c&=\varepsilon^c_u-\varepsilon^c_d=\left(\langle u|H_{0,c}|u\rangle-\langle d|H_{0,c}|d\rangle\right)=-\omega_c\label{app_delta_varepsilon_next1}.
\end{align}
By substituting Eqs.~\eqref{Eq_deltap_main}, \eqref{app_delta_varepsilon_next0}, and \eqref{app_delta_varepsilon_next1} into the definition of \( Q_k \) given in Eq.~\eqref{Q_k_and_J_k_main}, we obtain:
\begin{equation}
    Q_h= \omega_h\Delta p\;,\;
    Q_c=-\omega_c\Delta p\;,\;
    W=Q_h+Q_c=\left(\omega_h-\omega_c\right)\Delta p.
\end{equation}
Finally, we can calculate the efficiency from Eq. \eqref{etadisc_and_etacont_main} as 
\begin{equation}\label{eff_end_matter_disc}
    \eta = \frac{W}{Q_h}=1+\frac{Q_c}{Q_h}=1-\frac{\omega_c}{\omega_h}.
\end{equation}
The detailed calculation is given in the Sec. III A of the Supplementary material.
%\subsection{Continuous Otto engine}

Continuous Otto engine: The dynamical model of the continuous Otto engine is described by the Hamiltonian:
\begin{align}
    H_{0} &=  H_{0,h}+H_{0,c}=\omega_h \sigma_-^h\sigma_+^h  +  \omega_c \sigma_-^c\sigma_+^c  \\
    V_t &=g(\ketbra{u}{d}e^{- i \Omega t}+\text{H.C.}) =g \left(\sigma_-^h \sigma_+^c e^{- i \Omega t} + \text{H.C.} \right)
\end{align}
such that $\ket{u} = \ket{10}$ and $\ket{d} = \ket{01}$. Here $\text{H.C.}$ denote hermitian conjugate. The time evolution of the engine in the interaction picture is governed by the local master equation: 
\begin{eqnarray}\label{End_matter_LME}
    \mathcal{L}[\rho] &:=& \frac{d\rho}{dt} = -i[V_0,\rho_I(t)] + \mathcal{D}_h[\rho_I(t)] + \mathcal{D}_c[\rho_I(t)],\quad\nonumber\\ &&\text{where}\quad V_0 = g\Big(\ketbra{u}{d}+\text{H.C.}\Big).
\end{eqnarray}
Then, a closed set of equations can be obtained by imposing the stationary conditions $\langle \mathcal{L}^\dag[\Pi_{i,j}] \rangle = 0$ where $\Pi_{i,j}$ is the projector on the eigenstate of $\rho$ given in Eq. \eqref{App:Qubit_engine_state_without1} and $\langle\cdot\rangle$ denotes the expectation with respect to the stationary state. This set of equations, upon solving, yields
\begin{eqnarray}\label{app_expression_dotN}
    \langle \dot N \rangle &=& \left(\frac{2\Gamma_h \Gamma_c}{\Gamma_h + \Gamma_c}\right) \frac{\Delta p}{\left(1+\frac{ \Gamma_h \Gamma_c}{g^2}\right)},
\end{eqnarray}
where $\Delta p$ is the \emph{probability flow}  for the discrete version of this engine as given in Eq. \eqref{Eq_deltap_main} and
\begin{equation}
    \Gamma_k = (\gamma^k_+ + \gamma^k_-)/2\quad\text{with}\quad k\in\{h,c\}.
\end{equation}
 Recall that $\gamma^k_{+}$ and $\gamma^k_{-}$ denote the pumping and damping rates, respectively, associated with the $k^{\text{th}}$ heat bath that obeys the detailed balance condition $\gamma^k_{+}/\gamma^k_{-} = e^{-\beta_k \omega_k}.$ The detailed calculation of $\langle \dot N\rangle$ has been presented in Sec. III B of the supplementary material.
\begin{comment}
\begin{align}
    &-(\gamma_h^+ + \gamma_c^+) p_{00} + \gamma_h^- p_{10} +\gamma_c^- p_{01} = 0,\label{closed_eq_1}\\
     &-(\gamma_h^- + \gamma_c^-) p_{11} + \gamma_h^+ p_{01} +\gamma_c^+ p_{10} = 0,\label{closed_eq_2}\\
     &-(\gamma_h^+ + \gamma_c^-) p_{01} + \gamma_h^- p_{11} +\gamma_c^+ p_{00} + \langle \dot N \rangle = 0, \label{closed_eq_3}\\
     &-(\gamma_h^- + \gamma_c^+) p_{10} + \gamma_c^- p_{11} +\gamma_h^+ p_{00} - \langle \dot N \rangle = 0. \label{closed_eq_4}
\end{align}
\end{comment}

Substituting the value of $\langle \dot N \rangle $ from Eq. \eqref{app_expression_dotN}  to Eq. \eqref{Q_k_and_J_k_main} and employing Eq. \eqref{app_delta_varepsilon_next0} and Eq. \eqref{app_delta_varepsilon_next1}, we obtain:
\begin{align}
    J_h &= \omega_h \left(\frac{ 2\Gamma_h \Gamma_c\Delta p}{(\Gamma_h + \Gamma_c)\left(1+\frac{ \Gamma_h \Gamma_c}{g^2}\right)}\right)\label{app_Jh_otto_like},\\
    J_c &= -\omega_c \left(\frac{ 2\Gamma_h \Gamma_c\Delta p}{(\Gamma_h + \Gamma_c)\left(1+\frac{ \Gamma_h \Gamma_c}{g^2}\right)}\right)\label{app_Jc_otto_like},\\
    P &= J_h+J_c= \left(\omega_h-\omega_c\right)\left(\frac{ 2\Gamma_h \Gamma_c\Delta p}{(\Gamma_h + \Gamma_c)\left(1+\frac{ \Gamma_h \Gamma_c}{g^2}\right)}\right)\label{app_Power_otto_like}.
\end{align}
Finally, substituting $J_h$ and $J_c$ into Eq.~\eqref{etadisc_and_etacont_main}, we obtain the efficiency
\begin{equation}\label{eff_end_matt_cont}
    \eta =\frac{P}{J_h}=1+\frac{J_c}{J_h}= 1 - \frac{\omega_c}{\omega_h},
\end{equation}
which is same as Eq. \eqref{eff_end_matter_disc}. This confirms our claim that the efficiency remains invariant under the mapping proposed in Eq.~\eqref{Mapping_of_disc_and_cont}. The characteristic time \( \tau \) for such an engine can be obtained by taking the ratio of \( \Delta p \) from Eq.~\eqref{Eq_deltap_main} and \( \dot N \) from Eq.~\eqref{app_expression_dotN}, yielding
\begin{equation}\label{end_matt_tau_otto}
    \tau = \frac{\Delta p}{\langle \dot N \rangle} 
    = \left(1 + \frac{\Gamma_h \Gamma_c}{g^2}\right)
      \left(\frac{\Gamma_h + \Gamma_c}{2\Gamma_h \Gamma_c}\right).
\end{equation}
%\section{Example 2: Otto like engines with a qubit catalyst}\label{Examples2}
\textit{Appendix B: Example 2-Otto like engines with a qubit catalyst}
%\subsection{Discrete Otto like engines with a qubit catalyst}
Discrete Otto like engines with a qubit catalyst: The initial density matrix is given by the product of the catalyst state, taken to be a qubit, and the Gibbs states of the hot and cold systems, i.e.,
\begin{equation}\label{rho_cat}
    \rho = \rho_s \otimes \frac{e^{-\beta H_{0,h}}}{\Tr\left(e^{-\beta H_{0,h}}\right)} 
    \otimes \frac{e^{-\beta H_{0,c}}}{\Tr\left(e^{-\beta H_{0,c}}\right)},
\end{equation}
where \( H_{0,h} \) and \( H_{0,c} \) are the same Hamiltonians as in the previous scenario given in Eq.~\eqref{App:Qubit_engine_state_without1}, and the catalyst Hamiltonian is assumed to be trivial. We write the state of the catalyst as 
$\rho_s=p\ketbra{1}+(1-p)\ketbra{2}$ where $p$ depends on the unitary transformation governing the work stroke.

In the following, we consider the permutation $S$ governing the work stroke such that,  
\begin{equation}\label{app_defn_u1_d1_u2_d2}
    \ket{u_1} = \ket{200} \xleftrightarrow{S} \ket{110} = \ket{d_1},\quad
    \ket{u_2} = \ket{101} \xleftrightarrow{S} \ket{210} = \ket{d_2},
\end{equation}
and acts as identity on rest of the eigenvectors of $\rho$ in Eq. \eqref{rho_cat}. 

To compute the work produced by the engine and the heat exchanged with the respective heat baths, we use Eqs.~\eqref{Q_k_and_J_k_main} and \eqref{etadisc_and_etacont_main}, as in the case of the discrete Otto engine. To calculate the heat exchanged, we proceed by evaluating the corresponding energy differences as follows:
\begin{align}
    \Delta \varepsilon_1^h&=\left(\varepsilon_{u_1}^h-\varepsilon_{d_1}^h\right)=\Big(\langle u_1|H_{0,h}|u_1\rangle-\langle d_1|H_{0,h}|d_1\rangle\Big)=-\omega_h,\\
    \Delta \varepsilon_2^h&=\left(\varepsilon_{u_2}^h-\varepsilon_{d_2}^h\right)=\Big(\langle u_2|H_{0,h}|u_2\rangle-\langle d_2|H_{0,h}|d_2\rangle\Big)=-\omega_h,\\
    \Delta \varepsilon_1^c&=\left(\varepsilon_{u_1}^c-\varepsilon_{d_1}^c\right)=\Big(\langle u_1|H_{0,c}|u_1\rangle-\langle d_1|H_{0,c}|d_1\rangle\Big)=0,\\
    \Delta \varepsilon_2^c&=\left(\varepsilon_{u_2}^c-\varepsilon_{d_2}^c\right)=\Big(\langle u_2|H_{0,c}|u_2\rangle-\langle d_2|H_{0,c}|d_2\rangle\Big)=\omega_c.
\end{align}
Next, employing the catalysis condition in Eq. \eqref{catalsyis_disc_and_cond} we obtain 
\begin{eqnarray}\label{Eq:End_matt_deltaP}
    \Delta p_1=\Delta p_2&=& \frac{-1}{(1+e^{-\beta_h\omega_h})(1+e^{-\beta_c\omega_c})}\left[\frac{e^{-2\beta_h\omega_h}-e^{-\beta_c\omega_c}}{1+2e^{-\beta_h\omega_h}+e^{-\beta_c\omega_c}}\right]\nonumber\\&:=&\Delta p.
\end{eqnarray}
A detailed derivation of $\Delta p$, along with the proof of the equality $\Delta p_1 = \Delta p_2$, is provided in Sec.~IV~A of the Supplementary Material. This equality confirms that $S$, as defined in Eq.~\eqref{app_defn_u1_d1_u2_d2}, is indeed a simple permutation. Using the expression of $\Delta p$ in Eq. \eqref{Eq:End_matt_deltaP}, we have
\begin{align}
    Q_h&= -\omega_h \Delta p_1- \omega_h \Delta p_2=-\omega_h \Delta p- \omega_h \Delta p=-2\omega_h\Delta p,\label{app:Expression_of_Q_h_cat}\\
    Q_c&= \omega_c \Delta p_2=\omega_c\Delta p.
    \label{app:Expression_of_Q_c_cat}
\end{align}
Finally, the work and efficiency of the engine are obtained by substituting $Q_h$ and $Q_c$ from Eqs.~\eqref{app:Expression_of_Q_h_cat} and \eqref{app:Expression_of_Q_c_cat} into their respective definitions, yielding
\begin{align}\label{app_Work__eff_discrete}
    W &= Q_h + Q_c = \left(2\omega_h - \omega_c\right)\Delta p,\quad
    \eta = \frac{W}{Q_h} = 1 - \frac{\omega_c}{2\omega_h}.
\end{align}
Observe that the resulting efficiency is strictly greater than the Otto efficiency given in Eq.~\eqref{eff_end_matt_cont}, thereby capturing the catalytic enhancement in efficiency.

%\subsection{Continuous Otto-like Engines with a Qubit Catalyst}
Continuous Otto-like Engines with a Qubit Catalyst: In this section, we analyze the continuous Otto-like engines with a qubit catalyst. The Hamiltonian of the continuous Otto-like engine with a qubit catalyst is given by
\begin{eqnarray}\label{interaction_qubit}
H_0&=&H_{0,h}+H_{0,c}=\omega_h\sigma^h_{-}\sigma^h_{+}+\omega_c\sigma^c_{-}\sigma^c_{+},\\
    V_t&=&g\left(e^{-i\Omega_1t}\ketbra{u_1}{d_1}+e^{-i\Omega_2 t}\ketbra{u_2}{d_2}+\text{H. C.}\right),
\end{eqnarray}
where we have introduced $\ket{u_i}$ and $\ket{d_i}$ given in Eq. \eqref{app_defn_u1_d1_u2_d2}. Using the local master equation given in Eq.~\eqref{End_matter_LME}, one can derive the complete set of equations. Applying the catalysis condition for the continuous Otto-like engines from Eq.~\eqref{catalsyis_disc_and_cond}, we obtain the probabibility current according to Eq. \eqref{cat_condition_main} as
\begin{widetext}
\begin{equation}\label{Eq:End_matt_Ndot}
   \langle \dot N_1\rangle = \langle \dot N_2\rangle := \langle \dot N\rangle=\frac{\Delta p}{\left[\left(\frac{a_c+a_h}{1+a_c+2a_h}\right)\left(\alpha_2+\frac{A}{4g^2}\right)+\left(\frac{1+a_h}{1+a_c+2a_h}\right)\left(\phi_1+\frac{B}{4g^2}\right)+\left(\frac{\left(a_h^2-a_c\right)(\alpha_1+\phi_1+\xi_1+\alpha_2+\phi_2+\xi_2)}{(1+a_c)(1+a_h)(1+a_c+2a_h)}\right)\right]},\quad\text{where}
\end{equation} 
\begin{eqnarray}
&&\alpha_1:=\left(\frac{\gamma^c_{-}+\gamma^h_{-}}{\gamma^h_{+}(\gamma^c_{+}+\gamma^c_{-}+\gamma^h_{+}+\gamma^h_{-})}\right),\quad\quad\quad \alpha_2:=\left(\frac{\gamma^c_{-}+\gamma^h_{-}+\gamma^h_{+}}{\gamma^h_{+}(\gamma^c_{+}+\gamma^c_{-}+\gamma^h_{+}+\gamma^h_{-})}\right),\quad\quad\quad\phi_1:=\left(\frac{(\gamma^c_{-}+\gamma^h_{-})(\gamma^c_{+}+\gamma^h_{+})}{\gamma^c_{-}\gamma^h_{+}(\gamma^c_{+}+\gamma^c_{-}+\gamma^h_{+}+\gamma^h_{-})}\right),\label{defn:alpha}\\
    &&\phi_2:=\left(\frac{\gamma^c_{+}(\gamma^c_{-}+\gamma^h_{-})}{\gamma^c_{-}\gamma^h_{+}(\gamma^c_{+}+\gamma^c_{-}+\gamma^h_{+}+\gamma^h_{-})}\right),\quad\quad\quad \xi_1:=\left(\frac{\left(\gamma^c_{+}+\gamma^h_{+}\right)}{\gamma^c_{-}(\gamma^c_{+}+\gamma^c_{-}+\gamma^h_{+}+\gamma^h_{-})}\right),\quad\quad\quad \xi_2:=\left(\frac{\gamma^c_{+}}{\gamma^c_{-}(\gamma^c_{+}+\gamma^c_{-}+\gamma^h_{+}+\gamma^h_{-})}\right),\label{defn:xi}\\
    &&A:=\left(\gamma^h_{-}+\gamma^h_{+}+2\gamma^c_{+}-\frac{4\gamma^c_{-}\gamma^c_{+}}{\left(\gamma^h_{-}+\gamma^h_{+}+2\gamma^c_{-}\right)}\right),\quad\quad\quad B:=(\gamma^h_{-}+\gamma^h_{+}+\gamma^c_{-}+\gamma^c_{+}),\quad\quad a_h=e^{-\beta_h\omega_h},\quad\quad a_c=e^{-\beta_c\omega_c},\label{defn:AandB}
\end{eqnarray}
\end{widetext}
where $\Delta p$ is the probability flow associated with discrete version of the engine as derived in Eq. \eqref{Eq:End_matt_deltaP}. The detailed derivation of $\langle \dot N\rangle$ from Eq. \eqref{Eq:End_matt_Ndot} is given in Sec. IV B of the supplementary material. Using the expression for $\langle \dot N\rangle$ from Eq.~\eqref{Eq:End_matt_deltaP}, we obtain
\begin{align}
    J_h &= -\omega_h \langle \dot N_1\rangle - \omega_h \langle \dot N_2\rangle 
        = -\omega_h \langle \dot N\rangle - \omega_h \langle \dot N\rangle
        = -2\omega_h \langle \dot N\rangle,
    \label{app:Expression_of_J_h_cat}\\
    J_c &= \omega_c \langle \dot N_2\rangle 
        = \omega_c \langle \dot N\rangle.
    \label{app:Expression_of_J_c_cat}
\end{align}
Using $J_h$ and $J_c$ from Eqs.~\eqref{app:Expression_of_J_h_cat} and \eqref{app:Expression_of_J_c_cat}, one can obtain the power and the efficiency of the engine as:
\begin{align}\label{app_power__eff_cont}
    P &= J_h + J_c = \left(2\omega_h - \omega_c\right)\langle \dot N\rangle,\quad
    \eta= \frac{P}{J_h} = 1 - \frac{\omega_c}{2\omega_h}.
\end{align}
Note that the Otto-like engine with a qubit catalyst operates with the same efficiency as its discrete counterpart, given in Eq.~\eqref{app_Work__eff_discrete}, thereby capturing the exact catalytic enhancement in efficiency. 

Finally, the characteristic time can be obtained by dividing $\Delta p$ from Eq.~\eqref{Eq:End_matt_deltaP} with $\langle \dot N\rangle$ from 
Eq.~\eqref{Eq:End_matt_Ndot} that results the denominator of RHS of Eq. \eqref{Eq:End_matt_Ndot} i.e.
\begin{widetext}
\begin{align}\label{end_matt_tau_otto_cat}
    \tau_{\text{catalytic}}=\left(\frac{a_c+a_h}{1+a_c+2a_h}\right)\left(\alpha_2+\frac{A}{4g^2}\right)+\left(\frac{1+a_h}{1+a_c+2a_h}\right)\left(\phi_1+\frac{B}{4g^2}\right)+\left(\frac{\left(a_h^2-a_c\right)(\alpha_1+\phi_1+\xi_1+\alpha_2+\phi_2+\xi_2)}{(1+a_c)(1+a_h)(1+a_c+2a_h)}\right).
\end{align}
\end{widetext}

\onecolumngrid
\section*{Supplementary material: Equivalence of Discrete and Continuous Otto-Like Engines assisted by Catalysts: Mapping Catalytic  Advantages from the Discrete to the Continuous Framework}

%\tableofcontents

\section{Notation and useful identities}
\subsection{Notation and convention}

Before presenting the detailed calculations, we first introduce the notations used throughout. In expressions involving tensor products, we consistently follow the order $X^s \otimes Y^h \otimes Z^c$, where $X^s$ acts on the Hilbert space of the catalyst, $Y^h$ on the hot system, and $Z^c$ on the cold system. Similarly, we write vectors as $\ket{ijk}_{s,h,c} = \ket{i}_s \otimes \ket{j}_h \otimes \ket{k}_c$, and for brevity, we drop the subscripts and denote $\ket{ijk}_{s,h,c}$ simply as $\ket{ijk}$. In the absence of a catalyst, we consistently use the order $Y^h \otimes Z^c$, with $\ket{jk} := \ket{j}_h \otimes \ket{k}_c$.

Here, $\gamma^h_{+}$ and $\gamma^h_{-}$ is the pumping and damping rates associated with hot bath. Similarly, $\gamma^c_{+}$ and $\gamma^c_{-}$ is the pumping and damping rates associated with the cold bath. The parameter $g$ describes the strength of the interaction between the catalyst, the hot and cold qubit present in the working body of the engine. 
\begin{eqnarray}
    &&\frac{\gamma^h_{+}}{\gamma^h_{-}}:=e^{-\beta_h\omega_h}:=a_h,\quad\quad\quad \frac{\gamma^c_{+}}{\gamma^c_{-}}:=e^{-\beta_c\omega_c}=a_c\label{supp_defn:gibbs_factor}\\
    &&\sigma_{+}:=\ketbra{0}{1},\quad \quad\quad\sigma_{-}:=\ketbra{1}{0},\\
    && \Gamma_h = \frac{1}{2}(\gamma^h_+ + \gamma^h_-), \quad\quad\quad  \Gamma_c = \frac{1}{2}(\gamma^c_+ + \gamma^c_-)\\
    &&\alpha_1:=\frac{\gamma^c_{+}+\gamma^h_{+}}{\gamma^h_{-}(\gamma^h_{-}+\gamma^h_{+}+\gamma^c_{-}+\gamma^c_{+})},\quad\quad\quad \alpha_2:=\frac{\gamma^c_{+}+\gamma^h_{-}+\gamma^h_{+}}{\gamma^h_{-}(\gamma^h_{-}+\gamma^h_{+}+\gamma^c_{-}+\gamma^c_{+})}\label{supp_defn:alpha}\\
    &&\phi_1:=\frac{(\gamma^c_{+}+\gamma^h_{+})(\gamma^c_{-}+\gamma^h_{-})}{\gamma^h_{-}\gamma^c_{+}(\gamma^h_{-}+\gamma^h_{+}+\gamma^c_{-}+\gamma^c_{+})},\quad\quad\quad\phi_2:=\frac{\gamma^c_{-}(\gamma^c_{+}+\gamma^h_{+})}{\gamma^c_{+}\gamma^h_{-}(\gamma^h_{-}+\gamma^h_{+}+\gamma^c_{-}+\gamma^c_{+})}\label{supp_defn:phi}\\
    &&\xi_1:=\frac{(\gamma^c_{-}+\gamma^h_{-})}{\gamma^c_{+}(\gamma^h_{-}+\gamma^h_{+}+\gamma^c_{-}+\gamma^c_{+})}\quad\quad\quad \xi_2:=\frac{\gamma^c_{-}}{\gamma^c_{+}(\gamma^h_{-}+\gamma^h_{+}+\gamma^c_{-}+\gamma^c_{+})}\label{supp_defn:xi}\\
    &&A:=\left(\gamma^h_{-}+\gamma^h_{+}+2\gamma^c_{-}-\frac{4\gamma^c_{-}\gamma^c_{+}}{\gamma^h_{-}+\gamma^h_{+}+2\gamma^c_{+}}\right)\quad\quad\quad B:=(\gamma^h_{-}+\gamma^h_{+}+\gamma^c_{-}+\gamma^c_{+})\label{supp_defn:AandB}\\
    &&\mathcal{G}_A:=\left(\alpha_2+\frac{A}{4g^2}\right)\quad\quad\quad \mathcal{G}_B:=\left(\phi_1+\frac{B}{4g^2}\right)
\end{eqnarray}
\begin{table*}
\centering
\begin{tabular}{ |c|c|c| } 
 \hline
  & Discrete & Continuous \\ \hline 
 Probability / probability current & $\Delta p_i = p_{u_i} - p_{d_i}$ & $\langle \dot N_i \rangle = ig_i \langle \dyad{u_{i}}{d_{i}} - \dyad{d_{i}}{u_{i}} \rangle$ \\ \hline
 Heat / Heat current & $Q_k = \sum_i \Delta \varepsilon_{i}^k \Delta p_i$ & $J_k = \sum_i \Delta \varepsilon_{i}^k \langle \dot N_i \rangle$ \\ \hline
 Work / Power & $W = Q_h + Q_c$ & $P = J_h + J_c$ \\ \hline
  Clausius inequality & $\beta_h Q_h + \beta_c Q_c \le 0$ & $\beta_h J_h + \beta_c J_c \le 0$ \\ \hline
 Efficiency & $\eta = 1 + \frac{\sum_i \Delta \varepsilon_{i}^c \Delta p_i}{\sum_i \Delta \varepsilon_{i}^h \Delta p_i}$ &  $\eta = 1 + \frac{\sum_i \Delta \varepsilon_{i}^c \langle \dot N_i \rangle}{\sum_i \Delta \varepsilon_{i}^h \langle \dot N_i \rangle}$\\ \hline
  Catalysis condition & \ \ $\forall m,\;\; \sum_i \Delta \lambda_i^m \Delta p_i = 0$ \ \ & $\forall m,\;\; \sum_i \Delta \lambda_i^m \langle \dot N_i \rangle = 0$\\ \hline
\end{tabular}
\caption{Mapping between the thermodynamic quantities in the discrete and continuous engines.}
\label{supp_tab:discrete_continuous1}
\end{table*}
All notations used in this supplementary material are listed above. Each notation will be defined and referenced at the appropriate place in the text; however, for convenience, we encourage the reader to consult this list as needed.
\subsection{Identities}
In this section, we will introduce a series of identities, which are straightforward to prove and will be used frequently throughout the material.
\begin{eqnarray}
    &&\sigma_{+}\sigma_{-}=\ketbra{0}{0},\quad\sigma_{-}\sigma_{+}=\ketbra{1}{1},\quad\sigma_{+}\sigma_{-}\sigma_{+}=\sigma_{+},\quad\sigma_{-}\sigma_{+}\sigma_{-}=\sigma_{-}\\
    &&e^{i\alpha\frac{\sigma_z}{2}}\sigma_{-}e^{-i\alpha\frac{\sigma_z}{2}}=e^{-i\alpha}\sigma_{-}\quad;\quad e^{i\alpha\frac{\sigma_z}{2}}\sigma_{+}e^{-i\alpha\frac{\sigma_z}{2}}=e^{i\alpha}\sigma_{+}\\
    &&\mathcal{D}^{\dagger}(\sigma_{+}\sigma_{-})=\mathcal{D}^{\dagger}(\ketbra{0}{0})=-\gamma_{-}\ketbra{0}{0}+\gamma_{+}\ketbra{1}{1}=-\gamma_{-}\sigma_{+}\sigma_{-}+\gamma_{+}\sigma_{-}\sigma_{+}\label{supp_Eq:Don0}\\
    &&\mathcal{D}^{\dagger}(\sigma_{-}\sigma_{+})=\mathcal{D}^{\dagger}(\ketbra{1}{1})=\gamma_{-}\ketbra{0}{0}-\gamma_{+}\ketbra{1}{1}=\gamma_{-}\sigma_{+}\sigma_{-}-\gamma_{+}\sigma_{-}\sigma_{+}\label{supp_Eq:Don1}\\
    &&\mathcal{D}^{\dagger}(\sigma_{-})=\mathcal{D}^{\dagger}(\ketbra{1}{0})=-\frac{1}{2}(\gamma_{-}+\gamma_{+})\ketbra{1}{0}=-\frac{1}{2}(\gamma_{-}+\gamma_{+})\sigma_{-}\\
    &&\mathcal{D}^{\dagger}(\sigma_{+})=\mathcal{D}^{\dagger}(\ketbra{0}{1})=-\frac{1}{2}(\gamma_{-}+\gamma_{+})\ketbra{0}{1}=-\frac{1}{2}(\gamma_{-}+\gamma_{+})\sigma_{+}\\
\end{eqnarray}
where
\begin{equation}
    \mathcal{D}^{\dagger}(\mathcal{O})=\gamma_{+}\left(\sigma_{+}\mathcal{O}\sigma_{-}-\frac{1}{2}\left\{\sigma_{+}\sigma_{-},\mathcal{O}\right\}\right)+\gamma_{-}\left(\sigma_{-}\mathcal{O}\sigma_{+}-\frac{1}{2}\left\{\sigma_{-}\sigma_{+},\mathcal{O}\right\}\right).
\end{equation}

\section{Thermodynamic quantities}
Table \ref{supp_tab:discrete_continuous1} lists the relevant thermodynamic quantities and illustrates the correspondence between the discrete and continuous Otto-like engines.

\subsection{Heat current}
For the stationary state, we have
\begin{equation}
   \langle \mathcal{D}_k^\dag[H_{0,k}] \rangle = i\langle  [H_{0,k}, V_0] \rangle,
\end{equation}
which follows from the fact that $\langle \mathcal{L}^\dag[H_{0,k}] \rangle = \Tr[H_{0,k} \mathcal{L}[\rho_I^s]] = 0$. Furthermore, since 
\begin{equation}
    [H_{0,k}, V_0] = \sum_i \Delta  \varepsilon_{i}^k g_i ( \dyad{u_{i}}{d_{i}} - h.c.),
\end{equation}
finally, we get: 
\begin{equation} \label{supp_heat_cont}
    J_k = \sum_i \Delta \varepsilon_{i}^k \langle \dot N_i \rangle,
\end{equation}

\subsection{Master equation}
The system evolves under the local master equation:
\begin{equation}
    \dot \rho_S(t) = - i [H_0 + V_t, \rho_S(t)] + \mathcal{D}_h[\rho_S(t)] + \mathcal{D}_c[\rho_S(t)],
\end{equation}
such that
\begin{equation}
    \mathcal{D}^\dag_h[H_{0,c}] = \mathcal{D}^\dag_c[H_{0,h}] = 0.
\end{equation}
Generally, the local model of the dissipator is justified whenever $\| H_0 \| \gg g_i$.

The energy conservation with power $P(t)$ and heat current $J_k(t)$ is generally given by:
\begin{equation}
    \frac{d}{dt} \Tr[(H_0+V_t) \rho_S(t)] = -P(t) + J_h(t) + J_c(t). 
\end{equation}
where
\begin{align}
&P(t)  = -\Tr[\dot V_t \rho_S(t)], \quad J_{k}(t) = \Tr[\mathcal{D}_{k}^\dag [H_0+V_t] \rho_S(t)],
\end{align}
with $\dot V_t = -i\sum_i g_i \Omega_i \left(\dyad{u_{i}}{d_{i}} e^{- i \Omega_it} - h.c. \right)$. By introducing an interaction picture with a unitary transformation $\rho_I(t) = e^{-i H_0 t} \rho_S(t) e^{i H_0 t}$, we get
\begin{equation}
    \frac{d}{dt} {\rho_I(t)} = - i [V_0, \rho_I(t)] + \mathcal{D}_h[\rho_I(t)] + \mathcal{D}_c[\rho_I(t)]:=\mathcal{L}[\rho_I(t)].
\end{equation}
and 
\begin{align}
&P(t)  = -\Tr[\dot V_0 \rho_I(t)], \quad J_{k}(t) = \Tr[\mathcal{D}_k^\dag[H_{0,k}+V_0] \rho_I(t)].
\end{align}
%where we used the condition given in Eq. \eqref{supp_local_condition}.
\section{Discrete and continuous Otto-like engine (without catalyst)}
\subsection{Discrete Otto-like engine}
The initial density matrix is a product of hot and cold system's Gibbs states, i.e., 
\begin{equation}
    \rho = \frac{e^{-\beta H_{0,h}}}{\Tr[e^{-\beta H_{0,h}}]} \otimes \frac{e^{-\beta H_{0,c}}}{\Tr[e^{-\beta H_{0,c}}]},
\end{equation}
with a simplest two-level system Hamiltonian:
\begin{equation}
    H_{0,k} =  \omega_k \dyad{1}_k, \quad \text{with}\quad k\in\{h,c\},
\end{equation}
such that 
\begin{equation}\label{supp_App:Qubit_engine_state_without}
    \rho = \frac{1}{\mathcal{Z}_h \mathcal{Z}_c} \Big(\ketbra{00} + a_h \ketbra{10}+ a_c \ketbra{01} + a_ha_c \ketbra{11} \Big), 
\end{equation}
where $\mathcal{Z}_k = 1 + a_k$ and $a_k=e^{-\beta_k\omega_k}$ as introduced in Eq. \eqref{supp_defn:gibbs_factor}). %We consistently use the ordering \(|h,c\rangle\) to represent the state of the Otto-like engines, maintaining this fixed order throughout this section.
Then, we consider a single swap $S$ such that 
\begin{equation}\label{supp_app_u_and_d_Otto}
    \ket u := \ket{10} \xleftrightarrow{S}  \ket{01}=:\ket d,
\end{equation}
 namely 
\begin{align}
    S \ket{01} = \ket{10}, \quad S \ket{10} = \ket{01}, \\
    S \ket{11} = \ket{11}, \quad S \ket{00} = \ket{00}. 
\end{align}
Recall that from Table \ref{supp_tab:discrete_continuous1}, the transferred amount of heat $Q_h$ and $Q_c$ given as:
\begin{equation}\label{supp_app_Eq_heat_disc}
    Q_k = \sum_i \Delta \varepsilon_i^k \Delta p_i,
\end{equation}
where 
\begin{align}\label{supp_app_prob_difference}
    &\Delta p_i = p_{u_i} - p_{d_i}, \quad \Delta \varepsilon_i^k = \varepsilon_{u_i}^k - \varepsilon_{d_i}^k, \\
    &\varepsilon_{u_i}^k = \bra{u_i} H_{0,k} \ket{u_i}, \quad \varepsilon_{d_i}^k = \bra{d_i} H_{0,k} \ket{d_i}.
\end{align}
Therefore, for the state $\rho$ given in Eq. \eqref{supp_App:Qubit_engine_state_without} we can identify
\begin{equation}
    p_u=\langle u|\rho|u\rangle=\frac{a_h}{(1+a_h)(1+a_c)}\quad;\quad p_d=\langle d|\rho|d\rangle=\frac{a_c}{(1+a_h)(1+a_c)},
\end{equation}
that allows us to calculate
\begin{equation}\label{supp_app_Eq_deltap}
    \Delta p =p_u-p_d= \frac{ (a_h - a_c)}{(1+ a_h)(1+ a_c)}.
\end{equation}
Next, we calculate
\begin{eqnarray}
    \Delta \varepsilon^h&=& \varepsilon^h_u-\varepsilon^h_d=\left(\langle u|H_{0,h}|u\rangle-\langle d|H_{0,h}|d\rangle\right)=\left(\langle 10|H_{0,h}|10\rangle-\langle 01|H_{0,h}|01\rangle\right)=\omega_h,\label{supp_app_delta_varepsilon_next0}\\
    \Delta \varepsilon^c&=&\varepsilon^c_u-\varepsilon^c_d=\left(\langle u|H_{0,c}|u\rangle-\langle d|H_{0,c}|d\rangle\right)=\left(\langle 10|H_{0,c}|10\rangle-\langle 01|H_{0,c}|01\rangle\right)=-\omega_c\label{supp_app_delta_varepsilon_next1}.
\end{eqnarray}
Substituting the obtained expression of $\Delta p$, $\Delta \varepsilon^h$ and $\Delta \varepsilon^c$ from Eq. \eqref{supp_app_Eq_deltap} and Eq. \eqref{supp_app_delta_varepsilon}, we can calculate $Q_h$ and $Q_c$ from Eq. \eqref{supp_app_Eq_heat_disc} as follows:
\begin{eqnarray}\label{supp_app:OttoQ_cQ_h}
    Q_h &=& \Delta \varepsilon^h\Delta p=\frac{ \omega_h (a_h - a_c)}{(1+ a_h)(1+ a_c)} \nonumber\\
    Q_c &=& \Delta \varepsilon^c\Delta p =\frac{ -\omega_c (a_h - a_c)}{(1+ a_h)(1+ a_c)}.
\end{eqnarray}
Finally, the work and efficiency of the engine are given by the standard Otto formula. Substituting $Q_c$ and $Q_h$ from Eq.~\eqref{supp_app:OttoQ_cQ_h} into the definitions of work and efficiency, we obtain  
\begin{equation}\label{supp_app_work_per_cycle_eff_otto_like}
    W = Q_h + Q_c = \left(\omega_h - \omega_c\right)\Delta p=\frac{ (\omega_h-\omega_c) (a_h - a_c)}{(1+ a_h)(1+ a_c)}, 
    \qquad 
    \eta = 1 + \frac{Q_c}{Q_h} = 1 - \frac{\omega_c}{\omega_h}.
\end{equation}
%Physically, applying the operation $S$ lowers the energy of the system, such that the work is drawn by an external agent (which is justified by the fact that the process is unitary). To complete the thermodynamic cycle, the thermalization of the ``hot'' and ``cold'' systems is provided, such that $S\rho S^\dag \to \rho$, which triggers the corresponding energy flows equal to $Q_h$ and $Q_c$, interpreted as dissipative heat. 

\subsection{Continuous Otto engine}

The dynamical model of the continuous Otto engine is described by the Hamiltonian:
\begin{align}
    H_{0} &=  H_{0,h}+H_{0,c}=\omega_h \sigma_-^h\sigma_+^h  +  \omega_c \sigma_-^c\sigma_+^c  \\
    V_t &=g(\ketbra{u}{d}e^{- i \Omega t}+\text{Herm. Conj.}) =g \left(\sigma_-^h \sigma_+^c e^{- i \Omega t} + \text{Herm. Conj.} \right)
\end{align}
such that $\ket{u} = \ket{10}$ and $\ket{d} = \ket{01}$ as stated in Eq. \eqref{supp_app_u_and_d_Otto}.
From the main text, we recall that the time evolution of the engine in the interaction picture is governed by the local master equation: 
\begin{equation}
    \mathcal{L}[\rho] := \frac{d\rho}{dt} = -i[V_0,\rho_I(t)] + \mathcal{D}_h[\rho_I(t)] + \mathcal{D}_c[\rho_I(t)]\quad \text{where}\quad V_0 = g\Big(\ketbra{u}{d}+\ketbra{d}{u}\Big).
\end{equation}
%\begin{align}
   % &\Delta \varepsilon_{h} = \varepsilon_u^h - \varepsilon_d^h =  \omega_h, \\
   % &\Delta \varepsilon_{c} = \varepsilon_u^c - \varepsilon_d^c = - \omega_c, \\
   % &\Omega = (\Delta \varepsilon_{h} + \Delta \varepsilon_{c}) = \omega_h - \omega_c.
%\end{align}
The dissipator of the master equation is given by:
\begin{align}\label{supp_app_form_of_diss_Otto}
    &\mathcal{D}_k[\rho] = \gamma^k_+ \left(\sigma^k_{-}\rho\sigma^{k}_{+}-\frac{1}{2}\left\{\sigma^k_{+}\sigma^k_{-},\rho\right\}\right) + \gamma^k_- \left(\sigma^k_{+}\rho\sigma^{k}_{-}-\frac{1}{2}\left\{\sigma^k_{-}\sigma^k_{+},\rho\right\}\right)\quad\text{where}\quad k\in\{h,c\}.
\end{align}
Equivalently, the local master equation governing the evolution of an observable $\mathcal{O}_I$ in the interaction picture follows from the Heisenberg equation of motion:
\begin{equation}\label{supp_app_HEOM}
    \frac{d\mathcal{O}_I}{dt}=i[V_0,\mathcal{O}_I]+\mathcal{D}_h^{\dagger}(\mathcal{O}_I)+\mathcal{D}_c^{\dagger}(\mathcal{O}_I):=\mathcal{L}^{\dagger}[\mathcal{O}_I],
\end{equation}
where
\begin{eqnarray}
    \mathcal{D}^{\dagger}_{k}(\mathcal{O}_{I})=\gamma_{+}^{k}\left(\sigma_{+}^{k}\mathcal{O}_{I}\sigma_{-}^{k}-\frac{1}{2}\left\{\sigma_{+}^{k}\sigma_{-}^{k},\mathcal{O}_{I}\right\}\right)+\gamma^{k}_{-}\left(\sigma_{-}^{k}\mathcal{O}\sigma^{k}_{+}-\frac{1}{2}\left\{\sigma_{-}^{k}\sigma_{+}^{k},\mathcal{O}_{I}\right\}\right),
\end{eqnarray}
with $k\in\{h,c\}$. Within this model, we have
\begin{equation}\label{supp_app_capital_Gamma_k}
    \mathcal{D}_k^\dag[V_0] = - \Gamma_k V_0\quad \text{where} \quad \Gamma_k = (\gamma^k_+ + \gamma^k_-)/2\quad\text{with}\quad k\in\{h,c\}.
\end{equation}
The expectation value of 
$\mathcal{L}^\dag[V_0]$ with respect to the stationary state 
$\rho^s_I$ (where stationarity means $\mathcal{L}[\rho^s_I]=0$) can be expressed as
\begin{equation}
   \langle \mathcal{L}^\dag[V_0] \rangle 
   = \Tr\left(\mathcal{L}^{\dagger}[V_0]\,\rho^s_I\right)
   = \Tr\left(V_0\,\mathcal{L}[\rho^s_I]\right) 
   = -(\Gamma_h + \Gamma_c)\langle V_0 \rangle 
   = 0,
\end{equation}
where in the last step we used Eq.~\eqref{supp_app_capital_Gamma_k} and the fact that $[H_0,V_0]=0$. In this particular scenario, the definitions of the stationary heat currents $J_h$ and $J_c$, given in Table~\ref{supp_tab:discrete_continuous1}, reduce to
\begin{eqnarray}\label{supp_app_Jh_Jc_Otto_like}
    J_h = \sum_i \Delta \varepsilon_{i}^h \langle \dot N \rangle = \Delta \varepsilon^h \langle \dot N \rangle=\omega_h\langle \dot N \rangle\quad\text{and}\quad
    J_c = \sum_i \Delta \varepsilon_{i}^c \langle \dot N \rangle = \Delta \varepsilon^c \langle \dot N \rangle=-\omega_c\langle \dot N \rangle,
\end{eqnarray}
where we substitute the value $\Delta \varepsilon^h$ and $\Delta \varepsilon^c$ from Eq. \eqref{supp_app_delta_varepsilon_next0} and Eq. \eqref{supp_app_delta_varepsilon_next1}. From $J_h$ and $J_c$, we can calculate power as 
\begin{equation}\label{supp_app_power_Otto_like}
    P=J_h+J_c=\left(\omega_h-\omega_c\right)\langle \dot N \rangle,
\end{equation}
as given in Table \ref{supp_tab:discrete_continuous1}. It is straightforward to see that the efficiency is given by
\begin{equation}
    \eta = \frac{P}{J_h} = 1 + \frac{J_c}{J_h} = 1 - \frac{\omega_c}{\omega_h}.
\end{equation}
Next, we calculate the stationary values of the heat currents and the power of the engine by evaluating $\langle \dot N \rangle$. Let us introduce the operator
\begin{equation}\label{supp_Defn_of_Ndot}
    \dot N := ig\left(\ketbra{u}{d}-\ketbra{d}{u}\right)
    = ig\left(\sigma_-^h\sigma_+^c-\sigma_+^h\sigma_-^c\right),
\end{equation}
whose expectation value in the stationary state $\rho^s_{I}$ yields $\langle \dot N \rangle$ i.e.,
\begin{equation}
    \langle \dot N \rangle = ig \Tr[(\sigma_-^h\sigma_+^c - \text{Herm. Conj.}) \rho_I^s ].
\end{equation}
The value of $\langle \dot N \rangle$ can be obtained by evaluating the expectation value of $\mathcal{L}^\dag[\dot N]$ with respect to the stationary state $\rho^s_I$, starting from Eq.~\eqref{supp_app_HEOM}:
\begin{align}
    \langle \mathcal{L}^\dag [\dot N] \rangle 
    &= i \langle [V_0, \dot N] \rangle  
       + \langle \mathcal{D}_h^\dag[\dot N] \rangle 
       + \langle \mathcal{D}_c^\dag[\dot N] \rangle \notag \\
    &= 2 g^2 (p_{10} - p_{01}) 
       - (\Gamma_h + \Gamma_c) \langle \dot N \rangle = 0,
\end{align}
where we have used the fact that the expectation value of $\mathcal{L}^\dag [\dot N]$ with respect to the stationary state $\rho^s_{I}$ vanishes, i.e., $\langle \mathcal{L}^\dag [\dot N] \rangle = 0$. Here, the stationary value of occupation probabilities are defined as
\begin{align}
    p_{ij} &= \Tr[\dyad{ij}\,\rho_I^s] 
    \equiv \langle \Pi_{i,j} \rangle \quad\text{where}\quad \Pi_{i,j}=\ketbra{ij}{ij}.
\end{align}
Then, a closed set of equations can be obtained by imposing the stationary conditions $\langle \mathcal{L}^\dag[\Pi_{i,j}] \rangle = 0$:
\begin{align}
    \langle\mathcal{L}^{\dagger}[\Pi_{0,0}]\rangle&=-(\gamma^h_+ + \gamma^c_+) p_{00} + \gamma^h_- p_{10} +\gamma^c_- p_{01} = 0,\\
     \langle\mathcal{L}^{\dagger}[\Pi_{1,1}]\rangle &=-(\gamma^h_- + \gamma^c_-) p_{11} + \gamma^h_+ p_{01} +\gamma^c_+ p_{10} = 0,\\
      \langle\mathcal{L}^{\dagger}[\Pi_{0,1}]\rangle&=-(\gamma^h_+ + \gamma^c_-) p_{01} + \gamma^h_- p_{11} +\gamma^c_+ p_{00} + \langle \dot N \rangle = 0, \\
     \langle\mathcal{L}^{\dagger}[\Pi_{1,0}]\rangle&=-(\gamma^h_- + \gamma^c_+) p_{10} + \gamma^c_- p_{11} +\gamma^h_+ p_{00} - \langle \dot N \rangle = 0,
\end{align}
which, upon solving, yields
\begin{equation}
    \langle \dot N \rangle = \frac{\gamma^c_- \Gamma_h - \gamma^h_- \Gamma_c}{(\Gamma_c+\Gamma_h) \left(1 + \frac{\Gamma_c \Gamma_h}{g^2}\right)}.
\end{equation}
For a thermal bath satisfying the detailed balance condition $\gamma_+^k /\gamma_-^k = e^{-\beta_k  \omega_k}$ with $k\in\{h,c\}$, this can be rewritten as:
\begin{equation}\label{supp_app_expression_dotN}
    \langle \dot N \rangle = \frac{2\Gamma_h \Gamma_c}{(\Gamma_h + \Gamma_c)} \frac{\Delta p}{\left(1+\frac{ \Gamma_h \Gamma_c}{g^2}\right)}  =\frac{ 2\Gamma_h \Gamma_c(a_h - a_c)}{(1+ a_h)(1+ a_c)(\Gamma_h + \Gamma_c)\left(1+\frac{ \Gamma_h \Gamma_c}{g^2}\right)},
\end{equation}
where $\Delta p$ is the change in probability for the discrete version of this engine as given in Eq. \eqref{supp_app_Eq_deltap}. Substituting the value of $\langle \dot N \rangle $ from Eq. \eqref{supp_app_expression_dotN} to Eq. \eqref{supp_app_Jh_Jc_Otto_like} and Eq. \eqref{supp_app_power_Otto_like} we have
\begin{eqnarray}
    J_h &=& \omega_h \left(\frac{ 2\Gamma_h \Gamma_c(a_h - a_c)}{(1+ a_h)(1+ a_c)(\Gamma_h + \Gamma_c)\left(1+\frac{ \Gamma_h \Gamma_c}{g^2}\right)}\right)\label{supp_app_Jh_otto_like},\\
    J_c &=& -\omega_c \left(\frac{ 2\Gamma_h \Gamma_c(a_h - a_c)}{(1+ a_h)(1+ a_c)(\Gamma_h + \Gamma_c)\left(1+\frac{ \Gamma_h \Gamma_c}{g^2}\right)}\right)\label{supp_app_Jc_otto_like},\\
    P &=& J_h+J_c= \left(\omega_h-\omega_c\right) \left(\frac{2 \Gamma_h \Gamma_c(a_h - a_c)}{(1+ a_h)(1+ a_c)(\Gamma_h + \Gamma_c)\left(1+\frac{ \Gamma_h \Gamma_c}{g^2}\right)}\right)\label{supp_app_Power_otto_like}.
\end{eqnarray}

\subsection{Calculation of characteristics time}

We compute the characteristic time of the engine by taking the ratio of the work per cycle for the discrete Otto-like engine [Eq.~\eqref{supp_app_work_per_cycle_eff_otto_like}] to the power of the continuous Otto-like engine [Eq.~\eqref{supp_app_Power_otto_like}]. Specifically,  
\begin{align}
    P = \frac{W}{\tau_{\rm Otto}} 
    \quad \Longleftrightarrow \quad 
    \tau_{\rm Otto} = \frac{W}{P} = \frac{\Delta p}{\langle \dot N \rangle},
\end{align}
where \(\Delta p\) denotes the change in probability for the discrete engine [Eq.~\eqref{supp_app_Eq_deltap}] and \(\langle \dot N \rangle\) represents the probability current [Eq.~\eqref{supp_app_expression_dotN}]. Substituting these expressions yields  
\begin{equation}
    \tau_{\rm Otto} = \left(1 + \frac{\Gamma_h \Gamma_c}{g^2}\right) 
           \left(\frac{\Gamma_h + \Gamma_c}{2 \Gamma_h \Gamma_c}\right).
\end{equation}
We can express the characteristic time $\tau_{\rm Otto}$ using the relaxation time:
\begin{equation}
     \tau_k^{\rm eq}=\frac{2}{\gamma^k_{+}+\gamma^k_{-}}=\frac{1}{\Gamma_k},\quad \text{where} \quad k\in\{h,c\}.
\end{equation}
That gives:
\begin{eqnarray}
    \tau_{\rm Otto} 
    = \frac{\tau_h^{\rm eq} + \tau_c^{\rm eq}}{2} 
      \left(1 + \frac{1}{g^2 \tau_h^{\rm eq} \tau_c^{\rm eq}} \right).
\end{eqnarray}
\section{Discrete and continuous Otto-like engine with qubit catalyst}

\subsection{Discrete Otto-like engine with qubit catalyst}

Let us now analyze the discrete Otto like engine with a qubit catalyst. The initial density matrix is given by the product of state of the catalyst, the hot and the cold system's Gibbs states, i.e., 
\begin{equation}
    \rho = \rho_s\otimes\frac{e^{-\beta H_{0,h}}}{\Tr[e^{-\beta H_{0,h}}]} \otimes \frac{e^{-\beta H_{0,c}}}{\Tr[e^{-\beta H_{0,c}}]},
\end{equation}
with a two-level system Hamiltonian:
\begin{equation}
    H_{0,k} =  \omega_k \dyad{1}_k, \quad \text{with}\quad k\in\{h,c\},
\end{equation}
and the state of the catalyst
\begin{equation}
    \rho_s=p\ketbra{1}{1}+(1-p)\ketbra{2}{2}.
\end{equation}
Therefore, the combined state $\rho$ can be written as 
\begin{eqnarray}\label{supp_Eq_total_state_with_catalyst}
    \rho= &&\frac{p}{\mathcal{Z}_h \mathcal{Z}_c} \Big(\ketbra{100} + a_h \ketbra{110}+ a_c \ketbra{101} + a_ha_c \ketbra{111} \Big)\nonumber\\&+&\frac{(1-p)}{\mathcal{Z}_h \mathcal{Z}_c} \Big(\ketbra{200} + a_h \ketbra{210}+ a_c \ketbra{201} + a_ha_c \ketbra{211} \Big).
\end{eqnarray}
where $\mathcal{Z}_k = 1 + a_k$ and $a_k=e^{-\beta_k\omega_k}$ as introduced in Eq. \eqref{supp_defn:gibbs_factor}). % We consistently use the ordering \(|s,h,c\rangle\) to represent the state of the Otto-like engines with catalyst, maintaining this fixed order throughout. 
As argued in the main text, since the Hamiltonian of the catalyst does not affect the relevant thermodynamic quantities such as heat and work, we take the catalyst Hamiltonian to be trivial. 

Now, consider a permutation operator \( S \), composed of two swap operations acting on disjoint subspaces of the vector space on which \( \rho \) is supported. It  is defined as follows:
\begin{eqnarray}\label{supp_app_defn_u1_d1_u2_d2}
    &&\ket{u_1}=\ket{200}\xleftrightarrow{S}\ket{110}=\ket{d_1}\\&&\ket{u_2}=\ket{101}\xleftrightarrow{S}\ket{210}=\ket{d_2},
\end{eqnarray}
and acts as an identity on the rest of the basis vectors. The transferred probabilities $\Delta p_1$ and $\Delta p_2$ can be evaluated from their definitions, introduced in the main text, as follows:
\begin{eqnarray}
    \Delta p_1&=&p_{u_1}-p_{d_1}=\langle u_1|\rho|u_1\rangle-\langle d_1|\rho|d_1\rangle=\frac{1}{\mathcal{Z}_h\mathcal{Z}_c}\left[(1-p)-pa_h\right],\label{supp_app_delta_p1}\\
    \Delta p_2&=&p_{u_2}-p_{d_2}=\langle u_2|\rho|u_2\rangle-\langle d_2|\rho|d_2\rangle=\frac{1}{\mathcal{Z}_h\mathcal{Z}_c}\left[pa_c-(1-p)a_h\right]\label{supp_app_delta_p2}.
\end{eqnarray}
Next, we would like to obtain the value of $p$ from Eq. \eqref{supp_app_delta_p1} and Eq. \eqref{supp_app_delta_p2} employing the catalysis condition. Recall that the general form of the catalysis condition, as described in the main text, is given by (See also table \ref{supp_tab:discrete_continuous1}):
\begin{equation}\label{supp_cyclicity_disc}
    \forall m,\quad \sum_i \Delta \lambda_i^m \Delta p_i = 0, 
\end{equation}
where $\Delta \lambda_i^m = \lambda_{u_i}^m - \lambda_{d_i}^m$,
\begin{align}
    \lambda_{u_i}^m = \bra{u_i} \Pi_m \ket{u_i}, \quad \lambda_{d_i}^m = \bra{d_i} \Pi_m \ket{d_i},
\end{align}
and $\Pi_m$ denotes the $m^{\text{th}}$ eigenprojector of the catalyst. 

In this example, the catalyst is a qubit, and therefore $m = 1,2$. Inspecting the eigendecomposition of $\rho$ in Eq.~\eqref{supp_Eq_total_state_with_catalyst}, the projectors onto the first and second eigenspaces of the catalyst can be expressed as
\begin{eqnarray}
    \Pi_1 &:=& \ketbra{100} + \ketbra{101} + \ketbra{110} + \ketbra{111}\\
    \Pi_2 &:=& \ketbra{200} + \ketbra{201} + \ketbra{210} + \ketbra{211}.
\end{eqnarray}
This will allow us to calculate 
\begin{eqnarray}
    \Delta\lambda^{1}_{1}&=&\lambda^1_{u_1}-\lambda^1_{d_1}=\langle u_1|\Pi_1|u_1\rangle-\langle d_1|\Pi_1|d_1\rangle=-1\quad;\quad \Delta\lambda^{1}_{2}=\lambda^1_{u_2}-\lambda^1_{d_2}=\langle u_2|\Pi_1|u_2\rangle-\langle d_2|\Pi_1|d_2\rangle= 1\label{supp_app_Delta_p_eq1}\\
    \Delta\lambda^{2}_{1}&=&\lambda^2_{u_1}-\lambda^2_{d_1}=\langle u_1|\Pi_2|u_1\rangle-\langle d_1|\Pi_2|d_1\rangle=1\quad;\quad \Delta\lambda^{2}_{2}=\lambda^2_{u_2}-\lambda^2_{d_2}=\langle u_2|\Pi_2|u_2\rangle-\langle d_2|\Pi_2|d_2\rangle= -1\label{supp_app_Delta_p_eq2}.
\end{eqnarray}
Using Eq.~\eqref{supp_app_delta_p1}, Eq.~\eqref{supp_app_delta_p2}, Eq.~\eqref{supp_app_Delta_p_eq2}, and Eq.~\eqref{supp_app_Delta_p_eq1}, the catalysis condition in Eq.~\eqref{supp_cyclicity_disc} simplifies to
\begin{eqnarray}
    \Delta\lambda^{1}_{1}\Delta p_1+\Delta\lambda^{1}_{2}\Delta p_2&=&-\Delta p_1+\Delta p_2=0\quad\Longleftrightarrow \Delta p_1=\Delta p_2,\label{supp_catalysis_Eq_final1}\\
    \Delta\lambda^{2}_{1}\Delta p_1+\Delta\lambda^{2}_{2}\Delta p_2&=&\Delta p_1-\Delta p_2=0\quad\Longleftrightarrow \Delta p_1=\Delta p_2\label{supp_catalysis_Eq_final2}.
\end{eqnarray}
Note that the above Eq. \eqref{supp_catalysis_Eq_final1} and Eq. \eqref{supp_catalysis_Eq_final2} are equivalent. Thus, either equation can be used to determine the value of $p$. Using the explicit form of $\Delta p_1$ and $\Delta p_2$ from Eq. \eqref{supp_app_delta_p1} and Eq. \eqref{supp_app_delta_p2}, we can solve Eq. \eqref{supp_catalysis_Eq_final1} and obtain $p$ as follows:
\begin{equation}
   \Delta p_1=\Delta p_2 \quad\Longleftrightarrow\quad \frac{1}{\mathcal{Z}_h\mathcal{Z}_c}\left[(1-p)-pa_h\right]=\frac{1}{\mathcal{Z}_h\mathcal{Z}_c}\left[pa_c-(1-p)a_h\right]\quad\Longrightarrow\quad p=\frac{1+a_h}{1+2a_h+a_c}\label{supp_app_form_of_p}.
\end{equation}
Substituting the obtained expression of $p$ from Eq. \eqref{supp_app_form_of_p} into Eq. \eqref{supp_app_delta_p1}, gives 
\begin{equation}\label{supp_app_delta_p_cat}
    \Delta p_1=\frac{1}{\mathcal{Z}_h\mathcal{Z}_c}\left[\frac{(a_h+a_c)}{1+2a_h+a_c}-\frac{(1+a_h)a_h}{1+2a_h+a_c}\right]=\frac{1}{\mathcal{Z}_h\mathcal{Z}_c}\left[\frac{a_c-a_h^2}{1+2a_h+a_c}\right]=\frac{1}{(1+a_h)(1+a_c)}\left[\frac{a_c-a_h^2}{1+2a_h+a_c}\right]=\Delta p_2,
\end{equation}
where to write the final equality, we use the fact $\Delta p_1=\Delta p_2$ from Eq. \eqref{supp_catalysis_Eq_final1}.
Let us denote
\begin{equation}\label{supp_app_delta_p_cat_def}
    \frac{1}{\mathcal{Z}_h\mathcal{Z}_c}\left[\frac{a_c-a_h^2}{1+2a_h+a_c}\right]=\frac{1}{(1+a_h)(1+a_c)}\left[\frac{a_c-a_h^2}{1+2a_h+a_c}\right]:=\Delta p.
\end{equation}

Employing the expression for $\Delta p$ from Eq.~\eqref{supp_app_delta_p_cat_def}, we calculate the heats $Q_h$ and $Q_c$ using their definition given in Table \ref{supp_tab:discrete_continuous1}. In order to do so we first compute
\begin{eqnarray}\label{supp_app_delta_varepsilon}
    \Delta \varepsilon_1^h&=&\left(\varepsilon_{u_1}^h-\varepsilon_{d_1}^h\right)=\Big(\langle u_1|H_{0,h}|u_1\rangle-\langle d_1|H_{0,h}|d_1\rangle\Big)=\Big(\langle 200|H_{0,h}|200\rangle-\langle 110|H_{0,h}|210\rangle\Big)=-\omega_h,\\
    \Delta \varepsilon_2^h&=&\left(\varepsilon_{u_2}^h-\varepsilon_{d_2}^h\right)=\Big(\langle u_2|H_{0,h}|u_2\rangle-\langle d_2|H_{0,h}|d_2\rangle\Big)=\Big(\langle 101|H_{0,h}|101\rangle-\langle 210|H_{0,h}|210\rangle\Big)=-\omega_h,\\
    \Delta \varepsilon_1^c&=&\left(\varepsilon_{u_1}^c-\varepsilon_{d_1}^c\right)=\Big(\langle u_1|H_{0,c}|u_1\rangle-\langle d_1|H_{0,c}|d_1\rangle\Big)=\Big(\langle 200|H_{0,c}|200\rangle-\langle 110|H_{0,c}|210\rangle\Big)=0,\\
    \Delta \varepsilon_2^c&=&\left(\varepsilon_{u_2}^c-\varepsilon_{d_2}^c\right)=\Big(\langle u_2|H_{0,c}|u_2\rangle-\langle d_2|H_{0,c}|d_2\rangle\Big)=\Big(\langle 101|H_{0,c}|101\rangle-\langle 210|H_{0,c}|210\rangle\Big)=\omega_c.
\end{eqnarray}
Using the fact $\Delta p_1=\Delta p_2=\Delta p$, we have
\begin{eqnarray}
    Q_h&=&\sum_{i=1}^2 \Delta \varepsilon_i^h \Delta p_i= -\omega_h \Delta p_1- \omega_h \Delta p_2=-\omega_h \Delta p- \omega_h \Delta p=-2\omega_h\Delta p=\frac{2\omega_h}{(1+a_h)(1+a_c)}\left[\frac{a_h^2-a_c}{1+2a_h+a_c}\right],\label{supp_app:Expression_of_Q_h_cat}\\
    Q_c&=&\sum_{i=1}^2 \Delta \varepsilon_i^c \Delta p_i= \omega_c \Delta p_2=\omega_c\Delta p=\frac{-\omega_c}{(1+a_h)(1+a_c)}\left[\frac{a_h^2-a_c}{1+2a_h+a_c}\right].
    \label{supp_app:Expression_of_Q_c_cat}
\end{eqnarray}
Finally, the work and efficiency of the engine are obtained by substituting $Q_h$ and $Q_c$ from Eqs.~\eqref{supp_app:Expression_of_Q_h_cat} and \eqref{supp_app:Expression_of_Q_c_cat} into their respective definitions, yielding
\begin{equation}\label{supp_app_Work__eff_discrete}
    W = Q_h + Q_c = \left(2\omega_h - \omega_c\right)\Delta p=\frac{(2\omega_h-\omega_c)}{(1+a_h)(1+a_c)}\left[\frac{a_h^2-a_c}{1+2a_h+a_c}\right], 
    \qquad 
    \eta = 1 + \frac{Q_c}{Q_h} = 1 - \frac{\omega_c}{2\omega_h}.
\end{equation}

\subsection{Continuous Otto-like engine with qubit catalyst}
\subsubsection{Description and deriving a complete set of equations.}
The Hamiltonian of the continuous Otto engine with a qubit catalyst is given by
\begin{eqnarray}\label{supp_interaction_qubit}
H_0&=&H_{0,h}+H_{0,c}=\omega_h\sigma^h_{-}\sigma^h_{+}+\omega_c\sigma^c_{-}\sigma^c_{+},\\
    V_t&=&g\left(e^{-i\Omega_1t}\ketbra{u_1}{d_1}+e^{-i\Omega_2 t}\ketbra{u_2}{d_2}+\text{Herm. Conj.}\right),%=g\Big(e^{i\omega_ht}\ketbra{2}{1}\otimes\sigma^h_{+}\otimes\sigma^c_{+}\sigma^c_{-}+e^{i\Omega t}\ketbra{1}{2}\otimes\sigma^h_{+}\otimes\sigma^c_{-}+ \text{Herm. Conj.}\Big),
\end{eqnarray}
where we recall
\begin{eqnarray}
    \ket{u_1} &=& \ket{200}\quad;\quad \ket{d_1} = \ket{110},\\
    \ket{u_2} &=& \ket{101}\quad;\quad \ket{d_2} = \ket{210}.
\end{eqnarray}
from Eq. \eqref{supp_app_defn_u1_d1_u2_d2}. 
\begin{comment}
Using the values of $\varepsilon^k_i$ from Eq.~\eqref{supp_app_delta_varepsilon}, with $k \in \{h,c\}$ and $i=1,2$, we obtain
\begin{eqnarray}
    \Omega_1 &=& \Delta \varepsilon^h_1 + \Delta \varepsilon^c_1 = -\omega_h,\\
    \Omega_2 &=& \Delta \varepsilon^h_2 + \Delta \varepsilon^c_2 = -\omega_h + \omega_c.
\end{eqnarray}
\end{comment}
As in the continuous non-catalytic Otto-like engine, the time evolution in the interaction picture is governed by the local master equation:
\begin{equation}
    \mathcal{L}[\rho] := \frac{d\rho}{dt} = -i[V_0,\rho_I(t)] + \mathcal{D}_h[\rho_I(t)] + \mathcal{D}_c[\rho_I(t)],
\end{equation}
where the interaction Hamiltonian $V_0$ and dissipated part $ \mathcal{D}_{k}(\rho)$ is now given as
\begin{align}
    V_0 &= g\Big(\ketbra{u_1}{d_1}+\ketbra{u_2}{d_2}+\ketbra{d_1}{u_1}+\ketbra{d_2}{u_2}\Big),\quad\text{with}\\
     \mathcal{D}_{k}(\rho)&=\gamma_{+}^{k}\left(\sigma_{-}^{k}\rho\sigma_{+}^{k}-\frac{1}{2}\left\{\sigma_{+}^{k}\sigma_{-}^{k},\rho\right\}\right)+\gamma^{k}_{+}\left(\sigma_{+}^{k}\rho\sigma^{k}_{-}-\frac{1}{2}\left\{\sigma_{-}^{k}\sigma_{+}^{k},\rho\right\}\right)\quad\text{where}\quad k\in\{h,c\}.
\end{align}
The local master equation governing the evolution of an observable $\mathcal{O}_I$ in the interaction picture follows from the Heisenberg equation of motion:
\begin{align}
    \frac{d\mathcal{O}_I}{dt}&=i[V_0,\mathcal{O}_I]+\mathcal{D}_h^{\dagger}(\mathcal{O}_I)+\mathcal{D}_c^{\dagger}(\mathcal{O}_I):=\mathcal{L}^{\dagger}[\mathcal{O}_I]\quad\text{with},\\
    \mathcal{D}^{\dagger}_{k}(\mathcal{O}_I)&=\gamma_{+}^{k}\left(\sigma_{+}^{k}\mathcal{O}_I\sigma_{-}^{k}-\frac{1}{2}\left\{\sigma_{+}^{k}\sigma_{-}^{k},\mathcal{O}_I\right\}\right)+\gamma^{k}_{+}\left(\sigma_{-}^{k}\mathcal{O}_I\sigma^{k}_{+}-\frac{1}{2}\left\{\sigma_{-}^{k}\sigma_{+}^{k},\mathcal{O}_I\right\}\right)\quad\text{where}\quad k\in\{h,c\}.
\end{align}
Let us denote the projector onto the eigenvector of $H_0$ as
\begin{eqnarray}
     \Pi_{j,0,0}&:=&\ketbra{j}{j}\otimes\ketbra{0}{0}\otimes\ketbra{0}{0}=\ketbra{j}{j}\otimes\sigma^h_{+}\sigma^h_{-}\otimes \sigma^c_{+}\sigma^c_{-},\label{supp_defn:Pik00_rep}\\
    \Pi_{j,0,1}&:=&\ketbra{j}{j}\otimes\ketbra{0}{0}\otimes\ketbra{1}{1}=\ketbra{j}{j}\otimes\sigma^h_{+}\sigma^h_{-}\otimes\sigma^c_{-}\sigma^c_{+},\label{supp_defn:Pik01_rep}\\
    \Pi_{j,1,0}&:=&\ketbra{j}{j}\otimes\ketbra{1}{1}\otimes\ketbra{0}{0}=\ketbra{j}{j}\otimes\sigma^h_{-}\sigma^h_{+}\otimes\sigma^c_{+}\sigma^c_{-},\label{supp_defn:Pik10_rep}\\
    \Pi_{j,1,1}&:=&\ketbra{j}{j}\otimes\ketbra{1}{1}\otimes\ketbra{1}{1}=\ketbra{j}{j}\otimes\sigma^h_{-}\sigma^h_{+}\otimes\sigma^c_{-}\sigma^c_{+},\label{supp_defn:Pik11_rep}
\end{eqnarray}
with $j\in\{1,2\}$. We further introduce a set of operators that encapsulate the transition of the probability current.
\begin{eqnarray}
    \dot {N_1}&:=&ig\Big(\ketbra{u_1}{d_1}-\ketbra{d_1}{u_1}\Big)=ig\Big(\ketbra{200}{110}-\ketbra{110}{200}\Big)=ig\Big(\ketbra{2}{1}\otimes\sigma^h_{+}\otimes\sigma^c_{+}\sigma^c_{-}-\ketbra{1}{2}\otimes\sigma^h_{-}\otimes\sigma^c_{+}\sigma^c_{-}\Big),\label{supp_defn:X_k+100k10_rep}\\
%    X_{2,0,1:1,1,0}&:=&\ketbra{2}{1}\otimes\ketbra{0}{1}\otimes\ketbra{1}{0}-\ketbra{1}{2}\otimes\ketbra{1}{0}\otimes\ketbra{0}{1}=\ketbra{2}{1}\otimes\sigma^h_{+}\otimes\sigma^c_{-}-\ketbra{1}{2}\otimes\sigma^h_{-}\otimes\sigma^c_{+}\label{supp_defn:X_k+101k10_rep}\\
     \dot {N_2}&:=&ig\Big(\ketbra{u_2}{d_2}-\ketbra{d_2}{u_2}\Big)=ig\Big(\ketbra{101}{210}-\ketbra{210}{101}\Big)=ig\Big(\ketbra{1}{2}\otimes\sigma^h_{+}\otimes\sigma^c_{-}-\ketbra{2}{1}\otimes\sigma^h_{-}\otimes\sigma^c_{+}\Big),\label{supp_defn:X_101d10_rep}\\
    X&:=&ig\left(\ketbra{111}{201}-\ketbra{201}{111}\right)=ig\left(\ketbra{1}{2}\otimes\sigma^h_{-}\otimes\sigma^c_{-}\sigma^c_{+}-\ketbra{2}{1}\otimes\sigma^h_{+}\otimes\sigma^c_{-}\sigma^c_{+}\right).\label{supp_defn:X_k11k+101_rep}
\end{eqnarray}
This leads to the following set of equations, expressed in terms of the projectors $\Pi_{j,k,l}$ onto the eigenvectors of $H_0$ and the operators $\dot{N}_1$, $\dot{N}_2$, and $X$:
\begin{eqnarray}
    \mathcal{L}^{\dagger}[\Pi_{1,0,0}]&=&\frac{d\Pi_{1,0,0}}{dt}= -(\gamma^h_{+}+\gamma^c_{+})\Pi_{1,0,0}+\gamma^h_{-}\Pi_{1,1,0}+\gamma^c_{-}\Pi_{1,0,1}\label{supp_qubit100},\\
    \mathcal{L}^{\dagger}[\Pi_{2,0,0}]&=&\frac{d\Pi_{2,0,0}}{dt}= -\dot N_1-(\gamma^h_{+}+\gamma^c_{+})\Pi_{2,0,0}+\gamma^h_{-}\Pi_{2,1,0}+\gamma^c_{-}\Pi_{2,0,1},\label{supp_qubit200}\\
    \mathcal{L}^{\dagger}[\Pi_{1,0,1}]&=&\frac{d\Pi_{1,0,1}}{dt}= -\dot N_2-(\gamma^h_{+}+\gamma^c_{-})\Pi_{1,0,1}+\gamma^h_{-}\Pi_{1,1,1}+\gamma^c_{+}\Pi_{1,0,0},\label{supp_qubit101}\\
    \mathcal{L}^{\dagger}[\Pi_{2,0,1}]&=&\frac{d\Pi_{2,0,1}}{dt}= -(\gamma^h_{+}+\gamma^c_{-})\Pi_{2,0,1}+\gamma^h_{-}\Pi_{2,1,1}+\gamma^c_{+}\Pi_{2,0,0}\label{supp_qubit201},\\
    \mathcal{L}^{\dagger}[\Pi_{1,1,0}]&=&\frac{d\Pi_{1,1,0}}{dt}= \dot N_1-(\gamma^h_{-}+\gamma^c_{+})\Pi_{1,1,0}+\gamma^h_{+}\Pi_{1,0,0}+\gamma^c_{-}\Pi_{1,1,1},\label{supp_qubit110}\\
    \mathcal{L}^{\dagger}[\Pi_{2,1,0}]&=&\frac{d\Pi_{2,1,0}}{dt}= \dot N_2-(\gamma^h_{-}+\gamma^c_{+})\Pi_{2,1,0}+\gamma^h_{+}\Pi_{2,0,0}+\gamma^c_{-}\Pi_{2,1,1},\label{supp_qubit210}\\
    \mathcal{L}^{\dagger}[\Pi_{1,1,1}]&=&\frac{d\Pi_{1,1,1}}{dt}= -(\gamma^h_{-}+\gamma^c_{-})\Pi_{1,1,1}+\gamma^h_{+}\Pi_{1,0,1}+\gamma^c_{+}\Pi_{1,1,0}\label{supp_qubit111},\\
    \mathcal{L}^{\dagger}[\Pi_{2,1,1}]&=&\frac{d\Pi_{2,1,1}}{dt}= -(\gamma^h_{-}+\gamma^c_{-})\Pi_{2,1,1}+\gamma^h_{+}\Pi_{2,0,1}+\gamma^c_{+}\Pi_{2,1,0}\label{supp_qubit211},\\
    \mathcal{L}^{\dagger}[\dot N_2]&=&\frac{d\dot N_2}{dt}=-2g^2\left(\Pi_{2,1,0}-\Pi_{1,0,1}\right)-\frac{1}{2}\left(\gamma^h_{-}+\gamma^h_{+}+\gamma^c_{-}+\gamma^c_{+}\right)\dot{N_2},\label{supp_X2}\\
    \mathcal{L}^{\dagger}[\dot N_1]&=&\frac{d\dot N_1}{dt}=-2g^2\left(\Pi_{1,1,0}-\Pi_{2,0,0}\right)-\frac{1}{2}\left(\gamma^h_{-}+\gamma^h_{+}+2\gamma^c_{+}\right)\dot{N_1}-\gamma^c_{-}X,\label{supp_X1}\\
    \mathcal{L}^{\dagger}[X]&=&\frac{dX}{dt}=-\frac{1}{2}\left(\gamma^h_{-}+\gamma^h_{+}+2\gamma^c_{-}\right)X-\gamma^c_{+}\dot{N_1}\label{supp_X3}.
\end{eqnarray}
We proceed by observing that, upon adding Eq.~\eqref{supp_qubit100}, Eq.~\eqref{supp_qubit101}, Eq.~\eqref{supp_qubit110}, and Eq.~\eqref{supp_qubit111}, we obtain  
\begin{equation}\label{supp_firstlevel_of_catalyst}
     \mathcal{L}^{\dagger}[\Pi_1] = \frac{d\Pi_{1}}{dt} 
     = \frac{d\Pi_{1,0,0}}{dt} + \frac{d\Pi_{1,0,1}}{dt} 
     + \frac{d\Pi_{1,1,0}}{dt} + \frac{d\Pi_{1,1,1}}{dt} 
     = -\dot{N}_2 + \dot{N}_1,
\end{equation}
where $\Pi_{1}$ denotes the projector onto the first level of the catalyst, namely  
\begin{equation}
    \Pi_1 = \Pi_{1,0,0} + \Pi_{1,0,1} + \Pi_{1,1,0} + \Pi_{1,1,1}.
\end{equation}

We see that dissipative terms cancel each other out because the dissipators, \(\mathcal{D}_h\) and \(\mathcal{D}_c\), act solely on the hot and cold qubits, leaving the catalyst unaffected. Since for the stationary state we have $\mathcal{L}[\rho_I^s]=0$, therefore the expectation of any operator of the form $\mathcal{L}^{\dagger}[A]$ with respect to the steady state $\rho_I^s$ vanishes i.e.,
\begin{equation}
    \Tr\left(\mathcal{L}^{\dagger}[A]\rho_I^s\right)= \Tr\left(A\mathcal{L}[\rho_I^s]\right)=0.
\end{equation}
Therefore, using Eq. \eqref{supp_firstlevel_of_catalyst}, we can write 
\begin{equation}\label{supp_firstlevel_of_catalyst_expectation}
    \Tr\left(\mathcal{L}^{\dagger}[\Pi_1]\rho_I^s\right)= 0 \quad \Longleftrightarrow \quad \langle \dot N_2\rangle = \langle \dot N_1\rangle:=\langle \dot N\rangle,
\end{equation}
where we denote the expectation value of $\dot N_1$ and $\dot N_2$ at the stationary state $\rho^s_I$ by $\langle \dot N \rangle$. We recall from the main text that heat currents and the power in the stationary state can be expressed as (Please see Table \ref{supp_tab:discrete_continuous1})
\begin{equation}\label{supp_app_Jk_P} 
 J_k = \sum_i \Delta \varepsilon_{i}^k \langle \dot N_i \rangle\quad\quad P =J_h+J_c= \sum_i \Omega_i \langle \dot N_i \rangle\quad \text{where}\quad \Omega_i=\Delta \varepsilon_{i}^h+\Delta \varepsilon_{i}^c \quad \text{and}\quad k\in\{h,c\}.
\end{equation}
Using the relation $\langle \dot N_1 \rangle = \langle \dot N_2 \rangle = \langle \dot N \rangle$ from Eq. \eqref{supp_firstlevel_of_catalyst_expectation}, in this scenario, the stationary currents and the power given in Eq. \eqref{supp_app_Jk_P} reduce to 

\begin{eqnarray}
    J_h&=&\Delta\varepsilon^h_{1}\langle \dot {N}_1 \rangle+\Delta\varepsilon^h_{2}\langle \dot {N}_2 \rangle= \Delta\varepsilon^h_{1}\langle \dot {N} \rangle+\Delta\varepsilon^h_{2}\langle \dot {N} \rangle=-2\omega_h\langle \dot {N} \rangle\label{supp_power_heat_currents1}\\
    J_c&=&\Delta\varepsilon^c_{1}\langle \dot {N}_1 \rangle+\Delta\varepsilon^c_{2}\langle \dot {N}_2 \rangle= \Delta\varepsilon^c_{1}\langle \dot {N} \rangle+\Delta\varepsilon^c_{2}\langle \dot {N} \rangle=\omega_c\langle \dot {N} \rangle\label{supp_power_heat_currents2}\\
    P&=& J_h+J_c=\left(-2\omega_h+\omega_c\right)\langle \dot N\rangle\label{supp_power_heat_currents3}.
\end{eqnarray}
The efficiency can then be expressed as  
\begin{equation}
    \eta = \frac{P}{J_h} = 1 + \frac{J_c}{J_h}=1+\frac{\omega_c\langle \dot {N} \rangle}{-2\omega_h\langle \dot {N} \rangle} = 1 - \frac{\omega_c}{2\omega_h}.
\end{equation}
To evaluate the stationary heat currents $J_h$, $J_c$, and the power, we have to determine $\langle \dot N \rangle$. In order to do so, we aim to obtain a complete set of equations whose solution will give us $\langle \dot N \rangle$. In order to do so we first take the expectation of Eq. \eqref{supp_X3} at the stationary state $\rho^s_I$, which gives
\begin{eqnarray}
    &&\langle\mathcal{L}^{\dagger}[X]\rangle=\Tr(\mathcal{L}^{\dagger}[X]\rho^s_I)=0=-\frac{1}{2}\left(\gamma^h_{-}+\gamma^h_{+}+2\gamma^c_{-}\right)\langle X\rangle-\gamma^c_{+}\langle\dot{N_1}\rangle\\&\Longrightarrow&\quad \langle X\rangle=\frac{-2\gamma^c_{+}\;\langle\dot{N_1}\rangle}{\left(\gamma^h_{-}+\gamma^h_{+}+2\gamma^c_{-}\right)}=\frac{-2\gamma^c_{+}\;\langle\dot{N}\rangle}{\left(\gamma^h_{-}+\gamma^h_{+}+2\gamma^c_{-}\right)}\label{supp_app_X_in_terms_of_N1},
\end{eqnarray}
where to write the final equality, we use Eq. \eqref{supp_firstlevel_of_catalyst_expectation}.
Taking the expectation value of $\mathcal{L}^{\dagger}[\dot N_1]$ as given in Eq. \eqref{supp_X1}, with respect to stationary state $\rho_I^s$, we have
\begin{eqnarray}
    \langle\mathcal{L}^{\dagger}[\dot N_1]\rangle=\Tr(\mathcal{L}^{\dagger}[N_1]\rho^s_I)=0&=&-2g^2\left(p_{110}-p_{200}\right)-\frac{1}{2}\left(\gamma^h_{-}+\gamma^h_{+}+2\gamma^c_{+}\right)\langle\dot{N_1}\rangle-\gamma^c_{-}\langle X\rangle\\
    &=& -2g^2\left(p_{110}-p_{200}\right)-\frac{1}{2}\left(\gamma^h_{-}+\gamma^h_{+}+2\gamma^c_{+}\right)\langle\dot{N}\rangle-\gamma^c_{-}\langle X\rangle\label{supp_app_expectation_value_X},
\end{eqnarray}
where we denote expectation of $\Pi_{i,j,k}$ with respect to $\rho^s_I$:
\begin{equation}\label{supp_app_expectation_p_ijk}
    \langle \Pi_{i,j,k} \rangle := p_{ijk},
\end{equation}
and employ the fact $\langle \dot{N}_1 \rangle=\langle \dot{N}_2 \rangle=\langle \dot{N}\rangle$ from Eq. \eqref{supp_firstlevel_of_catalyst_expectation} to obtain Eq. \eqref{supp_app_expectation_value_X}.
Substituting $\langle X\rangle$ from Eq. \eqref{supp_app_X_in_terms_of_N1} to Eq. \eqref{supp_app_expectation_value_X}, we get
\begin{equation}\label{supp_app_iANdot}
    0=-2g^2\left(p_{110}-p_{200}\right)-\frac{1}{2}\underbrace{\left(\gamma^h_{-}+\gamma^h_{+}+2\gamma^c_{+}-\frac{4\gamma^c_{-}\gamma^c_{+}}{\left(\gamma^h_{-}+\gamma^h_{+}+2\gamma^c_{-}\right)}\right)}_{:=A}\langle \dot N\rangle = -2g^2\left(p_{110}-p_{200}\right)-\frac{A}{2}\langle \dot N\rangle.
\end{equation}

Finally, taking the expectation of Eqs.~\eqref{supp_qubit100}--\eqref{supp_X2} with respect to the stationary state $\rho^{s}_{I}$, together with Eq.~\eqref{supp_app_iANdot} we obtain the following complete set of equations:
\begin{eqnarray}
   \langle\mathcal{L}^{\dagger}[\Pi_{1,0,0}]\rangle&=& -(\gamma^h_{+}+\gamma^c_{+})p_{100}+\gamma^h_{-}p_{110}+\gamma^c_{-}p_{101}=0\label{supp_qubit100_exp},\\
    \langle\mathcal{L}^{\dagger}[\Pi_{2,0,0}]\rangle&=& -\langle \dot{N}\rangle-(\gamma^h_{+}+\gamma^c_{+})p_{200}+\gamma^h_{-}p_{210}+\gamma^c_{-}p_{201}=0,\label{supp_qubit200_exp}\\
    \langle\mathcal{L}^{\dagger}[\Pi_{1,0,1}]\rangle&=& -\langle \dot{N}\rangle-(\gamma^h_{+}+\gamma^c_{-})p_{101}+\gamma^h_{-}p_{111}+\gamma^c_{+}p_{100}=0,\label{supp_qubit101_exp}\\
    \langle\mathcal{L}^{\dagger}[\Pi_{2,0,1}]\rangle&=& -(\gamma^h_{+}+\gamma^c_{-})p_{201}+\gamma^h_{-}p_{211}+\gamma^c_{+}p_{200}\label{supp_qubit201_exp},\\
    \langle\mathcal{L}^{\dagger}[\Pi_{1,1,0}]\rangle&=& \langle \dot{N}\rangle-(\gamma^h_{-}+\gamma^c_{+})p_{110}+\gamma^h_{+}p_{100}+\gamma^c_{-}p_{111}=0,\label{supp_qubit110_exp}\\
    \langle\mathcal{L}^{\dagger}[\Pi_{2,1,0}]\rangle&=& \langle \dot{N}\rangle-(\gamma^h_{-}+\gamma^c_{+})p_{210}+\gamma^h_{+}p_{200}+\gamma^c_{-}p_{211}=0,\label{supp_qubit210_exp}\\
    \langle\mathcal{L}^{\dagger}[\Pi_{1,1,1}]\rangle&=& -(\gamma^h_{-}+\gamma^c_{-})p_{111}+\gamma^h_{+}p_{101}+\gamma^c_{+}p_{110}=0\label{supp_qubit111_exp},\\
    \langle\mathcal{L}^{\dagger}[\Pi_{2,1,1}]\rangle&=& -(\gamma^h_{-}+\gamma^c_{-})p_{211}+\gamma^h_{+}p_{201}+\gamma^c_{+}p_{210}=0\label{supp_qubit211_exp},\\
    \langle\mathcal{L}^{\dagger}[\dot N_2]\rangle&=&-2g^2\left(p_{210}-p_{101}\right)-\frac{1}{2}\underbrace{\left(\gamma^h_{-}+\gamma^h_{+}+\gamma^c_{-}+\gamma^c_{+}\right)}_{:=B}\langle\dot N\rangle=-2g^2\left(p_{210}-p_{101}\right)-\frac{B}{2}\langle \dot N\rangle,\label{supp_X2_exp}\\
    && -2g^2\left(p_{110}-p_{200}\right)-\frac{A}{2}\langle \dot N\rangle = 0,\label{supp_X1_exp}
\end{eqnarray}
where we use the fact $\langle \dot{N}_1 \rangle=\langle \dot{N}_2 \rangle=\langle \dot{N} \rangle$ from Eq. \eqref{supp_firstlevel_of_catalyst_expectation}, in writing Eq. \eqref{supp_qubit200_exp}, Eq. \eqref{supp_qubit101_exp}, Eq. \eqref{supp_qubit110_exp}, Eq. \eqref{supp_qubit210_exp} and Eq. \eqref{supp_X2_exp}. The solution of the above set of equations Eq. \eqref{supp_qubit100_exp}-Eq. \eqref{supp_X1_exp} give us $\langle \dot N\rangle$.

\subsubsection{Calculation of the stationary current and the power}
In this section, we compute the stationary heat currents $J_h$, $J_c$, and the power output $P$ of the engine by calculating $\langle\dot N\rangle$. 
\begin{comment}
Here, we have used 
\begin{equation}
    \Omega_1=\Delta\varepsilon^h_{1}+\Delta\varepsilon^c_{1}=-\omega_h\quad;\quad \Omega_2=\Delta\varepsilon^h_{2}+\Delta\varepsilon^c_{2}=-\omega_h+\omega_c,
\end{equation}
with  $\Delta\varepsilon^{k}_i$ from  Eq. \eqref{supp_app_delta_varepsilon}. 
\end{comment}
Let us express $p_{111}$ and $p_{211}$ using Eq.~\eqref{supp_qubit111_exp} and Eq.~\eqref{supp_qubit211_exp}, respectively, as follows:
\begin{comment}
by calculating the expectation value 
of Eq. \eqref{supp_X3} with respect to $\rho^{s}_{I}$ and using Eq. \eqref{supp_firstlevel_of_catalyst_expectation} which gives the following 
\begin{eqnarray}\label{supp_Eq:form_of_x3}
     0=-\frac{1}{2}\left(\gamma^h_{-}+\gamma^h_{+}+2\gamma^c_{+}\right)\langle X_{1,1,1:2,0,1}\rangle-\gamma^c_{-}x\quad\Rightarrow\quad  \langle X_{1,1,1:2,0,1}\rangle=\frac{-2\gamma^c_{-}x}{\left(\gamma^h_{-}+\gamma^h_{+}+2\gamma^c_{+}\right)}\label{supp_X3_exp}.
\end{eqnarray}
\end{comment}
\begin{eqnarray}\label{supp_p_111_and_p_211}
    p_{111}= \frac{\gamma^h_{+}p_{101}+\gamma^c_{+}p_{110}}{\gamma^c_{-}+\gamma^h_{-}} \quad\text{and}\quad p_{211}= \frac{\gamma^h_{+}p_{201}+\gamma^c_{+}p_{210
    }}{\gamma^c_{-}+\gamma^h_{-}}.
\end{eqnarray}
Substituting the expression of $p_{111}$ and $p_{211}$ from  Eq. \eqref{supp_p_111_and_p_211} to Eq. \eqref{supp_qubit101_exp} and Eq. \eqref{supp_qubit201_exp} which allows us to write $p_{101}$ and $p_{201}$  as follows:
\begin{eqnarray}
    p_{101}=\frac{\left(\gamma^c_{-}+\gamma^h_{-}\right)\Big(\gamma^c_{+}p_{100}-\langle \dot N\rangle\Big)+\gamma^c_{+}\gamma^h_{-}p_{110}}{\gamma^c_{-}\left(\gamma^c_{-}+\gamma^h_{-}+\gamma^h_{+}\right)}\quad\text{and}\quad p_{201}= \frac{\gamma^c_{+}\left(\gamma^c_{-}+\gamma^h_{-}\right)p_{200}+\gamma^c_{+}\gamma^h_{-}p_{210}}{\gamma^c_{-}\left(\gamma^c_{-}+\gamma^h_{-}+\gamma^h_{+}\right)}\label{supp_p201_in_terms_of_x_and_200}.
\end{eqnarray}
We proceed by recalling the detailed balance condition:
\begin{equation}
    \frac{\gamma^h_{+}}{\gamma^h_{-}}=e^{-\beta_h\omega_h}=a_h \quad  \quad \frac{\gamma^c_{+}}{\gamma^c_{-}}=e^{-\beta_c\omega_c}=a_c, 
\end{equation}
where $a_h$ and $a_c$ introduced in Eq. \eqref{supp_defn:gibbs_factor}. Next we substitute $p_{101}$ and $p_{201}$ from Eq. \eqref{supp_p201_in_terms_of_x_and_200} to Eq. \eqref{supp_qubit100_exp} and Eq. \eqref{supp_qubit200_exp} respectively, that gives $p_{100}$ and $p_{200}$ as follows:
\begin{eqnarray}
    p_{100}&=&\underbrace{\frac{\gamma^h_{-}}{\gamma^h_{+}}}_{=1/a_h}p_{110}-\langle \dot N\rangle\underbrace{\left(\frac{\gamma^c_{-}+\gamma^h_{-}}{\gamma^h_{+}(\gamma^c_{+}+\gamma^c_{-}+\gamma^h_{+}+\gamma^h_{-})}\right)}_{:=\alpha_1} = \frac{p_{110}}{a_h}-\langle \dot N\rangle\alpha_1,\label{supp_p100_in_terms_of_x_p_110}\\
    \quad p_{200}&=&\underbrace{\frac{\gamma^h_{-}}{\gamma^h_{+}}}_{1/a_h}p_{210}-\langle \dot N\rangle\underbrace{\left(\frac{\gamma^c_{-}+\gamma^h_{-}+\gamma^h_{+}}{\gamma^h_{+}(\gamma^c_{+}+\gamma^c_{-}+\gamma^h_{+}+\gamma^h_{-})}\right)}_{:=\alpha_2}=\frac{p_{210}}{a_h}-\langle \dot N\rangle\alpha_2\label{supp_p200_in_terms_of_x_p_210}.
\end{eqnarray}

Now, substituting $p_{100}$ and $p_{200}$ from Eq. \eqref{supp_p100_in_terms_of_x_p_110} and Eq. \eqref{supp_p200_in_terms_of_x_p_210} in Eq. \eqref{supp_p201_in_terms_of_x_and_200}, we obtain $p_{101}$ and $p_{201}$ as:
\begin{eqnarray}
    p_{101}&=& \underbrace{\frac{\gamma^c_{+}\gamma^h_{-}}{\gamma^c_{-}\gamma^h_{+}}}_{=a_c/a_h}p_{110}-\langle \dot N\rangle\underbrace{\left(\frac{(\gamma^c_{-}+\gamma^h_{-})(\gamma^c_{+}+\gamma^h_{+})}{\gamma^c_{-}\gamma^h_{+}(\gamma^c_{+}+\gamma^c_{-}+\gamma^h_{+}+\gamma^h_{-})}\right)}_{:=\phi_1}=\frac{a_c}{a_h}p_{110}-\langle \dot N\rangle\phi_1\label{supp_p101_in_terms_of_x_p_110}\quad\text{and}\\
    p_{201}&=& \underbrace{\frac{\gamma^c_{+}\gamma^h_{-}}{\gamma^c_{-}\gamma^h_{+}}}_{=a_c/a_h}p_{210}-\langle \dot N\rangle\underbrace{\left(\frac{\gamma^c_{+}(\gamma^c_{-}+\gamma^h_{-})}{\gamma^c_{-}\gamma^h_{+}(\gamma^c_{+}+\gamma^c_{-}+\gamma^h_{+}+\gamma^h_{-})}\right)}_{:=\phi_2}=\frac{a_c}{a_h}p_{210}-\langle \dot N\rangle\phi_2\label{supp_p201_in_terms_of_x_p_210}.
\end{eqnarray}
Next, substituting $p_{101}$ and $p_{201}$ from Eq. \eqref{supp_p101_in_terms_of_x_p_110} and Eq. \eqref{supp_p201_in_terms_of_x_p_210} to Eq. \eqref{supp_p_111_and_p_211} we get $p_{111}$ and $p_{211}$ as:
\begin{eqnarray}
    p_{111}&=&\underbrace{\frac{\gamma^c_{+}}{\gamma^c_{-}}}_{=a_c}p_{110}- \langle \dot N\rangle\underbrace{\left(\frac{\left(\gamma^c_{+}+\gamma^h_{+}\right)}{\gamma^c_{-}(\gamma^c_{+}+\gamma^c_{-}+\gamma^h_{+}+\gamma^h_{-})}\right)}_{=\xi_1} =a_cp_{110}-\langle \dot N\rangle\xi_1\label{supp_p111_in_terms_of_x_p_110} \quad\text{and}\quad\\
    p_{211} &=&  \underbrace{\frac{\gamma^c_{+}}{\gamma^c_{-}}}_{=a_c}p_{210}- \langle \dot N\rangle\underbrace{\left(\frac{\gamma^c_{+}}{\gamma^c_{-}(\gamma^c_{+}+\gamma^c_{-}+\gamma^h_{+}+\gamma^h_{-})}\right)}_{=\xi_2} =a_cp_{210}-\langle \dot N\rangle\xi_2\label{supp_p211_in_terms_of_x_p_210}. 
    \end{eqnarray}
Observe that substituting $p_{200}$ from Eq. \eqref{supp_p200_in_terms_of_x_p_210} to Eq. \eqref{supp_X1_exp} we obtain $p_{210}$ as:
\begin{equation}\label{supp_p210A}
    p_{210}=a_hp_{110}+\langle \dot N\rangle a_h\underbrace{\left(\alpha_2+\frac{A}{4g^2}\right)}_{:=\mathcal{G}_A}=a_hp_{110}+\langle \dot N\rangle a_h\mathcal{G}_A,
\end{equation}
whereas substituting  $p_{101}$ from Eq. \eqref{supp_p101_in_terms_of_x_p_110} to Eq. \eqref{supp_X2_exp}, we obtain $p_{210}$ as:
\begin{eqnarray}\label{supp_p210B}
    p_{210}=\frac{a_c}{a_h}p_{110}-\langle \dot N\rangle\underbrace{\left(\phi_1+\frac{B}{4g^2}\right)}_{:=\mathcal{G}_B}=\frac{a_c}{a_h}p_{110}-\langle \dot N\rangle\mathcal{G}_{B}.
\end{eqnarray}

Hence, equating Eq. \eqref{supp_p210A} and Eq. \eqref{supp_p210B} allows us to write $p_{110}$ in the following manner:
\begin{equation}\label{supp_semi_final_p110}
    p_{110}=\left(\frac{a_h(\mathcal{G}_{A}a_h+\mathcal{G}_{B})}{a_c-a_h^2}\right)\langle \dot N\rangle.
\end{equation}
Furthermore, substituting $p_{110}$ from Eq. \eqref{supp_semi_final_p110} to Eq. \eqref{supp_p210B} gives $p_{210}$ as 
\begin{eqnarray}\label{supp_semi_final_p210}
    p_{210}=\langle \dot N\rangle a_h\left(\frac{a_c\mathcal{G}_{A}+a_h\mathcal{G}_{B}}{a_c-a_h^2}\right).
\end{eqnarray}
Using the normalization condition, we can write: 
\begin{equation}\label{supp_normalization_qubit}
    p_{100}+p_{101}+p_{110}+p_{111}+p_{200}+p_{201}+p_{210}+p_{211}=1.
\end{equation}
Substituting $p_{100}$ from Eq.~\eqref{supp_p100_in_terms_of_x_p_110}, 
$p_{200}$ from Eq.~\eqref{supp_p200_in_terms_of_x_p_210}, 
$p_{101}$ from Eq.~\eqref{supp_p101_in_terms_of_x_p_110}, 
$p_{201}$ from Eq.~\eqref{supp_p201_in_terms_of_x_p_210}, 
$p_{111}$ from Eq.~\eqref{supp_p111_in_terms_of_x_p_110}, 
and $p_{211}$ from Eq.~\eqref{supp_p211_in_terms_of_x_p_210} into Eq.~\eqref{supp_normalization_qubit}, we obtain
\begin{equation}\label{supp_sum_of_p_110_and_p_210_Important1}
    p_{110}+p_{210}=\frac{1+\langle \dot N\rangle(\alpha_1+\phi_1+\xi_1+\alpha_2+\phi_2+\xi_2)}{\left(1+a_c+\frac{a_c}{a_h}+\frac{1}{a_h}\right)}.
\end{equation}

Finally, substituting the expression of $p_{110}$ from Eq. \eqref{supp_semi_final_p110} and $p_{210}$ from Eq. \eqref{supp_semi_final_p210} to Eq. \eqref{supp_sum_of_p_110_and_p_210_Important1}, we solve $\langle \dot N \rangle$ as 
\begin{eqnarray}
    \langle \dot N \rangle&=&\frac{-(a_h^2-a_c)}{(1+a_c)(1+a_h)(1+a_c+2a_h)\left[\left(\frac{a_c+a_h}{1+a_c+2a_h}\right)\mathcal{G}_A+\left(\frac{1+a_h}{1+a_c+2a_h}\right)\mathcal{G}_B+\left(\frac{\left(a_h^2-a_c\right)(\alpha_1+\phi_1+\xi_1+\alpha_2+\phi_2+\xi_2)}{(1+a_c)(1+a_h)(1+a_c+2a_h)}\right)\right]}\\
    &=&\frac{-(a_h^2-a_c)}{(1+a_c)(1+a_h)(1+a_c+2a_h)\left[\left(\frac{a_c+a_h}{1+a_c+2a_h}\right)\left(\alpha_2+\frac{A}{4g^2}\right)+\left(\frac{1+a_h}{1+a_c+2a_h}\right)\left(\phi_1+\frac{B}{4g^2}\right)+\left(\frac{\left(a_h^2-a_c\right)(\alpha_1+\phi_1+\xi_1+\alpha_2+\phi_2+\xi_2)}{(1+a_c)(1+a_h)(1+a_c+2a_h)}\right)\right]}\\
    &=& \frac{\Delta p}{\left[\left(\frac{a_c+a_h}{1+a_c+2a_h}\right)\left(\alpha_2+\frac{A}{4g^2}\right)+\left(\frac{1+a_h}{1+a_c+2a_h}\right)\left(\phi_1+\frac{B}{4g^2}\right)+\left(\frac{\left(a_h^2-a_c\right)(\alpha_1+\phi_1+\xi_1+\alpha_2+\phi_2+\xi_2)}{(1+a_c)(1+a_h)(1+a_c+2a_h)}\right)\right]}.
    \label{supp_station_x_value_qubit}
\end{eqnarray}
where we substitute $\mathcal{G}_A$ and $\mathcal{G}_B$ from Eq. \eqref{supp_p210A} and Eq. \eqref{supp_p210B} respectively, to write the second equality and we substitute the expression of probability flow $\Delta p$ from Eq. \eqref{supp_app_delta_p_cat_def}. Recall that we have introduced $A$ and $B$ in Eq. \eqref{supp_app_iANdot} and Eq. \eqref{supp_X2_exp} as a function of jump rates. Substituting the expression $\langle \dot N\rangle$ from Eq. \eqref{supp_station_x_value_qubit} to Eq. \eqref{supp_power_heat_currents1}, Eq. \eqref{supp_power_heat_currents2}, and Eq. \eqref{supp_power_heat_currents3}, we have 
\begin{eqnarray}
    J_h&=& \frac{2\omega_h(a_h^2-a_c)}{(1+a_c)(1+a_h)(1+a_c+2a_h)\left[\left(\frac{a_c+a_h}{1+a_c+2a_h}\right)\left(\alpha_2+\frac{A}{4g^2}\right)+\left(\frac{1+a_h}{1+a_c+2a_h}\right)\left(\phi_1+\frac{B}{4g^2}\right)+\left(\frac{\left(a_h^2-a_c\right)(\alpha_1+\phi_1+\xi_1+\alpha_2+\phi_2+\xi_2)}{(1+a_c)(1+a_h)(1+a_c+2a_h)}\right)\right]}\label{supp_app_Jh_cont},\\
    J_c&=& \frac{-\omega_c(a_h^2-a_c)}{(1+a_c)(1+a_h)(1+a_c+2a_h)\left[\left(\frac{a_c+a_h}{1+a_c+2a_h}\right)\left(\alpha_2+\frac{A}{4g^2}\right)+\left(\frac{1+a_h}{1+a_c+2a_h}\right)\left(\phi_1+\frac{B}{4g^2}\right)+\left(\frac{\left(a_h^2-a_c\right)(\alpha_1+\phi_1+\xi_1+\alpha_2+\phi_2+\xi_2)}{(1+a_c)(1+a_h)(1+a_c+2a_h)}\right)\right]}\label{supp_app_Jc_cont},\\
    P&=& \frac{(2\omega_h-\omega_c)(a_h^2-a_c)}{(1+a_c)(1+a_h)(1+a_c+2a_h)\left[\left(\frac{a_c+a_h}{1+a_c+2a_h}\right)\left(\alpha_2+\frac{A}{4g^2}\right)+\left(\frac{1+a_h}{1+a_c+2a_h}\right)\left(\phi_1+\frac{B}{4g^2}\right)+\left(\frac{\left(a_h^2-a_c\right)(\alpha_1+\phi_1+\xi_1+\alpha_2+\phi_2+\xi_2)}{(1+a_c)(1+a_h)(1+a_c+2a_h)}\right)\right]}\label{supp_app_power_cont}.
\end{eqnarray}

\subsection{Calculation of characteristics time}
In this section, we determine the characteristic time of the engine by taking the ratio between the work per cycle of the discrete Otto-like engine with a qubit catalyst (see Eq. \eqref{supp_app_Work__eff_discrete}) and the power output of the continuous Otto-like engine with a qubit catalyst (See Eq. \eqref{supp_app_power_cont}) i.e.,
\begin{eqnarray}
    \tau_{\rm Catalytic} &=& W/P=\Delta p/\langle \dot N\rangle\nonumber\\%\frac{1}{(1+a_c+2a_h)}\left[\left(1+\frac{a_h}{a_c}\right)\mathcal{G}_A+\left(1+a_h\right)\mathcal{G}_B+\frac{(\alpha_1+\phi_1+\xi_1+\alpha_2+\phi_2+\xi_2)(a_h^2-a_c)}{(1+a_c)(1+a_h)}\right]\\    
    %\frac{\left(\frac{\left(1+\frac{a_h}{a_c}\right)(A+2g^2\alpha_2)}{2}+\frac{(1+a_h)(B+2g^2\phi_1)}{2}+\frac{g^2(\alpha_1+\phi_1+\xi_1+\alpha_2+\phi_2+\xi_2)(a_c-a_h^2)}{(1+a_c)(1+a_h)}\right)}{g^2}\nonumber\\&=&
    &=&\left[\left(\frac{a_c+a_h}{1+a_c+2a_h}\right)\left(\alpha_2+\frac{A}{4g^2}\right)+\left(\frac{1+a_h}{1+a_c+2a_h}\right)\left(\phi_1+\frac{B}{4g^2}\right)+\left(\frac{\left(a_h^2-a_c\right)(\alpha_1+\phi_1+\xi_1+\alpha_2+\phi_2+\xi_2)}{(1+a_c)(1+a_h)(1+a_c+2a_h)}\right)\right].
\end{eqnarray}
Now we will write the characteristic time in terms of relaxation time with respect to hot and cold bath denoted as $\tau_h^{\rm eq}$ and $\tau_c^{\rm eq}$ where
\begin{equation}\label{supp_equality_above}
    \tau_k^{\rm eq}=\frac{2}{\gamma^k_{+}+\gamma^k_{-}}=\frac{1}{\Gamma_k},\quad \text{where} \quad k\in\{h,c\}.
\end{equation}
Using the above Eq. \eqref{supp_equality_above} we can obtain
\begin{equation}
    \gamma^k_{+}+\gamma^k_{-}=\frac{2}{\tau_k^{\rm eq}}\Longleftrightarrow  \gamma^k_{-}\left(1+\underbrace{\frac{\gamma^k_{+}}{\gamma^k_{-}}}_{=a_k}\right) = \gamma^k_{-}(1+a_k)=\frac{2}{\tau_k^{\rm eq}}\Longleftrightarrow \gamma^k_{-} = \frac{2}{\tau_k^{\rm eq}(1+a_k)}\quad \text{where} \quad k\in\{h,c\}.\label{supp_supp_Eq_gamma_k_m}
\end{equation}
Employing the fact $\gamma^k_{+}/\gamma^k_{-}=e^{-\beta_k\omega_k}=a_k$ , we can obtain from Eq. \eqref{supp_supp_Eq_gamma_k_m} 
\begin{equation}\label{supp_supp_Eq_gamma_k_p}
    \gamma^k_{+} = \frac{2a_k}{\tau_k^{\rm eq}(1+a_k)} \quad\text{for}\quad k\in\{h,c\}.
\end{equation}
Substituting the expression of $\gamma^k_{-}$ and $\gamma^k_{+}$ from Eq. \eqref{supp_supp_Eq_gamma_k_m} and Eq. \eqref{supp_supp_Eq_gamma_k_p}, we can simplify $\tau_{\rm Catalytic}$ as 
\begin{align}
    \tau_{\rm Catalytic}&=\frac{1}{2}\left[\left(\frac{1+ah(1+a_c+a_h)}{(1+a_h)(1+a_c+2a_h)}\right)\left(\frac{(\tau^{\rm eq}_c)^2}{\tau^{\rm eq}_c+\tau^{\rm eq}_h}\right)+\left(\frac{1+3a_h+a_c(3+a_h)}{(1+a_c)(1+a_c+2a_h)}\right)\left(\frac{(\tau^{\rm eq}_h)^2}{\tau^{\rm eq}_c+\tau^{\rm eq}_h}\right)+\left(\frac{2\tau^{\rm eq}_h\tau^{\rm eq}_c}{\tau^{\rm eq}_c+\tau^{\rm eq}_h}\right)\right]\nonumber\\&+\frac{1}{g^2}\left[\frac{1+a_h}{2(\tau_c^{\rm eq}+a_c\tau_c^{\rm eq}+2a_h\tau_c^{\rm eq})}+\frac{1}{2\tau_h^{\rm eq}}+\frac{a_c(a_c+a_h)}{(1+a_c+2a_h)(\tau_c^{\rm eq}+a_c\tau_c^{\rm eq}+2\tau_h^{\rm eq})}\right].
\end{align}
Considering a special case with $\tau^{\rm eq}_h=\tau^{\rm eq}_c=\tau_{\rm eq}$, we will have
\begin{equation}
    \tau_{\rm Catalytic} = \zeta\tau_{\rm eq}\left(1+\frac{\kappa}{g^2\tau_{\rm eq}^2}\right),
\end{equation}
where 
\begin{align}
    \zeta = \frac{1}{4}\left[3+\frac{1}{1+a_c}-\frac{1}{1+a_h}+\frac{1+3a_c}{(1+a_c)(1+a_c+2a_h)}\right]\;\;;\;\;
    \kappa =\frac{2(1+a_c)(1+a_h)\left[6+9a_h+a_c(5+3a_c+5a_h)\right]}{(3+a_c)\left[4+2a_c(4+a_c)+(1+a_c)(11+3a_c)a_h+2(4+3a_c)a_h^2\right]}.
\end{align}
Now, we will show that the characteristic time $\tau_{\rm Catalytic}$ is upper bounded by $\tau_{\rm Otto}$ i.e.,
\begin{equation}
   \tau_{\rm Catalytic} \leq \tau_{\rm Otto},
\end{equation}
which will highlight a catalytic improvement in the characteristic time. To prove this, it suffices to show that both $\zeta$ and $\kappa$ satisfy
\begin{equation}
\zeta \leq 1 
\quad \text{and} \quad 
\kappa \leq 1.
\end{equation}
To show $\zeta \leq 1$ we proceed by showing $1-\zeta\geq 0$:
\begin{eqnarray}
   1-\zeta=\frac{a_h+2a_ca_h(1+a_h)+a_c^2(2+a_h)}{4(1+a_c)(1+a_h)(1+a_c+2a_h)}\geq 0.
\end{eqnarray}
This gives $1\leq \zeta$. Next, we will show $\kappa \leq 1$ by showing $1-\kappa \geq 0$. We see that
\begin{equation}
    1-\kappa = \frac{\left(1-a_c\right)\left[2 (2 a_c+3) a_h^2+3 (a_c+1)^2 a_h+2 a_c (2 a_c+3)\right]}{(3+a_c)\left[4+2a_c(4+a_c)+(1+a_c)(11+3a_c)a_h+2(4+3a_c)a_h^2\right]}\geq 0, \quad\text{as}\quad 1\geq a_c.
\end{equation}

\end{document}